\documentclass[a4paper,aps,prd,singlecolumn,tightenlines,preprintnumbers,nofootinbib,showkeys,superscriptaddress,longbibliography,notitlepage,twocolumn]{revtex4-1}
\pdfoutput=1
\usepackage[dvipsnames]{xcolor}
\usepackage[export]{adjustbox}
\usepackage{tikz,contour} 
\usetikzlibrary{shapes,arrows,positioning,automata,backgrounds,calc,er,patterns}
\usepackage{fullpage}
\usepackage{relsize}
\usepackage[left=1.615cm,right=1.615cm,top=1.5cm,bottom=1.65cm]{geometry}
\usepackage{XCharter}
\usepackage{mathptmx}
\usepackage[T1]{fontenc}
\usepackage{pifont}
\usepackage{amsfonts,amsmath,amsthm,amssymb,amsbsy,bm,mathtools,latexsym,esvect}
\usepackage{enumitem}
\usepackage{hyperref}
\hypersetup{colorlinks=true,citecolor=red,linkcolor=NavyBlue,urlcolor=NavyBlue}
\usepackage[caption=false,labelformat=simple]{subfig}
 
\usepackage{natbib}

\makeatletter

\def\section#1{%
  \vspace{1.0\baselineskip}%
  \refstepcounter{section}%
  \noindent{\bfseries\Large #1}\par
  \vspace{0.5\baselineskip}%
}

\def\subsection#1{%
  \vspace{0.75\baselineskip}%
  \refstepcounter{subsection}%
  \noindent\textbf{\large #1}\par
  \vspace{0.35\baselineskip}%
}

\makeatother

\begin{document}

\title{\Large Kinematic budget of quantum correlations}
\author{Maaz Khan}
\email{maaz.khan@research.iiit.ac.in}
\affiliation{Center for Computational Natural Sciences and Bioinformatics, International Institute of Information Technology, Hyderabad 500 032, India}

\author{Subhadip Mitra}
\email{subhadip.mitra@iiit.ac.in}
\affiliation{Center for Computational Natural Sciences and Bioinformatics, International Institute of Information Technology, Hyderabad 500 032, India}
\affiliation{Center for Quantum Science and Technology, International Institute of Information Technology, Hyderabad 500 032, India}

\begin{abstract}
\noindent
The diversity of quantum correlations -- discord, entanglement, steering, and Bell nonlocality -- disappears at the kinematic level of observable second moments. By treating state purity as a finite resource, we introduce a local-unitary-invariant budget that splits these moments into local and nonlocal sectors. This maps quantum systems onto compact, two-dimensional manifolds whose topology is governed by purity and time-reversal symmetry. This dimensional reduction reveals a deep structural link: exceeding classical capacity limits requires the activation of intrinsically time-odd generators, providing a dimension-agnostic guarantee of negative partial transpose (NPT)  entanglement. For two qubits, this geometry is analytically solvable; a single boundary isolates classical correlations, while nested regions define thresholds for steering and Bell nonlocality, alongside bounds on non-stabiliser magic. Beyond two qubits, dimensional bottlenecks enforce the kinematic limits on correlations. Because this macroscopic representation is completely determined by global and marginal purities, it bypasses the exponential scaling of full-state tomography. Thus, whenever an $n$-partite state's correlations exceed the classical capacity limits, its NPT entanglement is certified by only $n+1$ purity measurements, with sample complexity independent of Hilbert space dimension. By coarse-graining over gauge-like first moments, this geometry acts as a thermodynamic phase diagram, exposing the hierarchy of quantum resources and their dynamic redistribution under decoherence. 
\end{abstract}
\maketitle

\begin{figure*}
    \centering
    \captionsetup[subfigure]{labelformat=empty}
    \subfloat[\qquad\quad{\bf a}]{
        \includegraphics[width=0.42\linewidth]{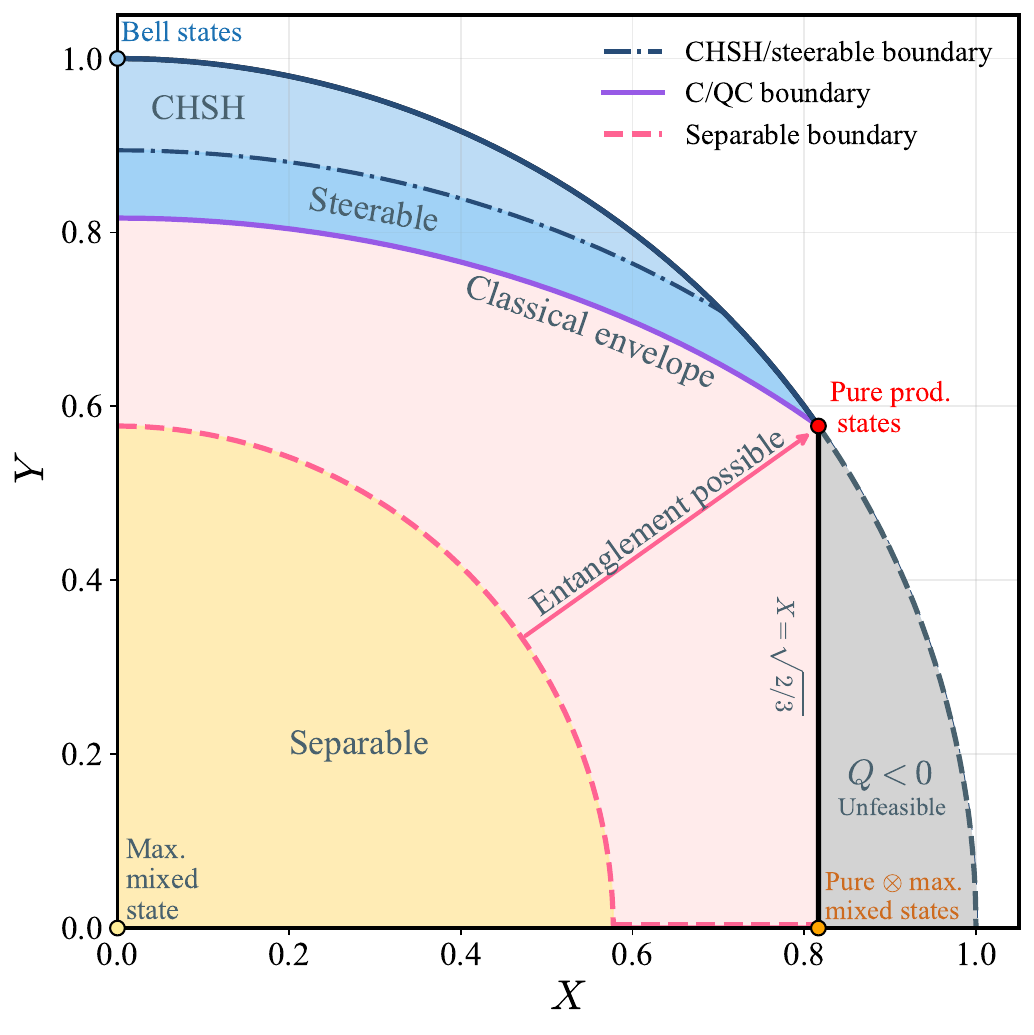}
        \label{fig:budgetXY_left}
    }\hspace{1cm}
    \subfloat[\qquad\quad{\bf b}]{
        \includegraphics[width=0.42\linewidth]{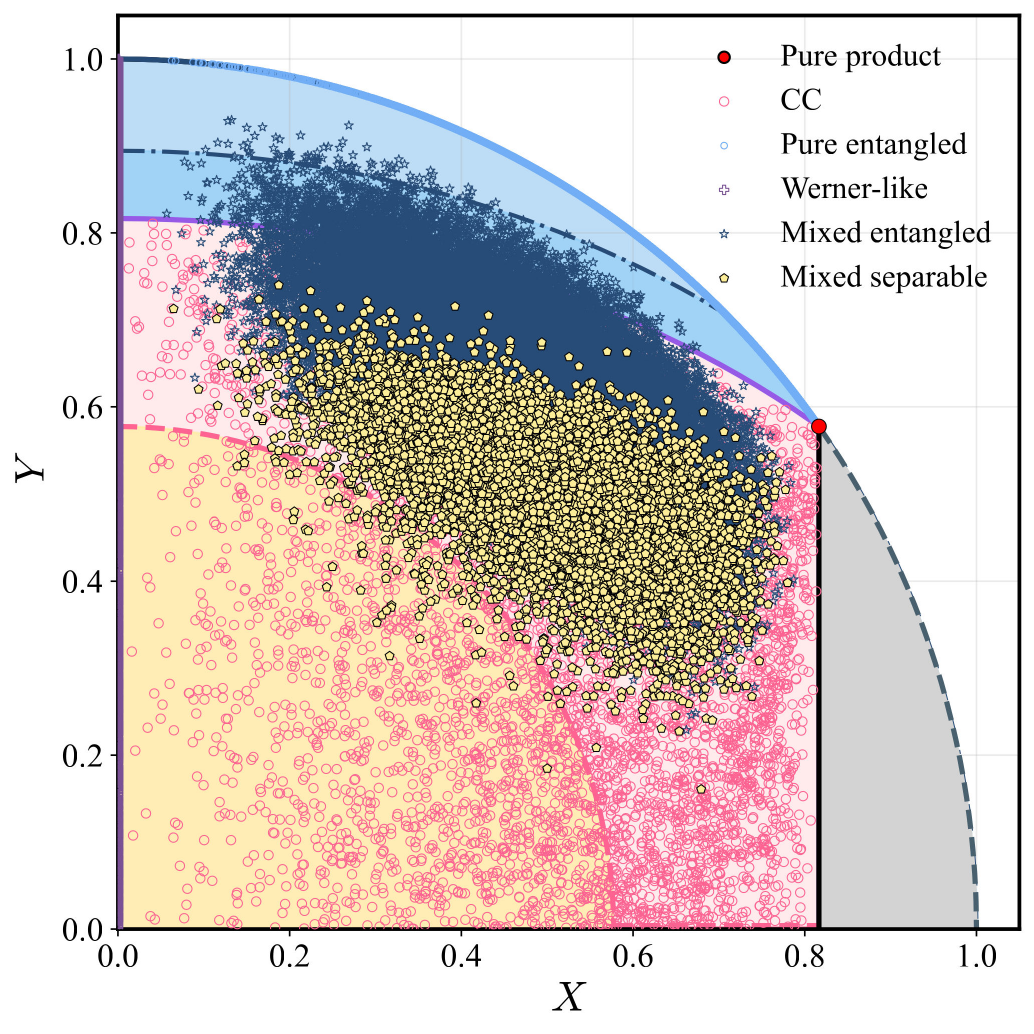}
        \label{fig:budgetXY_right}
    }
    \caption{\label{fig:budgetXY}{\bf Local basis-independent purity-budget geometry for two-qubit states.} ({\bf a}) The entire two-qubit state space, projected onto a single, compact geometric manifold where the kinematic bounds are exposed. The coordinates $X$ and $Y$ track the fundamental tug-of-war between local polarisation and nonlocal correlations. The grey region is unphysical, forbidden by the time-symmetric nature of quantum mechanics ($Q < 0$). Within the physical regime, exact analytic boundaries organise the space into the familiar correlation hierarchy. The classical envelope (purple curve) sets a hard capacity limit for all classically structured states; crossing this threshold guarantees nonclassicality, ensuring the state is both entangled and steerable. Additional curves delineate the threshold for guaranteed CHSH-Bell violations (blue dot-dashed) and the absolute separability limit (red dashed). ({\bf b}) Monte Carlo sampling of the state space. Distinct families of quantum states perfectly populate their analytically defined regions, confirming that this hole-free projection captures the entire physical space.}
\end{figure*}

\section{Introduction}
\noindent
Quantum correlations are typically described through separate frameworks for different quantum resources such as discord~\cite{Ollivier:2001fdq}, entanglement~\cite{Einstein:1935rr}, steering~\cite{Schrodinger:2008opj}, and Bell nonlocality~\cite{Bell:1964kc}, resulting in a fragmented picture even for two-qubit systems. We show that these apparently disparate structures unify at the kinematic level. Fundamentally, quantum mechanics is probabilistic and time-reversal symmetric. These two properties constrain observable correlations -- state purity ($P=\mathrm{tr}(\rho^2)$) imposes a conservation law on total second moments, and the overlap of a state with its time-reversed state ($Q=\mathrm{tr}(\rho\widetilde\rho)$) forces them into a rigid geometric structure. This unified structure is revealed through a unique two-dimensional projection of the state space onto local-unitary (LU)-invariant symmetric second moments. These are independent of local-basis choices and permutations of local subsystems. The correlation hierarchy appears naturally in the resulting geometry. Because second-moment quantities are quadratic functionals of the state, just as probabilities arise from squared amplitudes, they provide a natural language for experimentally accessible correlations. This change of perspective unifies quantum properties under a single geometric principle, revealing the kinematic action of noise on correlations and providing a compact diagnostic alternative to full-state tomography.

Purity quantifies the total second-moment content of a quantum state. It constitutes a finite resource budget that splits into a local contribution from all single-subsystem polarisations ($B_{\rm L}$) and a nonlocal contribution from net correlations ($B_{\rm NL}$). A fundamental trade-off exists between local and correlation degrees of freedom: for a fixed $P$, increasing the local allocation necessarily reduces the available nonlocal resources, and vice versa. This trade-off is quantified by two rationalised coordinates $(X,Y)$ that specify how the purity budget is distributed between local and nonlocal sectors (see Fig.~\ref{fig:budgetXY} for two-qubit systems). The second invariant, $Q$, establishes the positivity boundary of the feasible region; it constrains the allowed variation in second moments under time reflection. 

When quantum states are represented in $(B_{\rm L},B_{\rm NL})$ or $(X,Y)$ coordinates, they populate compact, hole-free regions in these LU-invariant planes. In this geometry, states admitting a fully classical description are bound by the classical (C) envelope. Systematically increasing the number of quantum subsystems generates a nested hierarchy of quantum-classical (QC) envelopes that define the capacity of symmetric generators at each tier. Crossing these limits necessitates the activation of asymmetric generators that are intrinsically odd under time reversal. Because local time reversal is equivalent to partial transposition (PT) in an LU-invariant framework, the asymmetric generators produce negative PT eigenvalues. The states above these boundaries are thus guaranteed to carry NPT entanglement across progressively more parties -- from bipartite NPT entanglement at the lowest tier up to genuine multipartite entanglement (GME) at the highest -- independently of their dimensions. While the invariants $P$ and $Q$ govern the boundaries universally, the geometry closes analytically for two qubits. Here, dimensional symmetry merges the classical and entanglement envelopes; crossing this single boundary guarantees both entanglement and steering. Consequently, this geometry maps the distinct regimes where Bell nonlocality is forbidden, permitted, or structurally guaranteed. The standard approach asks whether quantum behaviour is present; we invert this logic and ask instead how much correlation classical resources can support, and what must happen when they run out.

The representation is entirely macroscopic; for any physical system, its components are determined from the global and marginal purities alone. The coarse-graining exposes the kinematic constraints governing how noise and physical processes redistribute second moments. Open-system dynamics are trajectories traced on this plane. Decoherence typically destroys correlations, driving states along downward trajectories. Because pure states lie on the upper boundary, noise can probe the interior of the feasible region and validate the analytic limits of the budget. Furthermore, a state's position in this plane places strict bounds on quantum resources like discord and negativity, and constrains non-Clifford computational resources such as stabiliser R\'enyi-$2$ magic~\cite{Leone:2021rzd}.

The paper is structured as follows. Section~\ref{sec:II} introduces the budget geometry, mapping quantum systems onto the 2D budget plane. Section~\ref{sec:III} illustrates the geometric boundaries for quantum correlations and discord/entanglement enforcement; Section~\ref{sec:IV} shows the bounds on computational magic. Section~\ref{sec:V} explores how noise and system changes appear as geometric paths and outlines its use as a practical diagnostic tool, followed by a discussion in Sec.~\ref{sec:VI}.

\begin{figure*}
    \captionsetup[subfigure]{labelformat=empty}
    \centering
    \subfloat[{\bf a} Negativity]{

    \includegraphics[width=0.24\linewidth]{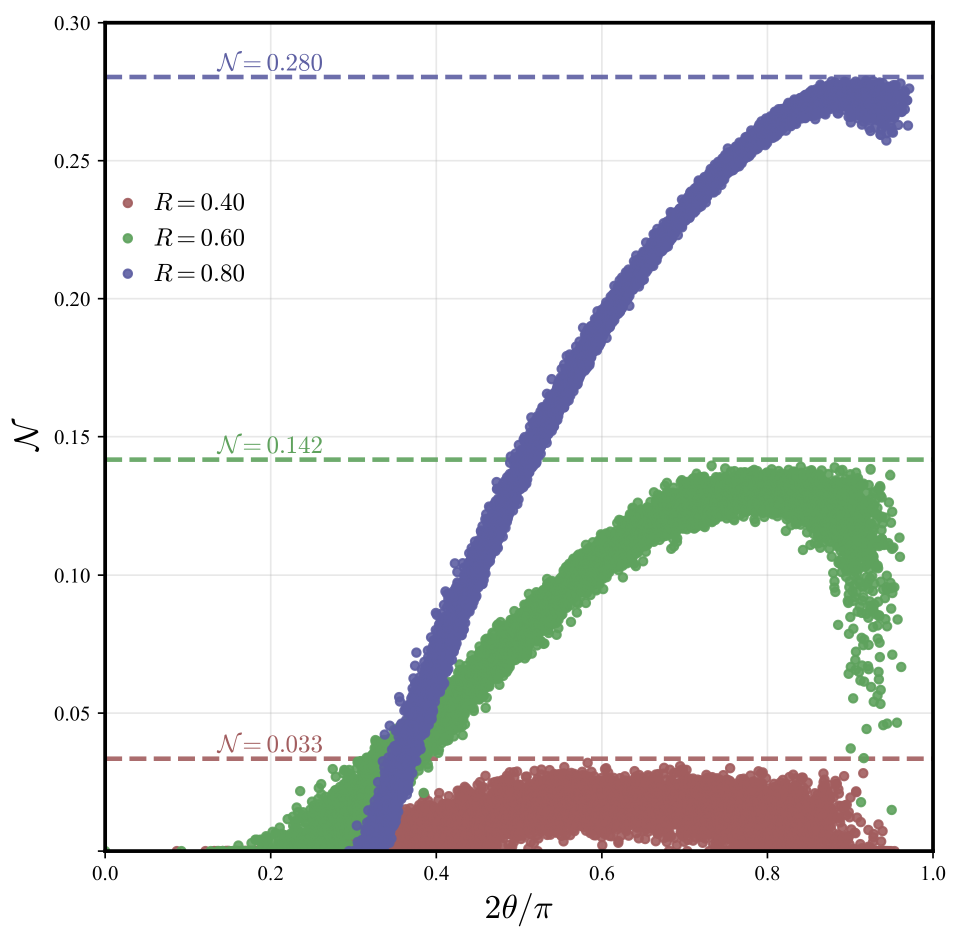}}\hfill
    \subfloat[{\bf b} Magic]{
    \includegraphics[width=0.24\linewidth]{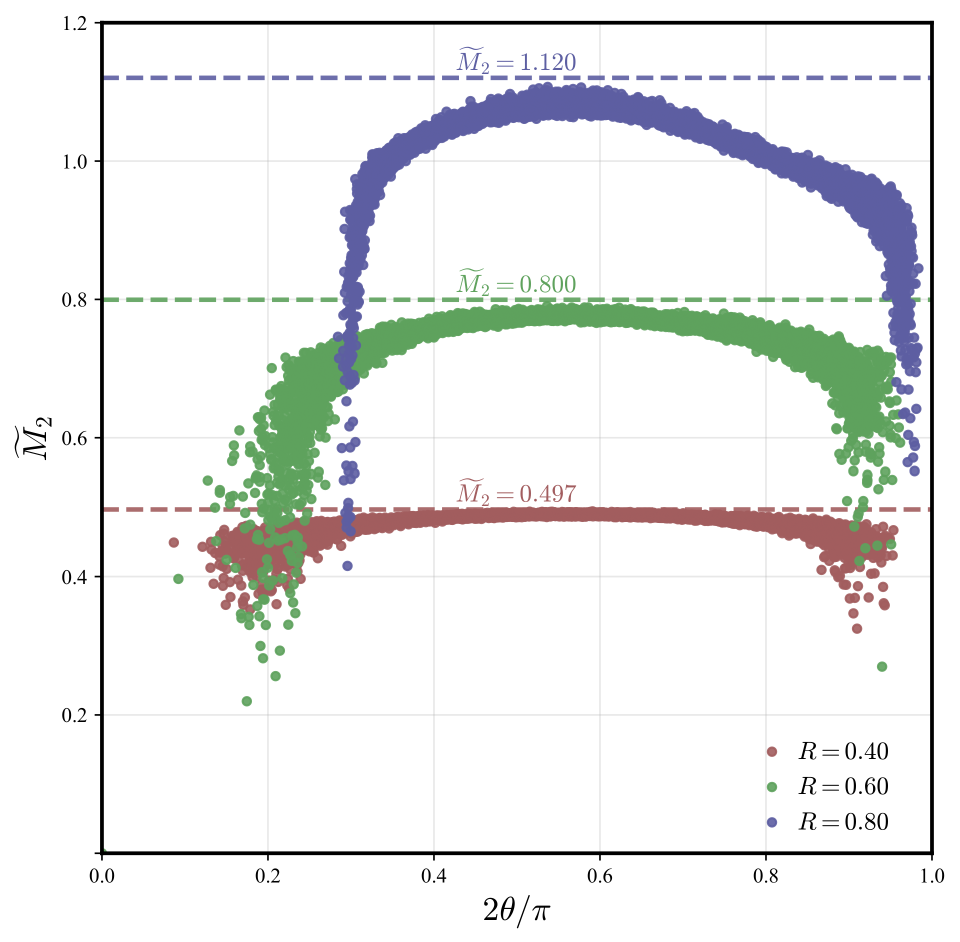}}\hfill
       \subfloat[{\bf c} $5$-qubit magic]{
    \includegraphics[width=0.24\linewidth]{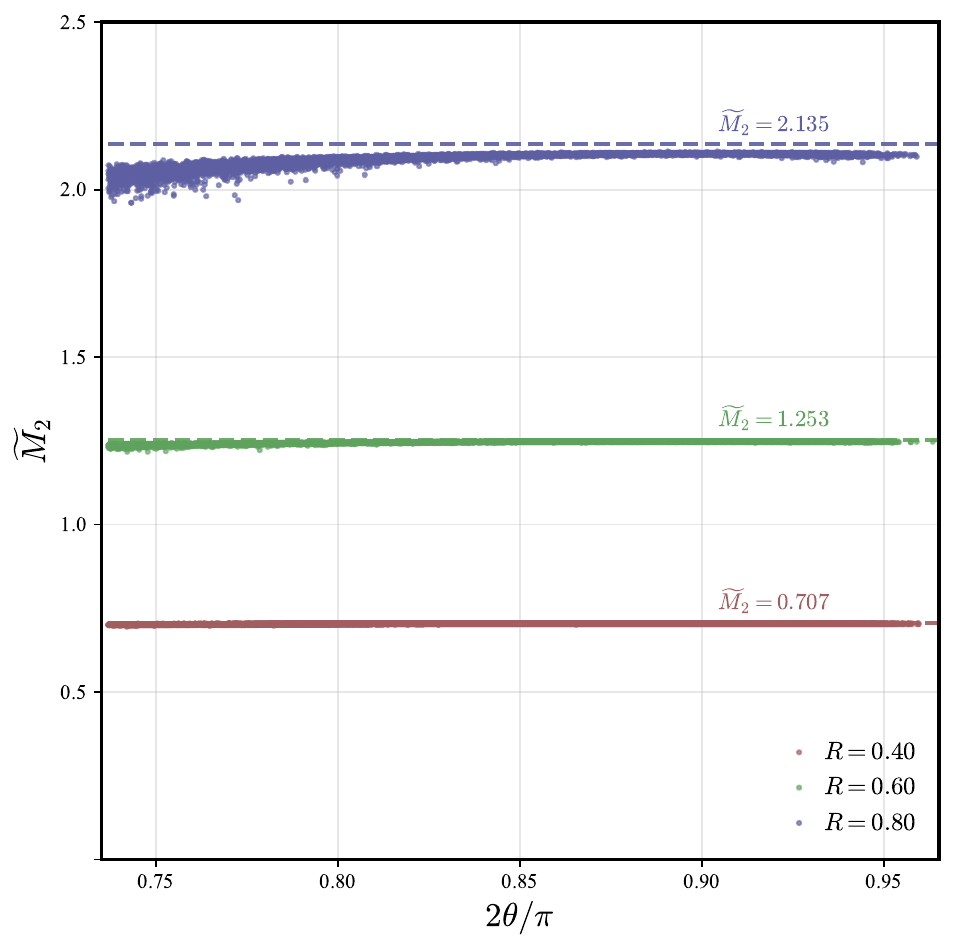}}\hfill
       \subfloat[{\bf d} $7$-qubit magic]{
    \includegraphics[width=0.24\linewidth]{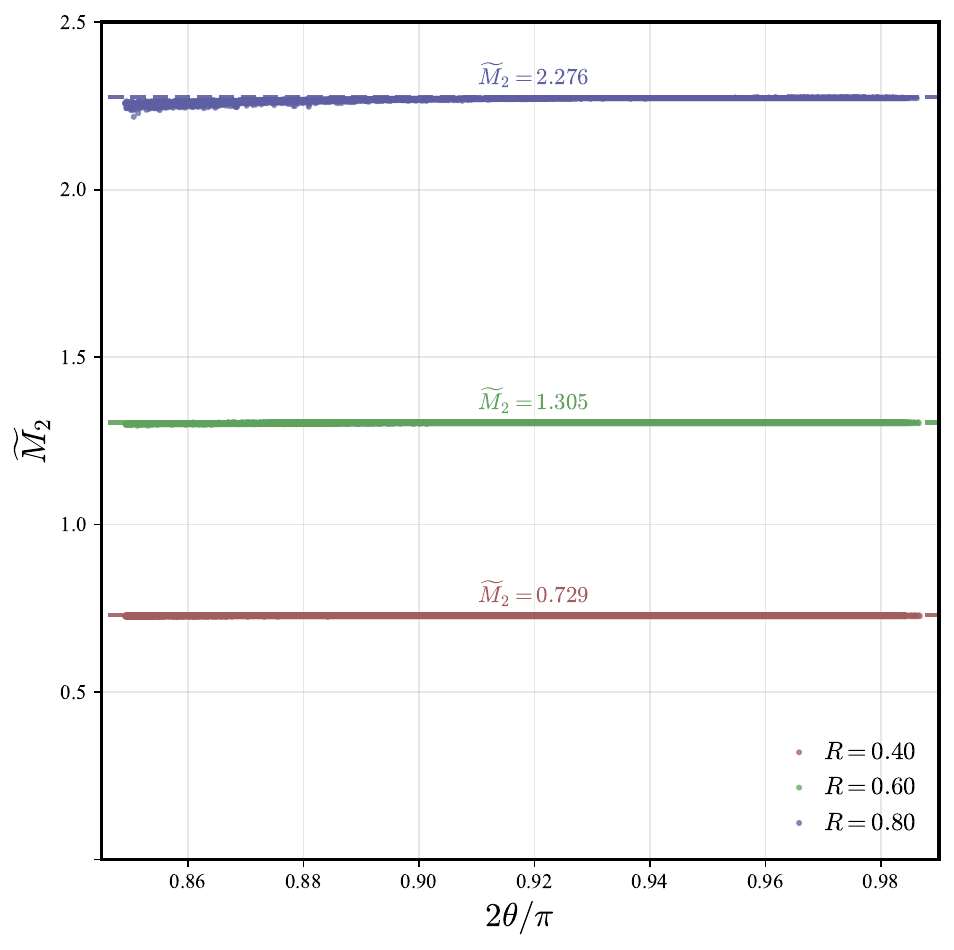}}
\caption{\label{fig:bounds}{\bf Purity budget-controlled maxima of entanglement and computational magic.} 
\textbf{a--b}, ({\bf a}) Maximum attainable entanglement negativity  and ({\bf b}) stabiliser R\'enyi-2 magic for two-qubit states at fixed purity. Both are plotted against the budget-allocation angle $\theta$, which tracks the internal trade-off between local and nonlocal budgets. Even when nonclassical resources are permitted, both exhibit a sharp, purity-dependent ceiling; entanglement exclusively emerges beyond the absolute separability threshold ($R=1/3$). \textbf{c--d}, Extension of the magic bounds to ({\bf c}) 5-qubit and ({\bf d}) 7-qubit systems. The expanded capacity of the Pauli spectrum in higher dimensions flattens the maximum magic profile, rendering the upper bounds largely insensitive to specific angular redistributions of the budget. Collectively, these envelopes establish the purity budget as a universal kinematic constraint -- indicating the onset of quantum resources and capping their maximum possible magnitude.} 
\end{figure*}

\section{The budget geometry}\label{sec:II}
\noindent
We formalise the purity-budget principle by decomposing the density matrix into orthogonal operator bases~\cite{Fano:1957zz}. We first define the projection for a general two-qubit state, 
\begin{align} 
    \rho=\frac{1}{4}\left(\mathbb{I}_2\otimes\mathbb{I}_2 +  r_i \sigma^i\otimes\mathbb{I}_{2}+\mathbb{I}_{2}\otimes s_i\sigma^i+t_{ij}\sigma^i\otimes\sigma^j\right),\label{eq:rhodef}
\end{align}
where repeated indices are summed over. Here, $r,\,s\in\mathbb{R}^3$ are the local Bloch vectors and $t\in\mathbb{R}^{3\times3}$ the correlation matrix. The purity-budget conservation principle can be framed precisely in terms of the local, nonlocal, and total budgets ($B_{\rm L}$, $B_{\rm NL}$, and $B$ respectively) as
\begin{align}
    4P-1  =\ \underbrace{\bigl(\|r\|^{2} + \|s\|^{2}\bigr)}_{B_{\rm L}\in[0,2]}\ \ + 
     \underbrace{\|t\|_{F}^{2}}_{B_{\rm NL}\in[0,3]} =\  B \in[0,3].
\end{align}
To avoid dealing with unequal ranges of $B_{\rm L}$ and $B_{\rm NL}$, we introduce the rationalised coordinates:
\begin{align}
X=\sqrt{\frac{B_{\rm L}}{3P}}=\sqrt{\frac{RB_{\rm L}}{B}}\ {\rm and}\ Y=\sqrt{\frac{B_{\rm NL}}{3P}}=\sqrt{\frac{RB_{\rm NL}}{B}},
\end{align}
where $R=B/3P\in \left[0,1\right]$ is the normalised budget-purity ratio. Both $B$ and $R$ are monotonically increasing functions of purity and each other: $R=
4/3-1/(3P)$.  For any fixed $P$, the rationalised coordinates form a circle:
\begin{align}
&X^{2}+Y^{2}=R, \quad 
X=\sqrt{R}\cos\theta,\quad Y=\sqrt{R}\sin\theta,
\label{eq:XYcircle2}
\end{align}
where $\theta\in[0,\pi/2]$ measures the distribution of the budget: $\tan\theta=\sqrt{B_{\rm NL}/B_{\rm L}}$, and the radius determines $P$. 

These coordinates offer a direct operational handle: $X$ quantifies the total local polarisation fraction of the budget, while $Y$ measures the correlation fraction. The reflection overlap $Q=\mathrm{tr}(\rho\widetilde\rho)$ governs the physical feasibility within this projection. In the two-qubit case, time-reversal symmetry acts cleanly on the budget sectors, aligning the $Q=0$ boundary with the vertical line $B_{\rm NL} = B_{\rm L} - 1$ (as $Q=(1-B_{\rm L}+B_{\rm NL})/4$ in this case). For such systems, every $(X,Y)$ point in the region $\{0\leqslant X\leqslant \sqrt{2/3}$, $0\leqslant Y\leqslant 1$, $R\leqslant1\}$ -- or equivalently, $\{1/4\leqslant P\leqslant1, Q\geqslant0\}$ -- corresponds to at least one valid density matrix. Consequently, there is no hole in this projection, a topological completeness we prove in the Methods section.  The negativity of $Q$ eliminates the $X^2>2/3$ region (the grey area in Fig.~\ref{fig:budgetXY}; this condition is equivalent to two-qubit positivity requirements~\cite{Morelli:2023khn} in the $(X,Y)$ plane, as we show in the Methods section). This boundary ($X^{2}=2/3$) acts as the positivity wall of the feasible set, as $Y$ becomes fully determined by $P$ alone. Pure-product states lie at the extremal point $(X^{2},Y^{2})=(2/3,1/3)$, while mixed product states on this boundary have $Y^{2}<1/3$. On this constant-$X$ boundary, the feasible set collapses to an effectively one-dimensional manifold. 

For an $n$-partite $d_1 \otimes d_2 \otimes \dots \otimes d_n$ system, we can define the total budget as $B = (D P - 1) \in [0, D-1]$, where $D = \prod_{k=1}^n d_k$, and its components as $B_{\rm L} = \sum_{k=1}^n (d_k P_k - 1)$ and $B_{\rm NL} = B - B_{\rm L}$, where $P_k$ is the marginal purity of the $k$th subsystem. The rationalised coordinates can be defined accordingly
\begin{equation}
X = \sqrt{\frac{B_{\rm L}}{(D-1)P}}, \quad Y = \sqrt{\frac{B_{\rm NL}}{(D-1)P}}.    
\end{equation}
For general higher-dimensional systems, the internal decomposition of $Q$ involves a more complex mixing of symmetric and asymmetric moments (see Supplementary Information). However, the topological principle remains the same: the vanishing of the reflection overlap ($Q=0$) defines the outer positivity boundary in this projection, which remains hole-free. 

For all quantum systems, the collection of density matrices creates a sharp and connected region in the $(X,Y)$ plane, and this second-moment projection is complete for feasibility. Mathematically, the $(X,Y)$ plane is a complete geometric orbit space for all symmetric LU-invariant second moments. In other words, no basis-independent quadratic information that does not distinguish between subsystems is lost in this representation, and every feasible point corresponds to at least one valid state. Thus, the geometry is fully revealing at the kinematic level. The $(X,Y)$ representation plays a practical role analogous to the Bloch sphere for a single qubit, not as a map of the full state space, but as a representation of the kinematically relevant second-moment degrees of freedom.

\section{The correlation hierarchy and entanglement guarantee}\label{sec:III}
\noindent
\subsection{The two-qubit case}
\noindent
The Peres-Horodecki criterion~\cite{Peres:1996dw,Horodecki:1997vt} is necessary and sufficient to detect two-qubit separable states. Under PT, a state can pick up at most one negative eigenvalue~\cite{Sanpera:1998fb}. This manifests in the $(X,Y)$ plane as follows: a state \emph{can} show negativity \emph{if and only if} $R>1/3$. In other words, all two-qubit states with $R\leqslant 1/3$ are absolutely separable. Beyond $R=1/3$,  the maximum negativity is bounded: $\mathcal N\leqslant \sqrt{9R/(64-48R)}-1/4$ (see Fig.~\ref{fig:bounds}).

Since classically simulatable states are restricted to local diagonal generators, the correlation matrix of any state with a classical subsystem (CC or QC) is rank-one in its natural basis. Maximising the nonlocal budget under this classical constraint thus gives an absolute correlation capacity limit: $B_{\rm NL} \leqslant 1$. This defines the C envelope, $B_{\rm NL}=1$, which in rationalised coordinates becomes
\begin{align}
Y_{\max}(X)=\sqrt{\frac{2}{3}-\frac12 X^{2}}.
\label{eq:classicalstructure}
\end{align}
This is also a physical demarcation as it defines the absolute correlation capacity of the symmetric generator sector. Maintaining positivity above the C envelope requires involving asymmetric, off-diagonal generators (e.g., $\sigma_y$). Within our LU-invariant framework, PT is equivalent to local time reversal. While a physical local time reversal flips all time-odd generators, LU invariance ensures that only flips that cannot be undone by LU rotations affect the budget. The effective operation is therefore local complex conjugation, under which only the imaginary, asymmetric generators are odd. The mandatory activation of these intrinsically time-odd generators instantly guarantees PT-negativity. Physically, if reversing a subsystem's local time arrow mathematically breaks the joint state, the state must be genuinely entangled, as its quantum correlations depend on a shared arrow of time (causal consistency). Therefore, any geometric excursion above the C envelope acts as a basis-independent entanglement witness, guaranteeing nonclassical correlations.

The C envelope also clarifies the behaviour of the basis-independent geometric discord $D_G$~\cite{Dakic:2010xfz}:
\begin{align*}
\frac{1}{4}(B_{\rm NL}-t_{\max}^2)\leqslant D_G \leqslant  \frac13 B = \frac{R}{4-3R}, 
\end{align*}
where $t_{\rm max}$ is the largest singular value of the correlation matrix. Outside the C envelope, at least two singular values of the $t$ matrix are nonzero, guaranteeing a strictly positive $D_G$. The absolute upper bound depends solely on the LU-invariant radius $R$, demonstrating how the budget geometry simultaneously identifies where discord must emerge and constrains its maximum magnitude.

For two-qubit systems, states above the C envelope are also necessarily steerable. The three-measurement linear steering (LS$_3$) has a natural expression in the budget projection: $B_{\rm NL} > 1$, which directly follows from the definition of $S_3$ in Ref.~\cite{Costa:2016pas}. Because the C envelope is exactly the $B_{\rm NL}=1$ line, the kinematic boundaries for guaranteed entanglement and LS$_3$ steering perfectly coincide in this geometry (see Fig.~\ref{fig:budgetXY}), assigning an additional operational meaning to the geometric threshold.

Finally, the purity-budget geometry fully characterises Clauser-Horne-Shimony-Holt (CHSH) nonlocality~\cite{Clauser:1969ny}. A CHSH violation occurs if and only if the sum of the two largest squared singular values satisfies $t_p^2+t_q^2>1$~\cite{Horodecki:1995nsk}. In the $(X,Y)$ projection, this is mathematically impossible on or within the C envelope. Above it, CHSH violation becomes possible, but remarkably, the geometry dictates a threshold for guaranteed nonlocality: any state with $B_{\rm NL}>3/2$ (e.g., a Werner state with $p>1/\sqrt2$) must violate the inequality. This maps to a distinct guaranteed-violation ellipse in the budget plane: $Y > \sqrt{4/5-3X^2/5}$.

\begin{figure*}
    \captionsetup[subfigure]{labelformat=empty}
    \centering
    \subfloat[{\bf a} Qubit-qutrit phase space]{
    \includegraphics[width=0.3\textwidth]{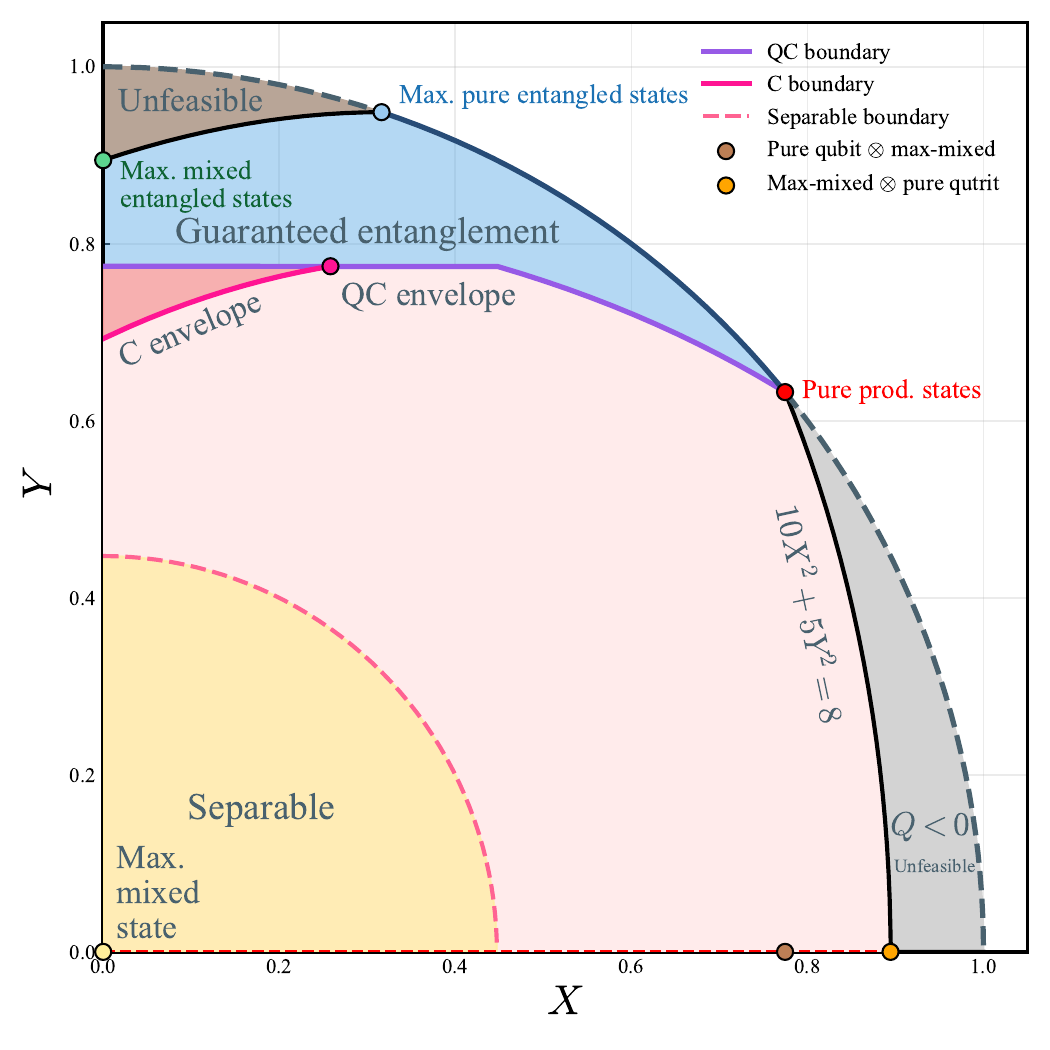}\label{fig:23system}}\hfill
    \subfloat[{\bf b} Qutrit-qutrit phase space]{
    \includegraphics[width=0.3\textwidth]{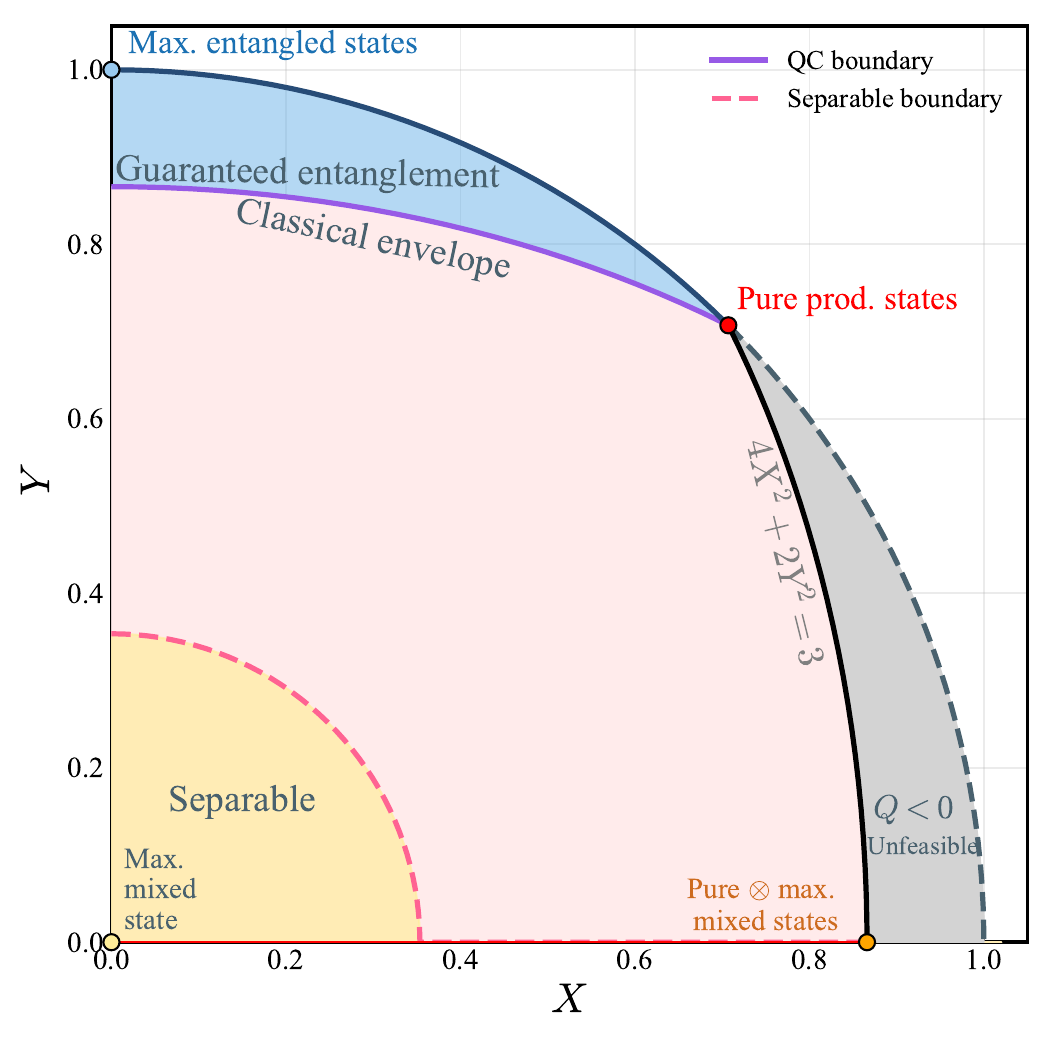}}\hfill
    \subfloat[{\bf c} Three-qubit phase space]{
    \includegraphics[width=0.3\textwidth]{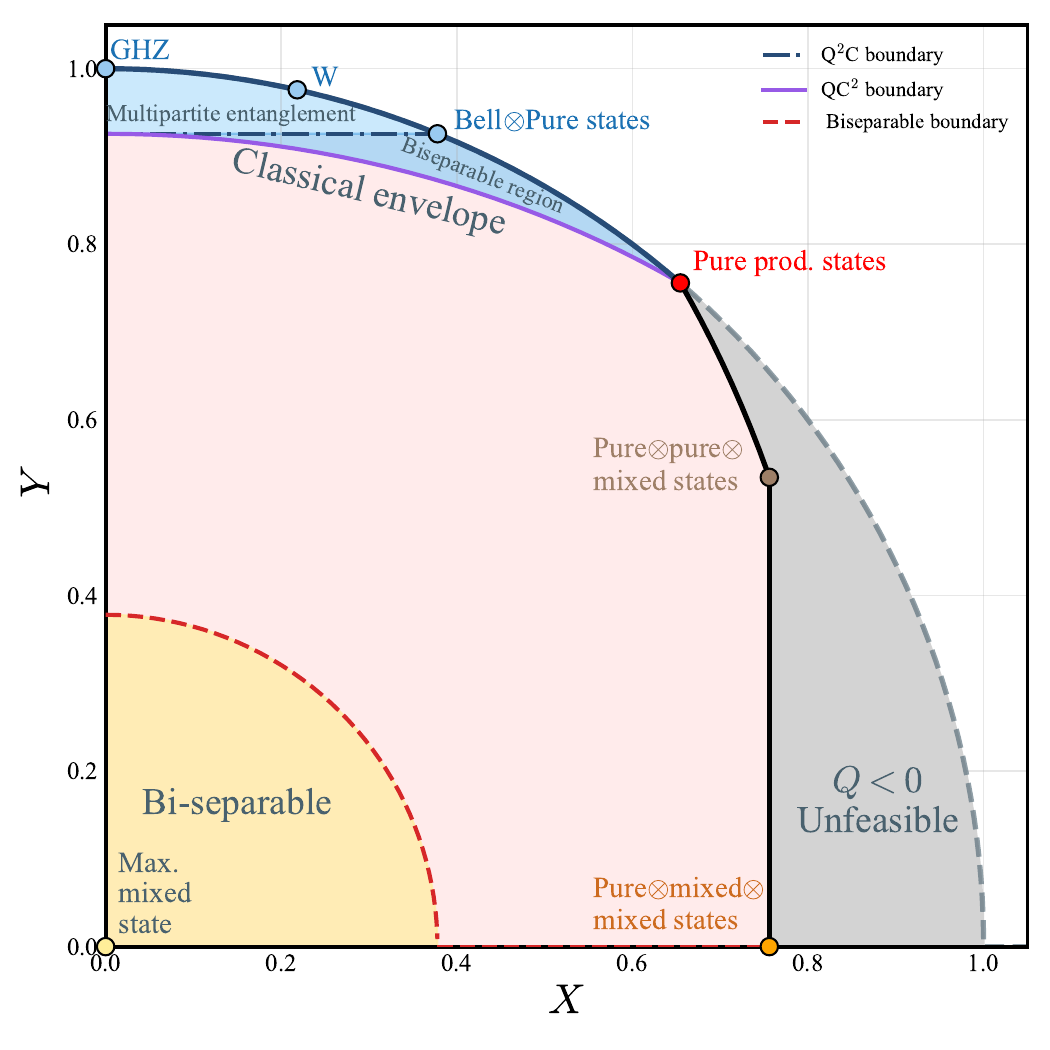}}\\
    \subfloat[{\bf d} Qubit-qutrit: Guaranteed NPT entanglement]{
    \includegraphics[width=0.235\textwidth]{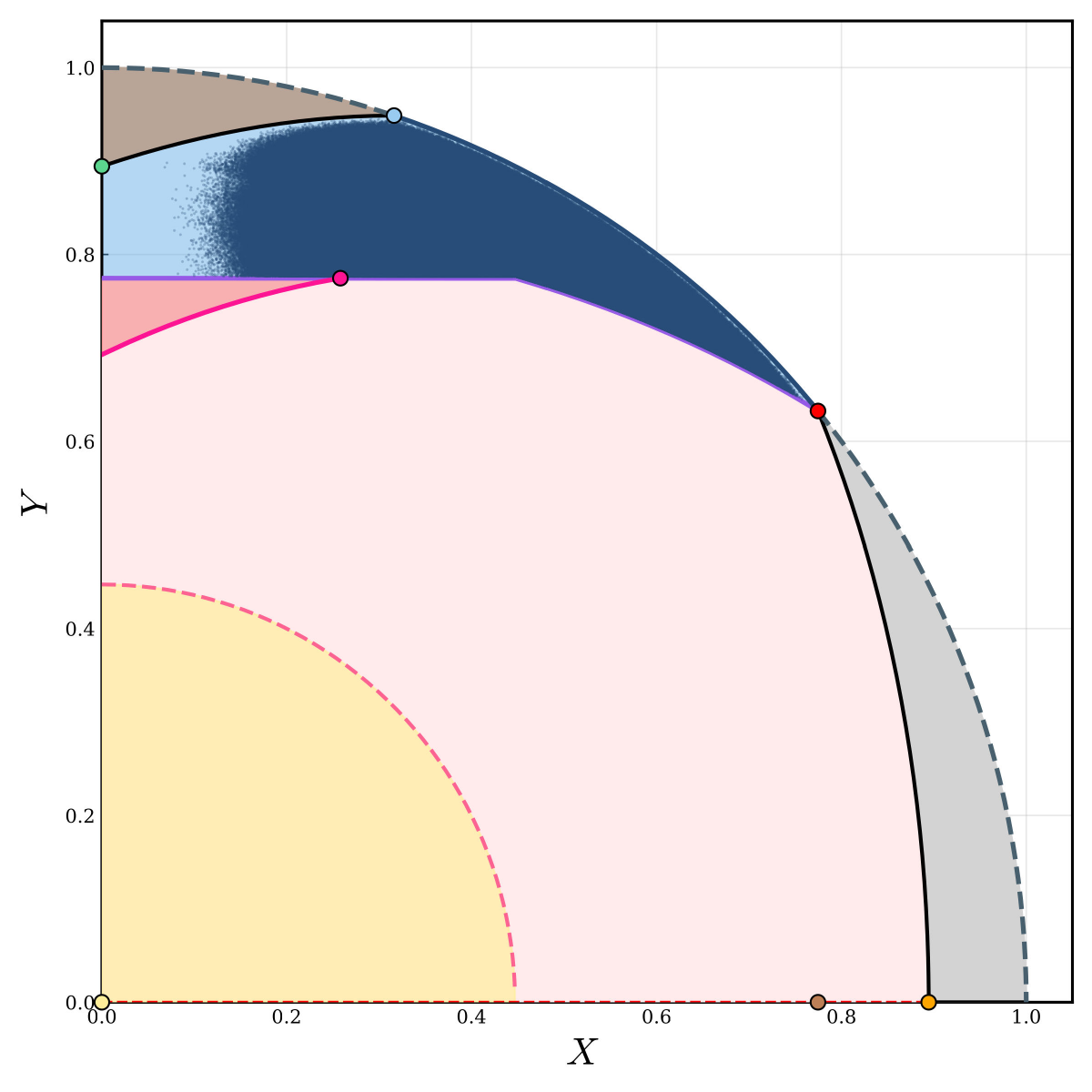}}\hfill
    \subfloat[{\bf e} Qutrit-qutrit: Guaranteed NPT entanglement]{
    \includegraphics[width=0.235\textwidth]{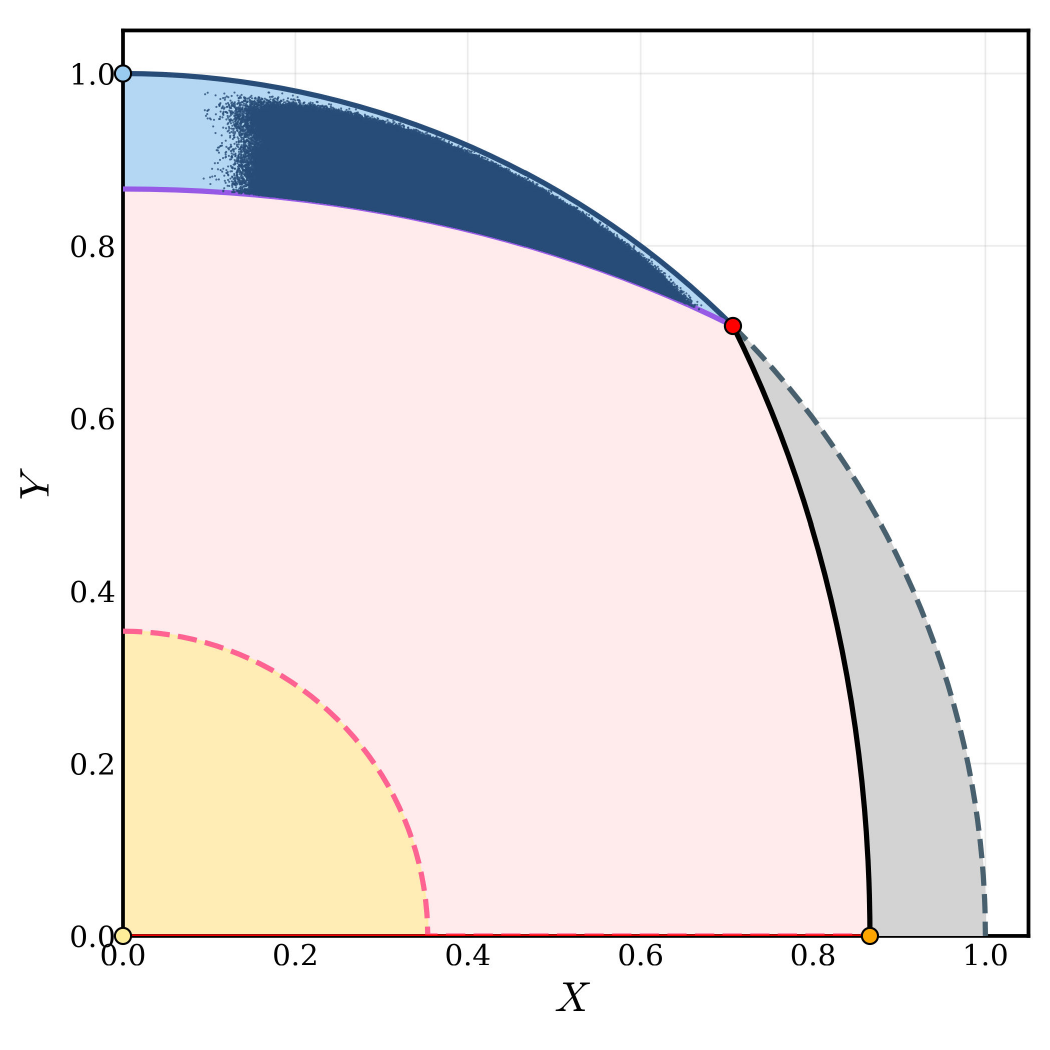}}\hfill
    \subfloat[{\bf f} Three-qubit: Bipartite NPT entanglement]{
    \includegraphics[width=0.235\textwidth]{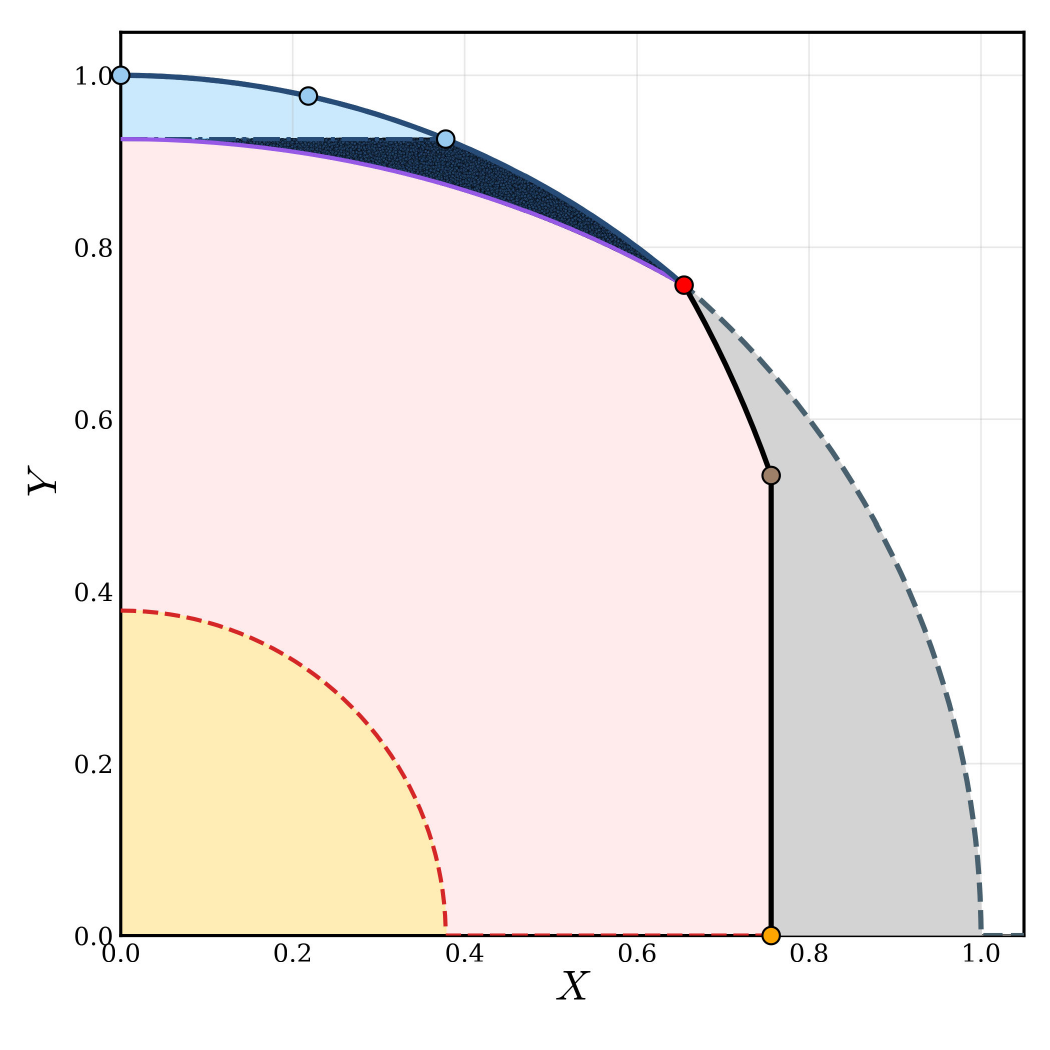}}\hfill
    \subfloat[{\bf g} Three-qubit: Genuine multipartite NPT entanglement]{
    \includegraphics[width=0.235\textwidth]{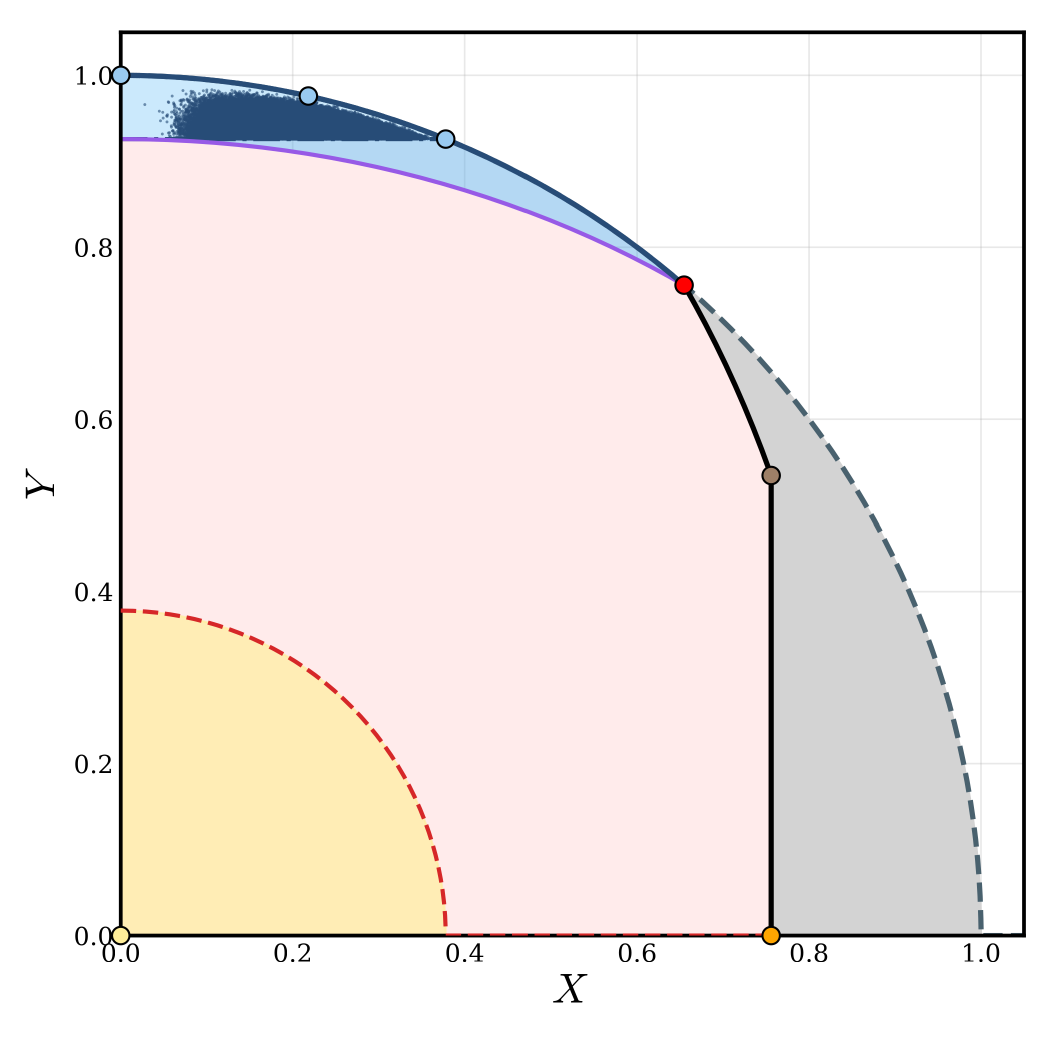}}
    \caption{\label{fig:Entangled}{\bf Dimensional bottlenecks and guaranteed NPT entanglement.} {\bf a}--{\bf c}, The purity-budget projection for ({\bf a}) qubit-qutrit, ({\bf b}) qutrit-qutrit, and ({\bf c}) three-qubit systems (the explicit construction of these geometries is given in Supplementary Information). The grey region remains unphysical ($Q < 0$). Dimensional asymmetry introduces a strict capacity bottleneck: dictated by the Schmidt decomposition, the marginals of a pure state must share the same spectrum. Consequently, the smaller qubit permanently caps the qutrit's entanglement capacity, forcing the correlation limit strictly below the $Y=1$ apex. For multipartite systems (e.g.,{\bf c}), sequential local depolarisation traces a piecewise, multi-segmented positivity wall. {\bf d}--{\bf g}, Zones of guaranteed NPT entanglement. The classical-quantum envelopes set the absolute correlation capacity of the time-symmetric generator sector. Traversing above these boundaries mathematically forces the activation of intrinsically time-odd, asymmetric generators to maintain global state positivity. The mandatory presence of these generators structurally guarantees negative eigenvalues under partial transposition. Any geometric excursion into these shaded regions, therefore, acts as a universal, basis-independent witness for NPT entanglement. For three qubits, this structural hierarchy precisely isolates entanglement classes: exceeding the QC$^2$ envelope guarantees bipartite NPT entanglement ({\bf f}), while crossing the higher Q$^2$C boundary precludes any single-party classical description, structurally forcing genuine multipartite NPT entanglement ({\bf g}) (e.g., the W state~\cite{Dur:2000zz}).}
\end{figure*}
\begin{figure*}
    \captionsetup[subfigure]{labelformat=empty}
    \centering
    \subfloat[{\bf a} Bell state]{
    \includegraphics[width=0.42\textwidth]{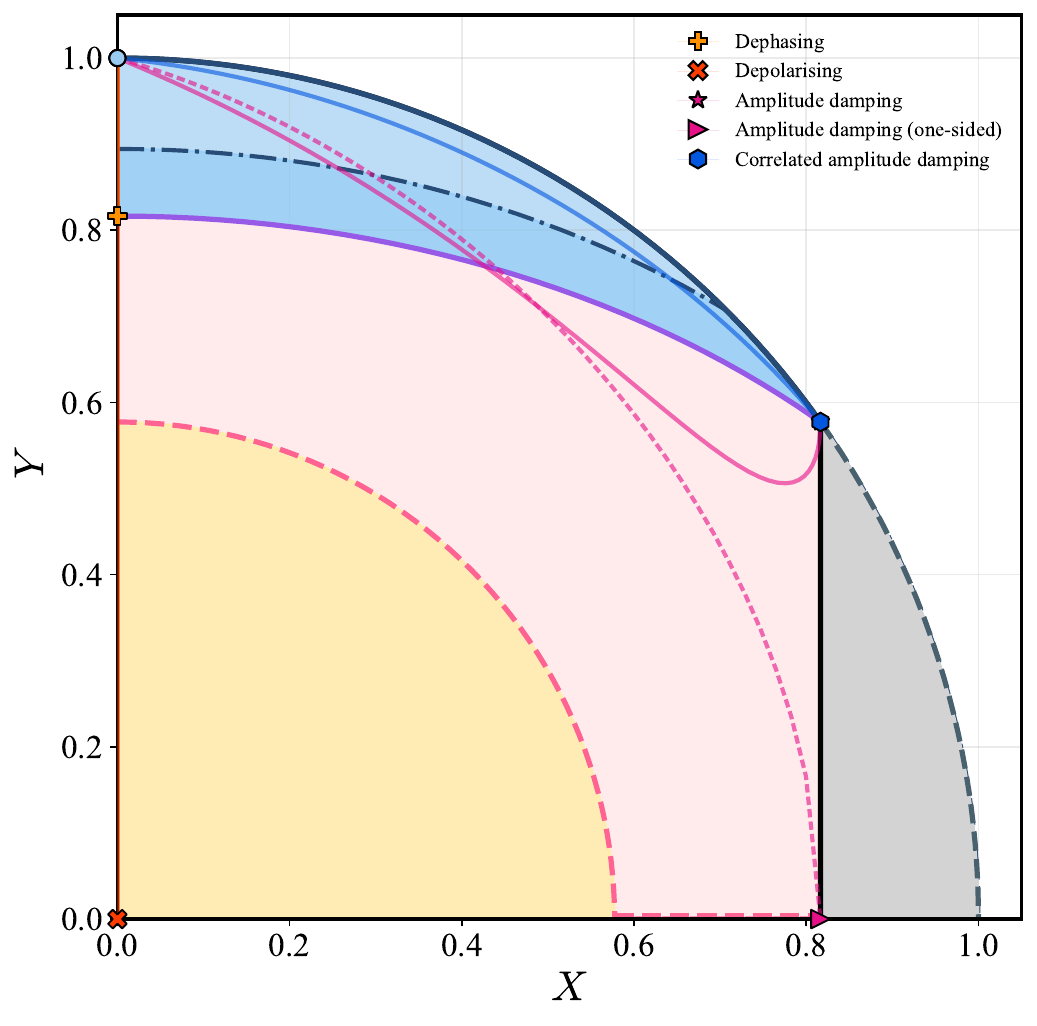}}\hspace{1cm}
    \subfloat[{\bf b} Pure-product state]{
    \includegraphics[width=0.42\textwidth]{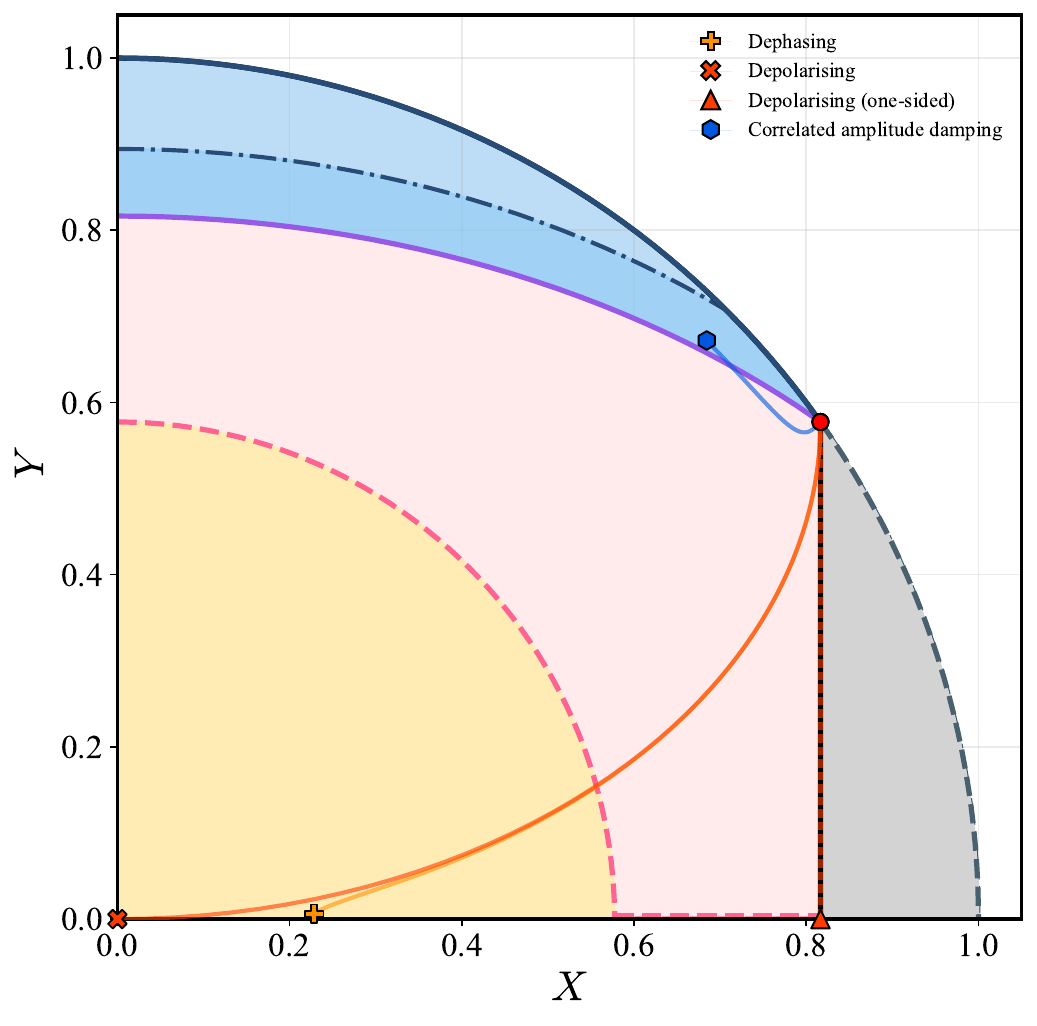}}\\
    \subfloat[{\bf c} Mixed-entangled state]{
    \includegraphics[width=0.42\textwidth]{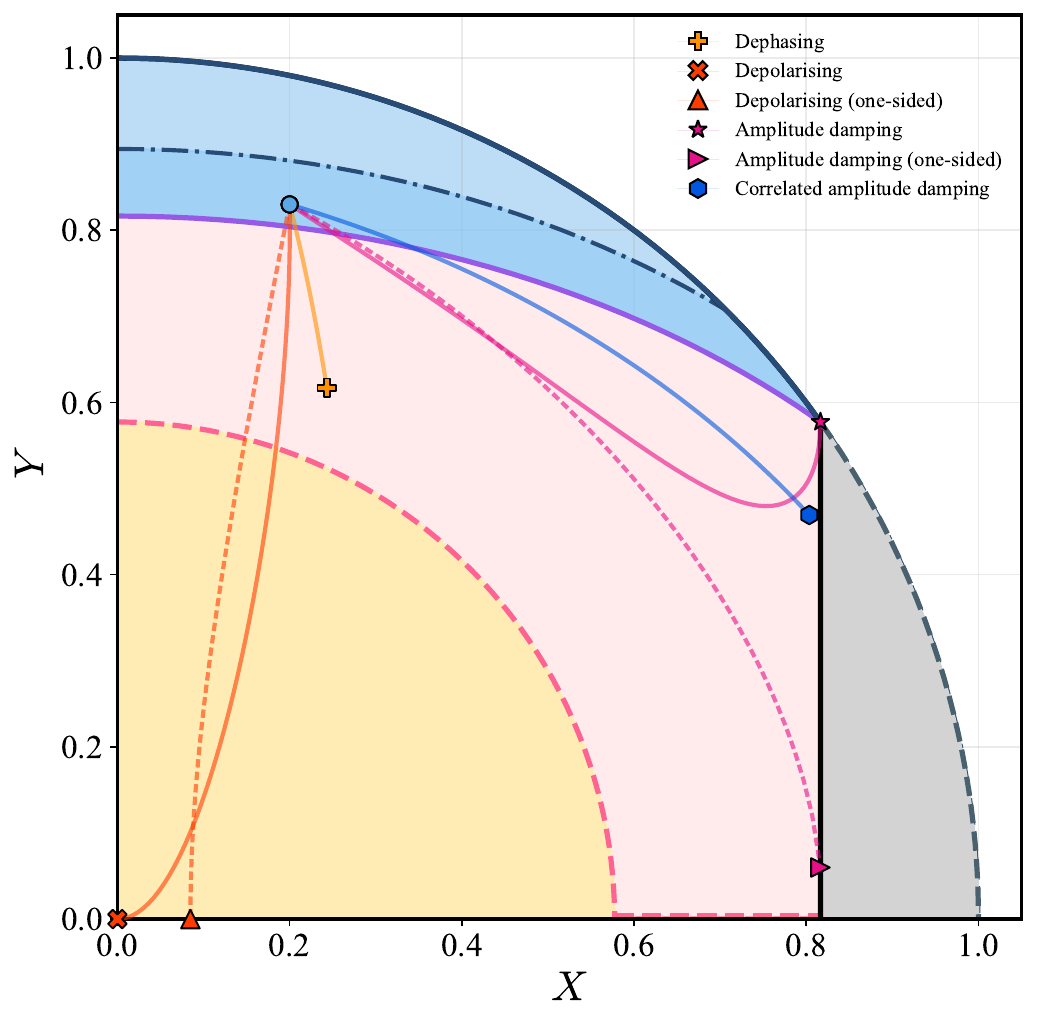}}\hspace{1cm}
    \subfloat[{\bf d} Mixed-separable state]{
    \includegraphics[width=0.42\textwidth]{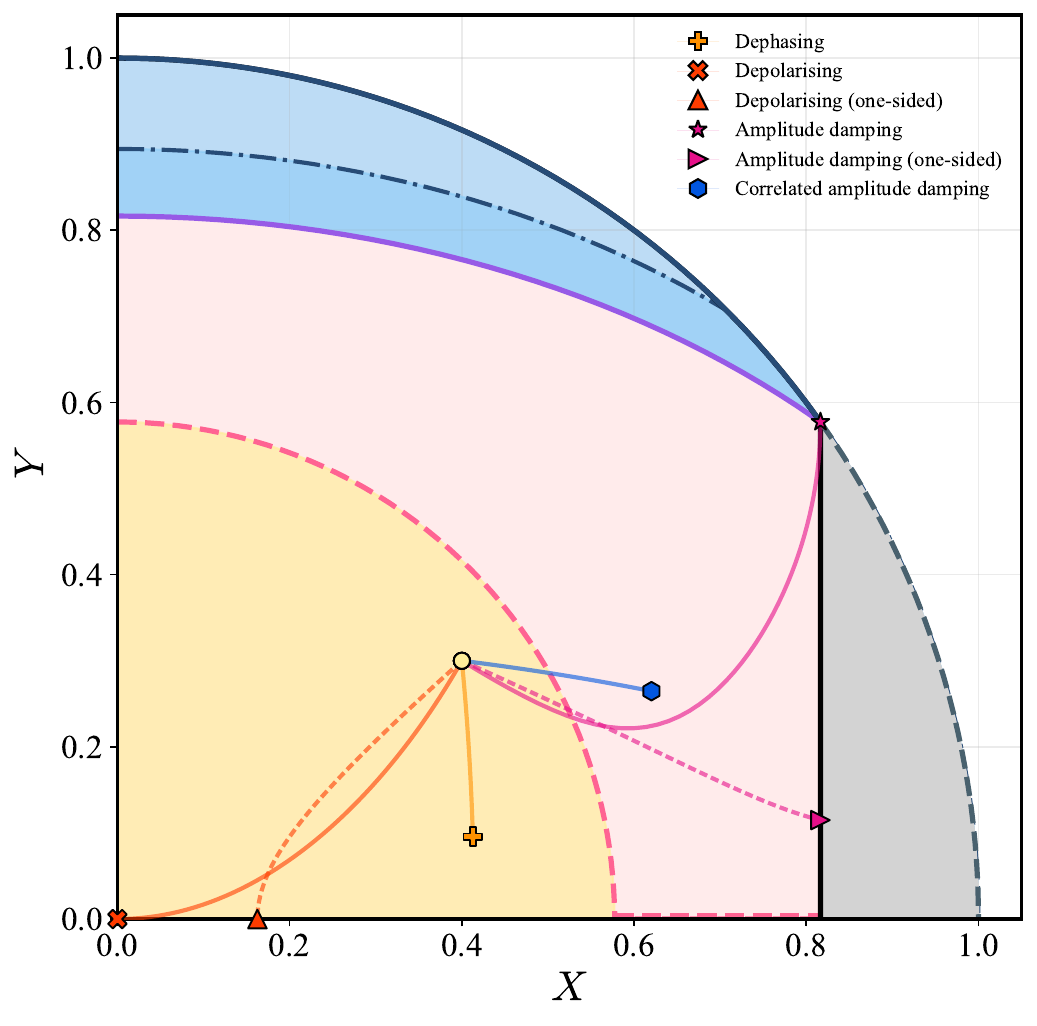}}
    \caption{\label{fig:noise} {\bf Kinematic flows of decoherence.} {\bf a}--{\bf d}, Parametric decoherence trajectories in the $(X,Y)$ plane for ({\bf a}) a maximally entangled Bell state, ({\bf b}) a pure-product state, ({\bf c}) a mixed-entangled state, and ({\bf d}) a mixed-separable state. The kinematic constraints and flow patterns generalise to higher dimensions beyond two-qubit systems. As noise strength increases, the resulting kinematic flows map how specific quantum channels redistribute the purity budget. Local noise degrades nonclassicality: depolarisation contracts states radially inward toward the origin, while dephasing drives them vertically downward to selectively suppress nonlocal correlations. Conversely, correlated noise redistributes these resources. By dissipating local polarisation faster than nonlocal correlations, correlated amplitude damping enables states to cross into steerable or entangled regimes despite an overall loss of purity. These trajectories reveal an empirical arrow of decoherence: under natural open-system dynamics, global purity ($P$) and time-reflection overlap ($Q$) cannot increase simultaneously (also see Fig.~\ref{fig:arrow}).}
\end{figure*}

\begin{figure*}
    \captionsetup[subfigure]{labelformat=empty}
    \centering
    \subfloat[{\bf a} Bell state]{
    \includegraphics[width=0.32\textwidth]{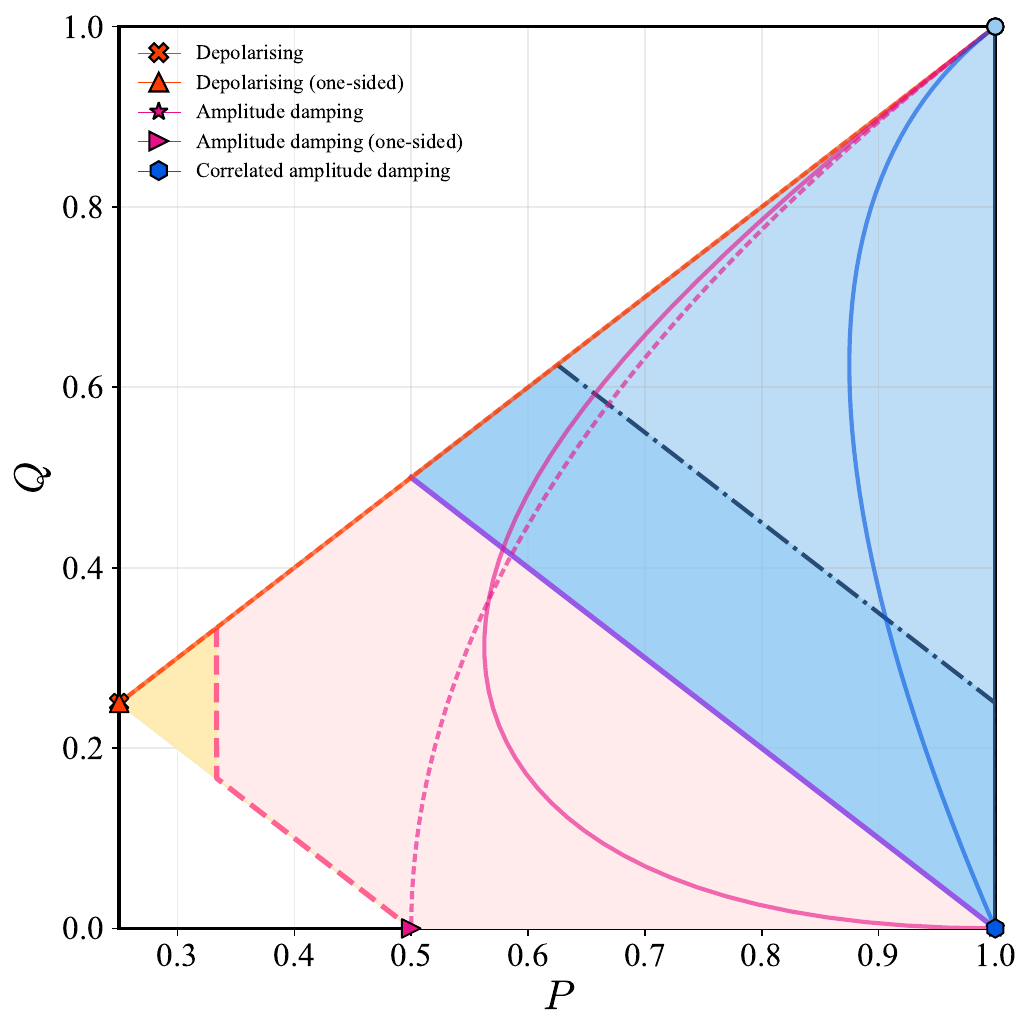}}\hfill
    \subfloat[{\bf b} CHSH state]{
    \includegraphics[width=0.32\textwidth]{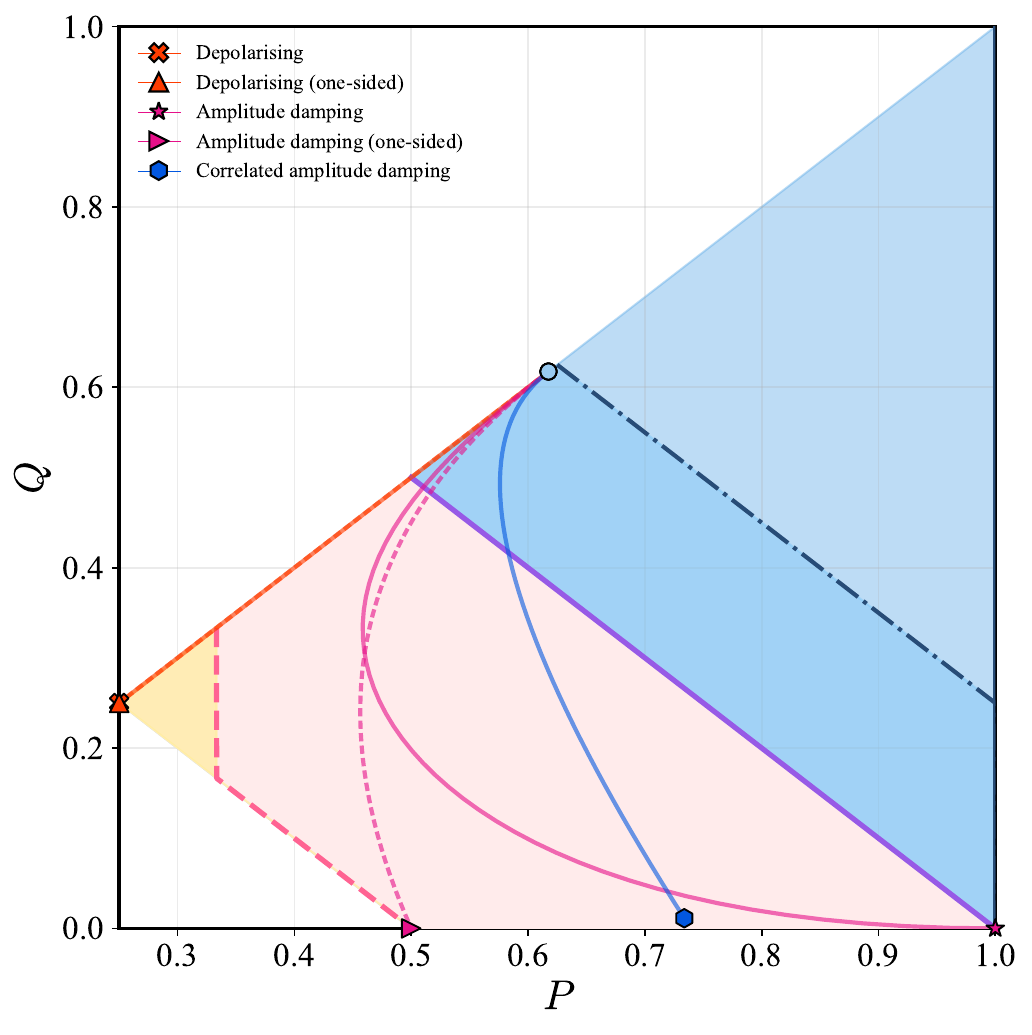}}\hfill
    \subfloat[{\bf c} Mixed-entangled state]{
    \includegraphics[width=0.32\textwidth]{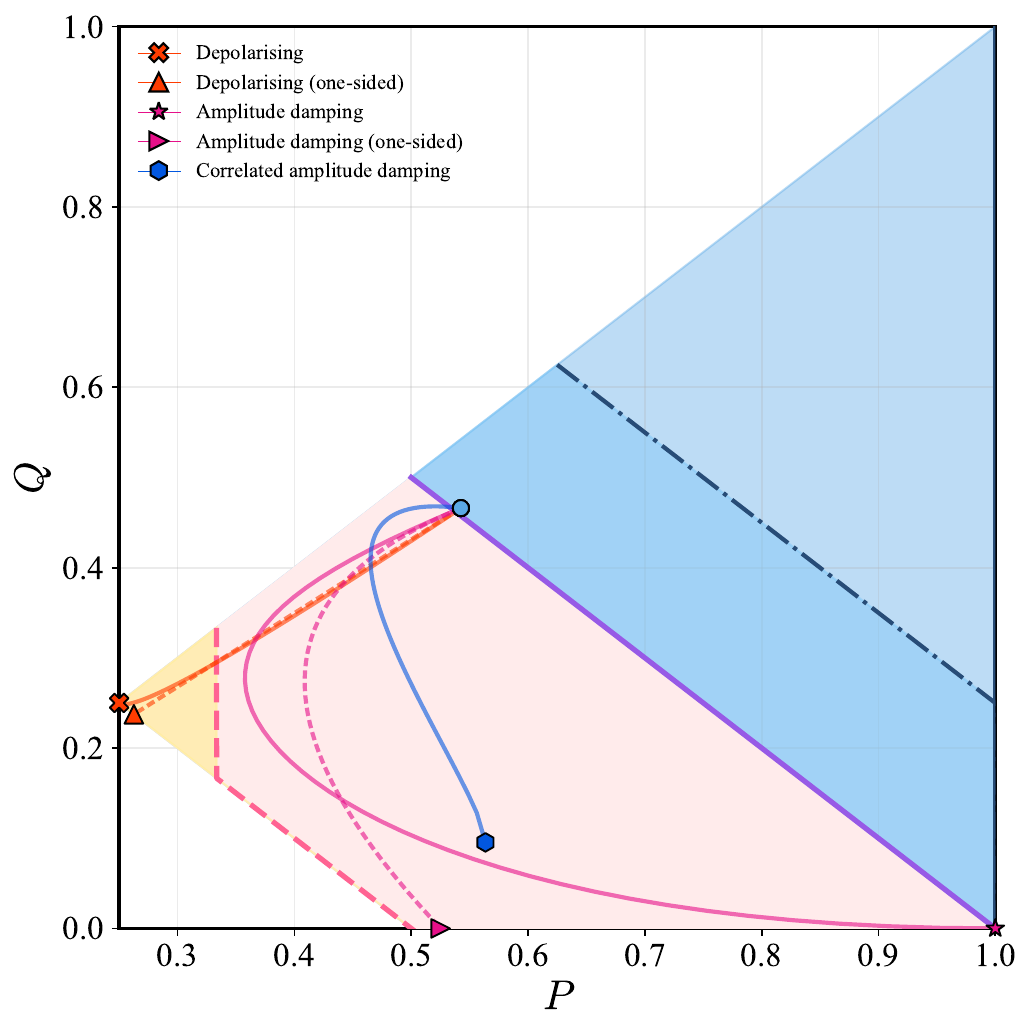}}\\
    \subfloat[{\bf d} Pure-product state]{
    \includegraphics[width=0.32\textwidth]{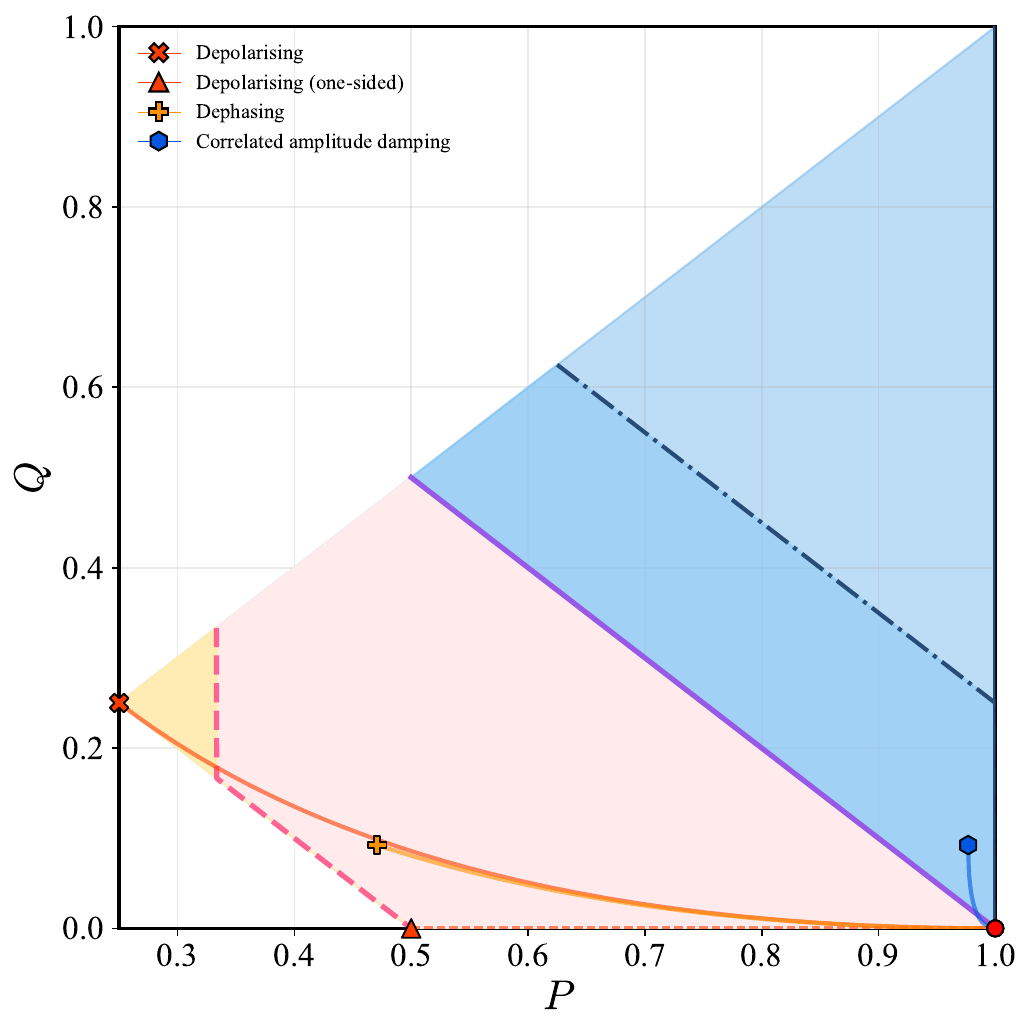}}\hfill
    \subfloat[{\bf e} Mixed-separable state]{
    \includegraphics[width=0.32\textwidth]{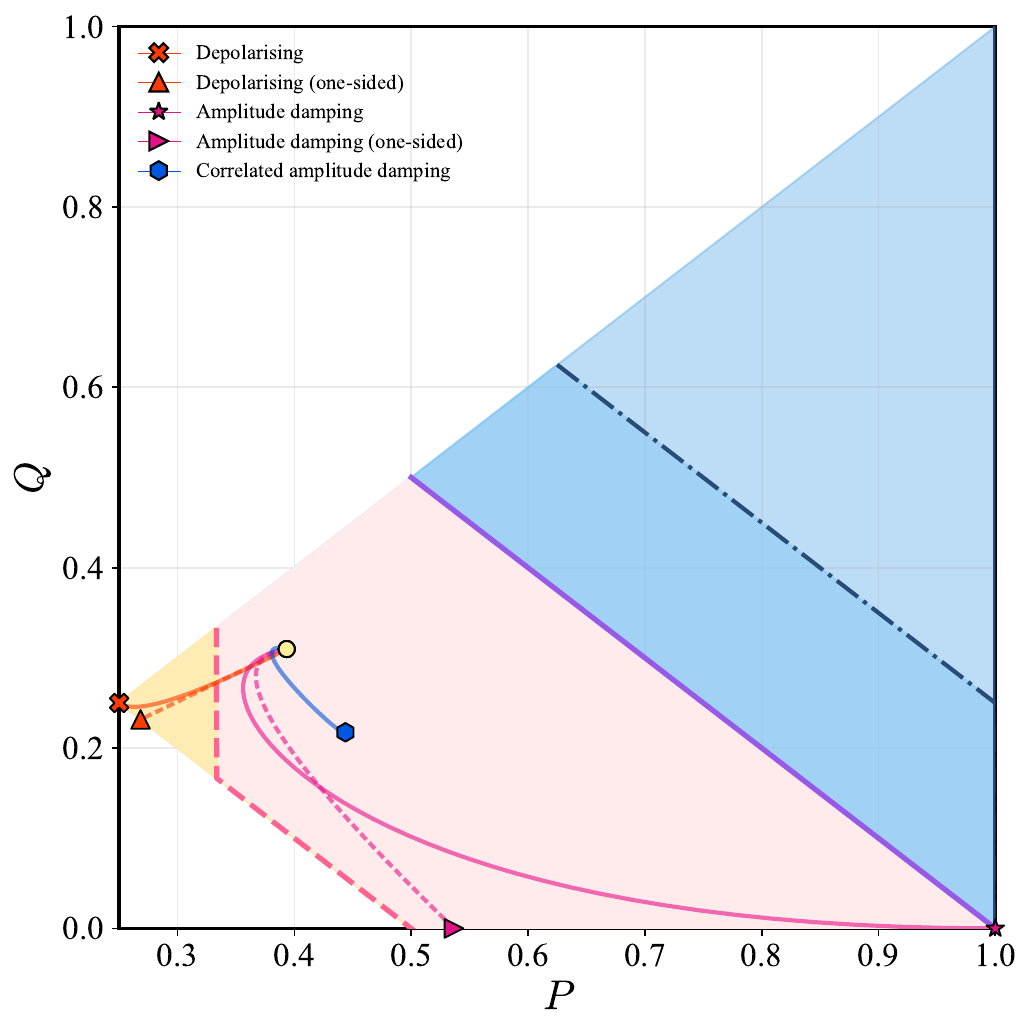}}\hfill
    \subfloat[{\bf f} Classical state]{
    \includegraphics[width=0.32\textwidth]{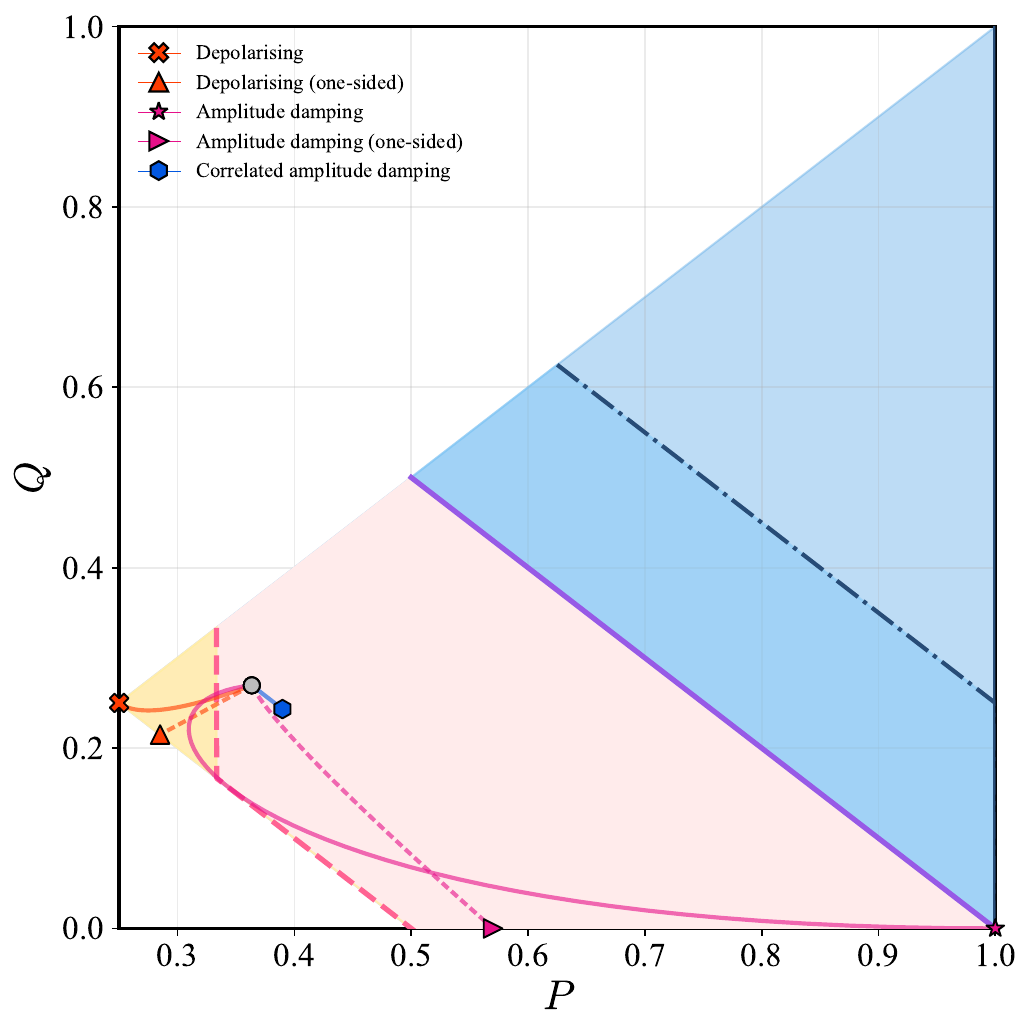}}
    \caption{\label{fig:arrow}\textbf{The empirical arrow of decoherence in the macroscopic $(P,Q)$ plane.} \textbf{a--f}, Open-system dynamics manifest as constrained flows, illustrated here by plots in the global purity ($P$) vs the time-reflection overlap ($Q$) plane for six distinct initial states: (\textbf{a}) a Bell state, (\textbf{b}) a CHSH state, (\textbf{c}) a mixed-entangled state, (\textbf{d}) a pure-product state, (\textbf{e}) a mixed-separable state, and (\textbf{f}) a classical state. Under natural, memoryless open-system dynamics, $P$ and $Q$ cannot increase simultaneously. Physically, environments that extract entropy to cool the system (increasing $P$) inevitably polarise the state, destroying time-reversal symmetry and lowering $Q$. Conversely, scrambling environments that restore this symmetry inherently inject entropy, thereby lowering $P$. This \emph{no-go} rule acts as a macroscopic, geometric arrow of decoherence, permanently restricting the accessible regions of the second-moment space during natural evolution in all two-qubit systems and beyond.}
\end{figure*}


\subsection{Enforced entanglement and discord beyond two qubits}
\noindent
For systems larger than $2 \otimes 2$ or $2 \otimes 3$, positive PT no longer guarantees separability due to bound entanglement, though a general notion of biseparability exists for composite systems~\cite{Zyczkowski:1998yd,Gurvits:2002yud}. However, a negative PT (NPT) remains a strict entanglement witness~\cite{Horodecki:2009zz}. Because maximising $B_{\rm NL}$ under classical constraints establishes a hard capacity limit that forces the activation of intrinsically time-odd asymmetric generators, we can generalise our geometric approach to detect NPT entanglement across arbitrary dimensions.

Consider an $n$-partite system with dimensions $d_1 \leqslant d_2 \leqslant \dots \leqslant d_n$. By systematically restricting $n-m$ subsystems to the classical (diagonal) sector (Q$^m$C$^{n-m}$), we can generate a nested hierarchy of kinematic envelopes. For any specific $m$, we construct the maximum-correlation boundary in the $(B_{\rm L}, B_{\rm NL})$ plane by evaluating all possible classical-quantum partitions to identify capacity bottlenecks (hinges). The convex hull of these points defines the absolute kinematic limit for that $m$-quantum tier. In other words, tracing this kinematic boundary requires extracting the maximum nonlocal correlation capacity for any given $B_{\rm L}$. For a specific classical-quantum configuration, this is achieved by minimising the marginal purities to maximise the global purity. However, a joint state's purity is strictly capped by its most mixed classical marginal. Consequently, the global purity encounters a hard structural ceiling at $1/d^{\rm C}_{\min}$ (where $d^{\rm C}_{\min}$ is the dimension of the smallest classical subsystem in that permutation), as its diagonal generators cannot support any more correlation. By calculating this capacity saturation limit across all possible structural permutations, we map out every local extreme (the hinges). Finally, by applying the convex hull rule from the pure-state peak down to the maximum correlation floor on the $B_{\rm NL}$ axis, we trace the absolute outermost boundary. (The full algorithmic construction is presented in the Methods section.)

Any state that exceeds this envelope requires asymmetric generators, thereby guaranteeing NPT entanglement across at least $m+1$ subsystems. Because the dimensions are ordered, we can systematically bound the combinations that produce the highest correlation floors. Exceeding the outermost envelope of the $m=n-1$ tier precludes any single-party classical description, guaranteeing genuine multipartite NPT entanglement. To understand the physical significance of these correlation capacity limits, we can focus on the smallest classical subsystem, which dictates the correlation bottleneck in any permutation. Our arguments are based on group theory: we can map these structural limits to the time-reversal symmetries of the local $SU(d^{\rm C}_{\min})$ Lie algebra. For any qudit, its generators mathematically partition into a symmetric sector (comprising the commuting diagonal $H$ and off-diagonal real-symmetric $S$ generators) and an intrinsically time-odd, asymmetric sector (the off-diagonal imaginary-antisymmetric $A$ generators). Now, an LU-invariant framework cannot distinguish between the maximum correlation capacities of the $H$ and $S$ sectors, as all real symmetric matrices are diagonalisable via local orthogonal rotations. Because the classical constraint restricts the subsystem to a real density matrix, bounding $B_{\rm NL}$ only requires optimising over the real orthogonal subgroup $SO(d)$, which bounds the entire $H \oplus S$ sector but explicitly cannot rotate into the imaginary $A$ sector. Therefore, the $A$ sector remains an independent, unbound resource required to cross the envelope. Hence, maximising $B_{\rm NL}$ over the strictly diagonal $H$ subspace yields the same absolute supremum as maximising it over the entire time-symmetric sector $H \oplus S$. (Because of this, a purely classical envelope $C\equiv C^n$ cannot be isolated this way; we obtain it differently in the Methods section. However, for symmetric systems ($d_1=d_2=\cdots=d_n$), the C envelope completely overlaps with the QC$^{n-1}$ envelope.) Consequently, the Q$^m$C$^{n-m}$ envelope represents the fundamental correlation capacity of the entire time-symmetric generator space for that partition. This completes the structural proof: crossing any Q$^m$C$^{n-m}$  envelope is a sufficient condition for PT negativity in the relevant partition -- a dimension-agnostic, basis-independent geometric witness for NPT entanglement.

Just crossing this boundary means one more subsystem becomes quantum; it mathematically necessitates accessing the asymmetric generators of the newly quantum sector. Above the envelope, the presence of these $A$ generators becomes LU-irreducible. Because the $A$-sector generators are intrinsically time-odd, their irreducible presence manifests through negative eigenvalues under PT (equivalently, local time reversal), signalling entanglement between the partitions. Thus, traversing above the kinematic envelopes calculated by the algorithm serves as a universal, dimension-agnostic proof of genuine NPT entanglement. Interestingly, for qubits ($SU(2)$), the algebra contains only a single $A$ generator ($\sigma_y$). Consequently, crossing the QC boundary instantly spans the full non-commuting algebra, ensuring that such states are also guaranteed to be steerable.

At the base of the hierarchy, the all classical tier (the C envelope) defines the absolute capacity limit for classically correlated networks; any excursion above this outermost line acts as a guaranteed, basis-independent witness for quantum discord.

\section{Kinematic bounds on quantum magic}\label{sec:IV}
\noindent
Just as the purity budget constrains spatial correlations, it caps the capacity of a state to hold non-Clifford computational resources. The stabiliser R\'enyi-$2$ magic $\widetilde M_2$~\cite{Leone:2021rzd,Leone:2024lfr} (a resource monotone with broad applications spanning from quantum computing to high-energy physics~\cite{White:2024nuc}) quantifies how far a state deviates from those that can be simulated using only Clifford operations and Pauli measurements. Because this magic depends on the fourth moment of the Pauli spectrum, the total second-moment budget bounds its range:
\begin{equation*}
M_B-\log_2 \left(1 + x_{\max}^2 B\right) \leqslant \widetilde M_2 \leqslant M_B-\log_2\left(1 + B^2/n_{\rm P}\right),
\end{equation*}
where $M_B=\log_2(1+B)$, $x_{\max}$ is the Pauli expectation value with the largest magnitude, and $n_{\rm P}$ is the total number of non-identity Pauli observables ($n_{\rm P}=15$ for two qubits). Figure~\ref{fig:bounds} illustrates this upper limit using Monte Carlo samples for three fixed $R$ values. 

In general, every point except the origin admits at least one state with nonzero magic. This follows directly from the LU invariance of the purity budget. The local unitary group is always continuous for any qudit, whereas the Clifford group is discrete. Therefore, applying a continuous local non-Clifford rotation to any zero-magic stabiliser state preserves its exact $(X, Y)$ coordinate while necessarily misaligning its Pauli spectrum from any discrete stabiliser configuration, thereby injecting non-zero magic. Hence, the entangled states above the QC$^{n-1}$ envelope can be computationally powerful as they have a high purity budget, elevating their capacity for non-stabiliser magic.

\begin{table*}
\begin{centering}
{\small\renewcommand\baselinestretch{1.6}\selectfont \begin{tabular*}{\textwidth}{@{\extracolsep{\fill}} llr} 
\hline\hline
& {\bf Two qubits ($\mathbf{2\otimes2}$)}
& {\bf Beyond two qubits} \\
\noalign{\vspace{-5pt}}&&{\bf ($\mathbf{d_1\otimes d_2\otimes \cdots \otimes d_n}$)}\\
\hline
\multicolumn{3}{l}{\bf Geometry \& feasibility}\\\hline
$B$, $B_{\rm L}$, $B_{\rm NL}$, $X$, $Y$ 
& \multicolumn{2}{c}{Defined from global and marginal purities.}\\
Feasible $(B_{\rm L},B_{\rm NL})$ or $(X,Y)$ region 
& \multicolumn{2}{c}{Convex, compact, hole-free. Fully characterised by $P$ and $Q$.} \\\hline
\multicolumn{3}{l}{\bf Reference states}\\\hline
Maximally mixed states 
& \multicolumn{2}{c}{Lie at the origin $(X,Y)=(0,0)$.}\\
Pure states 
& \multicolumn{2}{c}{Lie on the $R=1$ circle.} \\
Pure-product states 
& \multicolumn{2}{c}{Lie on the right-most edge of the $R=1$ line; zero entanglement.} \\
\hline
\multicolumn{3}{l}{\bf Classical correlations}\\\hline
Classical envelope
& \multicolumn{2}{c}{Coincides with the QC$^{n-1}$ envelope for symmetric states.}\\
& $Y=\sqrt{2/3-X^2/2}$. Classical correlations  & Confined by the diagonal generators.\\
\noalign{\vspace{-5pt}}&confined by a rank-$1$ correlation tensor.\\
\hline
\multicolumn{3}{l}{\bf Nonclassical resources}\\\hline
Discord 
& \multicolumn{2}{c}{Guaranteed above the C$^n$ envelope.}\\
Separability
& Absolute separability boundary: $R=1/3$ 
& PPT/NPT boundary (bipartite);\\
\noalign{\vspace{-5pt}}
& (necessary and sufficient). 
& partition-dependent multipartite notions~\cite{Gurvits:2002yud}.\\
Entanglement 
& \multicolumn{2}{c}{Guaranteed above the QC$^{n-1}$ envelope.}\\
Steering 
& Exact analytic boundary (LS$_3$); 
& Guaranteed-violation regions \\
\noalign{\vspace{-5pt}}& exclusion and guarantee coincide. &defined by inequalities.\\
Bell nonlocality 
& Exact analytic CHSH bounds. 
& Guaranteed-violation regions\\ 
\noalign{\vspace{-5pt}}& &  defined by inequalities.\\
Magic (non-stabiliser) 
& Exact kinematic bounds at fixed purity
& Exact kinematic bounds at fixed purity\\
\noalign{\vspace{-5pt}}& (R\'enyi-$2$ magic; $n_{\rm P}=15$). 
& (R\'enyi-$2$ magic; general $n_{\rm P}$). \\
\hline
\multicolumn{3}{l}{\bf Open-system dynamics}\\\hline
Decoherence trajectories
& \multicolumn{2}{c}{Universal kinematic flows: local noise degrades resources,}\\
\noalign{\vspace{-5pt}}& \multicolumn{2}{c}{while correlated noise geometrically redistributes them.}\\
Purity-preserving process
& \multicolumn{2}{c}{Conserves total budget.}\\
Arrow of decoherence
& \multicolumn{2}{c}{Empirical rule: natural dynamics cannot simultaneously}\\
\noalign{\vspace{-5pt}}(the \emph{no-go} rule)
& \multicolumn{2}{c}{increase global purity ($P$) and time-reflection overlap ($Q$).}\\
\hline\hline
\end{tabular*}}
\end{centering}
\caption{\label{tab:general} \textbf{The hierarchy of quantum correlations and resources in the purity-budget geometry.} This table contrasts the exact analytic boundaries derived for two-qubit systems with the universal kinematic constraints applicable to arbitrary dimensions. In all regimes, the feasible $(X, Y)$ projection is completely determined by $P$ and $Q$. Traversing beyond the classical envelope ensures the failure of classical descriptions and the emergence of nonclassical resources. While two-qubit systems permit a unified analytical closure for entanglement, steering, and Bell-nonlocality, the underlying purity budget universalises these physical constraints across all dimensions, including strict caps on non-stabiliser magic. Mapping open-system dynamics as geometric flows within this space exposes an empirical no-go principle: natural noise cannot simultaneously increase $P$ and $Q$, as shown in Fig.~\ref{fig:noise}.}
\end{table*}

\section{Operational impact}\label{sec:V}
\noindent
\subsection{Decoherence as budget flow}
\noindent
Open-system dynamics manifest universally as purity-budget flows in the $(X,Y)$ plane (see Fig.~\ref{fig:noise} for two-qubit systems and the Appendix for higher-dimensional systems). Radial motion dictates changes in global purity, while angular motion reflects the redistribution of second-moment content between local ($B_{\rm L}$) and nonlocal ($B_{\rm NL}$) sectors. For purity-preserving unital dynamics, the total second-moment budget is conserved, confining evolution to a fixed-purity shell. Generic non-unital processes, conversely, induce radial change (typically contraction) of the budget. Thus, the budget plays the role of a conserved charge for purity-preserving dynamics, while its dissipation signals decoherence at the level of second moments. 

Decoherence is naturally geometric in this projection. Noise moves states along kinematically permitted directions. Observing these flows reveals an empirical arrow of decoherence: under natural open-system dynamics, global purity $P$ and the time-reflection overlap $Q$ never increase simultaneously. Physically, this reflects a fundamental thermodynamic conflict: memoryless environments that extract entropy (cooling) inevitably polarise the system, destroying time-reversal symmetry (lowering $Q$); conversely, environments that scramble the system can restore symmetry but inherently inject entropy (lowering $P$) -- see Fig.~\ref{fig:arrow}. However, it is possible to engineer a process that increases both $P$ and $Q$ simultaneously (see Appendix~\ref{app:b} for an example). While $Q$ is uniquely determined at each $(X,Y)$ point for two qubits, higher-dimensional systems feature a degenerate $Q$ space. Nevertheless, the absolute feasibility boundary remains universally anchored by the $Q=0$ limit, the point at which the global state's diagonal populations can no longer shield its asymmetric coherences. This kinematic wall macroscopically restricts noise trajectories, ensuring states cannot be pushed into unphysical regimes.

Figure~\ref{fig:noise} illustrates these constraints through representative trajectories generated by local and correlated noise channels~\cite{Nielsen:2012yss,Breuer:2002pc}. The examples are chosen to expose distinct geometric mechanisms rather than to model specific implementations. Three general, experimentally relevant features are immediately visible:
\begin{itemize}
    \item[--] {\bf Local noise degrades:} Local noise produces inward or downward motion, kinematically preventing the generation of nonlocal resources.
    
    \item[--] {\bf Correlated noise redistributes:} Correlated amplitude damping~\cite{Xiao:2016dou} can suppress local quadratic weight faster than nonlocal weight, effectively converting local polarisation into correlations and vice versa.

    \item[--] {\bf Activation requires accessibility:} Entanglement generation from near the origin must cross the separability boundary. In contrast, activation from pure-product states, which lie on the positivity wall and can only move along it (preserving product structure), necessarily requires reducing local polarisation to increase correlations.
\end{itemize}
The budget geometry identifies which noise-induced transformations are possible, which are forbidden, and which are naturally preferred. This reframes correlated noise as a potential resource and provides an operational framework for diagnosing and exploiting noise.

\subsection{Experimental diagnostics}
\noindent
Since the kinematically feasible $(X,Y)$ region captures the entire second-moment state space, the projection naturally functions as a highly scalable diagnostic tool. Characterising nonclassical resources in many-body systems typically requires full state tomography, an approach that scales exponentially with system size. In contrast, the $(X,Y)$ coordinates require only $n+1$ purity measurements (one global and $n$ marginals). These can be extracted directly without full state reconstruction, using either two-copy overlap measurements, such as SWAP tests~\cite{Nakazato:2012lkq,Nguyen:2021oct}, or modern randomised measurement protocols like classical shadows~\cite{Huang:2020tih,vanEnk:2011xlo}. Using two-copy overlaps, for instance, the bounded nature of the observables ensures the purities can be estimated with an additive error $\varepsilon$ using $O(1/\varepsilon^{2})$ experimental shots, yielding a sample complexity completely decoupled from the dimension of the Hilbert space.

This is sufficient to locate any state in the budget plane and determine its position relative to the analytic feasibility and resource boundaries, without any optimisation over measurement settings or local bases. For example, it lets us determine whether classical descriptions are adequate, whether entanglement is structurally guaranteed, or, for two qubits, whether Bell nonlocality remains possible. In this way, the budget geometry provides an immediate, basis-independent macroscopic assessment of quantum resources, guiding experimental effort by dictating exactly when more measurement-intensive protocols are actually necessary.

\section{Discussion}\label{sec:VI}
\noindent
Quantum mechanics is a theory of observable correlations. Any physical constraint or structural hierarchy must ultimately manifest as a restriction on observable moments or their allowed variations. In any composite quantum system, the state's purity sets a finite resource budget for all second-moment quantities. When we partition this total budget into local and nonlocal contributions, their absolute limits are governed by the purity and the reflection overlap. While the exact geometric silhouette of this space changes with the system's dimensions, its topological integrity -- a completely filled, compact manifold -- is structurally guaranteed. The mapping from the density matrix to its Fano parameters is strictly linear, and the subsequent mapping to the budget components is quadratic and continuous. Since the rationalised coordinates are merely a continuous, one-to-one reparametrisation of the budgets, the unbroken topological nature of the physical state space is faithfully preserved. Thus, fixing the outer boundaries fixes the entire feasible region; no hidden positivity constraints can carve holes in the interior.

The two-qubit budget geometry provides an explicit, exactly solvable realisation of this universal structure. What distinguishes the two-qubit case is not the underlying physics, but the mathematical symmetry that allows all these constraints to close analytically. Yet, the core mechanism -- capacity bottlenecks -- persists in higher dimensions (see Table~\ref{tab:general}). For instance, classical correlations are always fundamentally bottlenecked by the maximal rank of their correlation tensors. This dimensional restriction universally induces a classical envelope in the budget plane, beyond which classical descriptions must fail.

Conceptually, this framework functions like a thermodynamic state space. Microscopic information, such as specific amplitudes and phases, is not directly observable; under local unitaries, it behaves as a gauge-like degree of freedom. By coarse-graining over these microscopic details, the invariant second moments emerge as macroscopic state variables. Viewed this way, the purity-budget geometry serves as a macroscopic phase diagram for quantum correlations. It provides the lowest-order, experimentally accessible description of a quantum system. Just as a thermodynamic state space dictates which phases of matter are possible under given macroscopic limits, this geometric projection determines exactly which operational behaviours and quantum resources are permitted in principle.

The experimental consequence is immediate: locating any $n$-partite state relative to its capacity envelope requires only $n+1$ purity measurements.  This geometric approach thus shifts the characterisation of quantum systems from a microscopic, tomographic challenge to a macroscopic, thermodynamic one. Because a state's position in the budget plane is entirely fixed by its global and marginal purities, assessing its structural nonclassicality requires a measurement complexity that is decoupled from the dimension of the Hilbert space. By reframing open-system dynamics as structured, constrained flows within these rigid kinematic bounds, the purity budget dictates which correlation redistributions are possible, forbidden, or inevitable. Ultimately, this geometry offers a fundamental, experimentally accessible language that identifies second-moment feasibility as the lowest-order organising principle of quantum mechanics, bridging the gap between abstract resource theories and immediate experimental application. 

Earlier second-moment approaches, such as the covariance-matrix separability criterion~\cite{deVicente:2006vka}, spin-squeezing-based entanglement witnesses~\cite{Guhne:2008qic}, and purity-based concurrence bounds~\cite{Mintert:2004wfj}, extract entanglement information as basis-dependent inequality families or scalar witnesses, each requiring optimisation over measurement settings or state-specific inputs. By contrast, the purity-budget geometry constructs the complete LU-invariant orbit space of symmetric second moments, whose capacity envelopes are structural limits rather than witness inequalities, and whose transgression certifies resources without any choice of basis. Recent work has established that quantifying entanglement in two-qubit systems requires maximally difficult measurements -- type $4$ in the randomised measurement hierarchy~\cite{Eisfeld:2026xjg}. Our result establishes the complementary floor for arbitrary $n$-partite systems: certifying its presence requires only $n+1$ purity measurements, independent of local dimensions. Together, these results define the complete measurement complexity landscape for entanglement -- a hard lower bound on what is sufficient, and a hard upper bound on what is necessary. Similarly, recent advances have provided efficient multicopy protocols to estimate stabiliser entropies directly~\cite{Haug:2023ffp}, but our purity-budget bounds provide an immediate, lower-overhead macroscopic diagnostic to constrain the maximum possible magic before deploying these advanced estimation techniques.

Finally, we note that the purity-budget framework certifies NPT entanglement but does not isolate bound entanglement, which hides in the time-symmetric sector and remains invisible to it. This is not an incidental limitation; it reflects the boundary of what second-moment information can establish. It raises a new question: what additional measurement, beyond purity, suffices to detect bound entanglement? The budget geometry provides the natural language in which to answer it, since bound entanglement is precisely the entanglement that does not require intrinsically time-odd generators.

\section{Methods}\label{sec:methods}
\noindent
We prove the claims made in the main text.

\subsection{Feasible and experimentally accessible}
\noindent
Because the two-qubit case is simple, we prove that the $(X, Y)$ projection is gapless with an explicit construction. We choose a canonical representative for any point by using the LU invariance of these coordinates. By aligning the local $z$-axes with the principal axes of the correlation matrix (i.e., $t_{ij}={\rm diag}(t_1,t_2,t_3)$), $\rho$ adopts a block-diagonal structure in the Bell basis. Its physical viability is governed by two pairs of eigenvalues:
\begin{equation}
    \lambda_{1,2} = \frac{1+t_3\pm\Delta_+}{4}, \quad \lambda_{3,4} = \frac{1-t_3\pm\Delta_-}{4}
\end{equation}
where $\Delta_\pm = \sqrt{(r_3\pm s_3)^2+(t_1\mp t_2)^2}$ shows the competition between local polarisation and nonlocal correlations. 

A continuous sweep of the parameter space $(\mu, \alpha)$ maps onto the entire budget region. We define the canonical parameters as:
\begin{equation}
    t_3 = \mu,\  t_1 = -t_2 = \mu \sin \alpha,\  r_3 = \mu \cos \alpha, \  s_3 = \cos \alpha. 
\end{equation}
These define a mapping from the unit rectangle $[0,1] \times [0, \pi/2]$ to the $(X, Y)$ plane. The topological integrity of this projection is confirmed by its boundaries: $\mu=1$ corresponds to the pure states, while $\alpha=0$ maps to the $Q=0$ boundary (see Fig.~\ref{fig:budgetXY}). As the map is continuous and the image of the parameter-space boundary encloses the budget region, every interior point corresponds to at least one valid, positive density matrix by construction. The fragile eigenvalue $\lambda_2$ vanishes for $\alpha = 0$, identifying the $Q=0$ boundary as the absolute edge of the physical state space where the density matrix remains singular, balanced on the threshold of non-positivity. Furthermore, the parametrisation shows that the $(X,Y)$ projection is not abstract; one can experimentally construct it using one-sided depolarising noise. The angle $\alpha$ is twice the Schmidt angle. By choosing it, one can start with a pure state and apply one-sided depolarisation. This process traces a downward trajectory as $\mu$ decreases monotonically to zero, eventually reaching the $X$ axis.\medskip

\noindent{\bf Relations with known two-qubit positivity conditions:} The necessary and sufficient conditions to determine the positivity of two-qubit density matrices derived by Morelli et al. are essentially two bounds on $B_{\rm NL}$~\cite{Morelli:2023khn}:
\begin{align}
    B_{\rm NL} \leqslant\ &\ 3+B_{\rm L} -4\|r\|\|s\|-4|\|r\|-\|s\||,\nonumber\\
    \sqrt{B_{\rm NL}} \geqslant\ &\ \|r\|+\|s\|-1.
\end{align}
In the purity-budget projection, the upper limit is satisfied for $B_{\rm L}+B_{\rm NL}\leqslant 3$ and the lower one is equivalent to $Q\geqslant0$. 
\begin{enumerate}
    \item[\emph{1.}] Let $d=|\|r\|-\|s\||$. Then, $4\|r\|\|s\|=2(B_{\rm L} - d^2)$, and $ B_{\rm NL}^{\max}=3+B_{\rm L} -2(B_{\rm L} - d^2)-4d = 1-B_{\rm L} + 2(d-1)^2$. Since $d$ can vary between $0$ and $1$ for a fixed $B_{\rm L}$ and the term $2(d-1)^2$ is maximised at $d=0$, the limit translates to $B_{\rm NL}^{\max}\leqslant 3-B_{\rm L}$, with equality at $\|r\|=\|s\|=\sqrt{B_{\rm L}/2}$. Thus, the upper limit is always satisfied for $B_{\rm L}+B_{\rm NL}\leqslant 3$. 

    \item[\emph{2.}] If $\|r\|+\|s\| <1$, then the lower limit is trivially satisfied for $\sqrt{B_{\rm NL}}\geqslant0$, irrespective of the value of $B_{\rm L}$. For $\|r\|+\|s\| \geqslant 1$ and a given $B_{\rm L}$, the lower limit on $B_{\rm NL}$ comes from $B_{\rm NL}\geqslant \min(\|r\|+\|s\|-1)^2=\min(B_{\rm L}+2\|r\|\|s\|-2\|r\|-2\|s\|+1)=B_{\rm L}-1$.
\end{enumerate}
 
Morelli et al.'s upper limit is satisfied by any $(X,Y)$ within $0\leqslant R\leqslant 1$ (equivalently, $0\leqslant B\leqslant3$), and the lower limit eliminates the same region as the non-negativity of $Q$ requirement in the $X$-$Y$ plane.\smallskip

\noindent{\bf Feasibility beyond two qubits:}
Since the map from $\rho$ to the budget components is quadratic and continuous, the hole-free nature of the $(B_{\rm L},B_{\rm NL})$ projection is topologically guaranteed. The rationalised coordinates merely stretch the projection; hence, they cannot create any gap inside. Therefore, for any arbitrary system, we simply need to identify the boundaries. It is possible to do this physically, as the projection remains experimentally accessible for arbitrary systems. We can always choose a pure state by the Schmidt angle and apply single-sided depolarising noise. However, in general systems, there will be multiple $Q=0$ lines; we need to choose the outermost one to trace the complete projection. This is always possible if we start depolarising from the smallest subsystem of a fully factorised, $Q=0$ pure-product state. Rather than solving a high-dimensional mathematical optimisation, this sequential depolarisation uses a natural physical process to organically expose the boundary. It works as a universal algorithm irrespective of the system's size or type. We illustrate this explicitly in Appendix~\ref{app:a}.\medskip

\noindent{\bf Generalising $Q$ -- Time-reversal symmetry and the edge of positivity:} In the fundamental spin basis defined by the angular momentum operators $\vec{J} = (J_x, J_y, J_z)$, the time-reversal operation $\Theta$ inverts the angular momentum vector, $\Theta \vec{J} \Theta^\dagger = -\vec{J}$. For qubits ($d=2$), the traceless operator algebra is spanned entirely by generators that are odd under time reversal. Saturating this odd sector defines the positivity wall, creating the vertical boundary at $X^2 = 2/3$. For higher-dimensional systems ($d \geqslant 3$), the algebra expands to include generators that are even under time reversal. The $Q \geqslant 0$ condition, therefore, enforces a trade-off between the time-odd and time-even parts of the kinematic budget. For a symmetric qudit-qudit system ($d\geqslant 3$), this boundary generalises to:
\begin{equation*}
2(d^2-1)X^2 + 4(d-1)Y^2 = d(d+1).
\end{equation*}

This geometric boundary also acts as the dynamic trajectory for one-sided noise. Because the reflection overlap factorises for product states, $Q(\rho_A \otimes \rho_B) = Q(\rho_A)Q(\rho_B)$, global $Q$ vanishes if one subsystem is fixed in a pure state polarised entirely in the time-odd sector ($Q_{\text{local}}=0$ for a local subsystem). If this odd anchor is coupled to a subsystem undergoing local depolarisation, the joint state slides along the maximal $Q=0$ ellipse.

For multipartite systems ($d_1 \otimes d_2 \otimes \dots \otimes d_n$), tracing the outermost feasibility boundary requires sequentially depolarising the subsystems, starting with the smallest. In the qubit-qutrit ($2 \otimes 3$) case, fixing the qutrit as the odd anchor while depolarising the qubit pushes the boundary out to $X^2 \approx 0.8$. Depolarising the qutrit first collapses the boundary to a vertical wall at $X^2 = 0.6$. The maximal kinematic space is therefore obtained by preserving the purity of the largest odd sector until the end.

\subsection{Two-qubit hierarchy}
\noindent {\bf Bounds on geometric discord:} 
For measurement on the $i$th subsystem, the geometric discord is defined as~\cite{Dakic:2010xfz}: 
\begin{align*}
D_G^{(i)} =\frac{1}{2}\big(\|v^{(i)}\|^2+\|t\|_F^2-k_{\max}^{(i)}\big),    
\end{align*} 
with $v^{(1,2)}=r,\,s$, and $k_{\max}^{(i)}$ denoting the largest eigenvalue of the real symmetric matrices, $K^{(1)}= rr^{\rm T} + tt^{\rm T}$ or $K^{(2)}= ss^{\rm T} + t^{\rm T}t$. Since $k_{\max}^{(i)}\ge \frac{1}{3}\mathrm{tr}(K^{(i)})$, 
\begin{align*}
D_G^{(i)} \leqslant \frac{1}{3}\big(\|v^{(i)}\|^2+\|t\|_F^2\big) \leqslant \frac13 B = \frac{R}{4-3R}.  
\end{align*}
This is an absolute upper as no global unitary can affect it. 

If we are allowed some local information, we get a lower bound and a tighter upper bound. The Rayleigh-Ritz theorem applied to $K^{(i)}$ gives $k_{\max}^{(i)} =\max_{\hat n^{(i)}\in\mathbb R^3,\ |\hat n^{(i)}|=1} \hat n^{(i)T} K^{(i)} \hat n^{(i)}$. In the basis where $t=\mathrm{diag}(t_1,t_2,t_3)$, we have $\hat n^{(i)T} K^{(i)} \hat n^{(i)}= (\hat n^{(i)}\cdot v^{(i)})^2 + \sum_{j=1}^3 t_j^2 (n^{(i)}_j)^2$. Thus, we have 
\begin{align*}
D_G^{(i)}(\rho)
=&\ \frac{1}{4}\Big(
\|v^{(i)}\|^2+B_{\rm NL}\nonumber\\
&\ -\max_{|\hat n^{(i)}|=1}\big[(\hat n^{(i)}\cdot v^{(i)})^2+\sum_j t_j^2 (n^{(i)}_j)^2\big]
\Big).\label{eq:RayleighRitzDG}
\end{align*}
This means that the part of the total budget $\|V^{(i)}\|^2+B_{\rm NL}$ available for classicalisation of subsystem $i$ minus the maximal amount a local measurement on $i$ can extract is captured by geometric discord. Now, let $t_{\max}^2=\max\{t_1^2,t_2^2,t_3^2\}$. Two elementary inequalities apply for any unit vector $\hat n$: $(\hat n^{(i)}\cdot v^{(i)})^2\leqslant\|v^{(i)}\|^2$ and $\sum_j t_j^2 (n^{(i)}_j)^2\leqslant t_{\max}^2$. Thus, $t_{\max}^2\ \leqslant\ k_{\max}^{(i)}\ \leqslant\ \|v^{(i)}\|^2 + t_{\max}^2$. Substituting, we get
\begin{align*}
\frac{1}{4}(B_{\rm NL}-t_{\max}^2)\ \leqslant&\ D_G^{(i)}\ \leqslant\ 
\frac{1}{4}(\|v^{(i)}\|^2+B_{\rm NL}-t_{\max}^2).
\end{align*}

\noindent
{\bf Bound on negativity:} Let the spectrum of $\rho^{T_A}$ be $\{\overline\lambda_1,\overline\lambda_2,\overline\lambda_3,-x\}$ (with $\overline\lambda_i\geqslant0$ and $x>0$). Let $\lambda= \overline\lambda_1+\overline\lambda_2+\overline\lambda_3 = 1+x$. For fixed $\lambda$, the sum of squares $\sum_{i=1}^3(\overline\lambda_i)^2$ is minimised when all $\overline\lambda_i$ are all equal, i.e., $\lambda/3$. Thus, $3(P-x^2) \geqslant\ (1+x)^2$. For an admissible root $x>0$, the largest possible $x$ is the upper root:
\begin{align*}
\qquad\mathcal N =x\leqslant \max\left(0, \sqrt{9R/(64-48R)}-1/4 \right).
\end{align*}

\noindent{\bf CHSH violation:}
Let $t_p\geqslant t_q\geqslant t_r\geqslant 0$ be the singular values of the correlation matrix. Any redistribution among them does not affect the budgets since $B_{\rm NL}=t_p^2+t_q^2+t_r^2$, but it affects CHSH violation as the condition for that is $t_p^2+t_q^2>1$. We consider two extreme possibilities. \emph{Best redistribution:}  If $t_r=0$, then CHSH violation $\Longleftrightarrow t_p^2+t_q^2 = B_{\mathrm{NL}} > 1$, or equivalently, $Y > \sqrt{2/3-X^2/2}$. This implies that a CHSH violation is possible only above the C envelope. \emph{Worst redistribution:} If $t_p=t_q=t_r=\sqrt{B_{\mathrm{NL}}/3}$, then $t_p^2+t_q^2 = 2B_{\mathrm{NL}}/3$. In this case, CHSH\ violation $\Longleftrightarrow B_{\mathrm{NL}} > 3/2$, or equivalently, $Y > \sqrt{4/5-3X^2/5}$. Above this line, all points violate the CHSH identity.

For general higher-dimensional systems, the purity-budget geometry continues to constrain Bell correlations. While exact analytic thresholds analogous to the two-qubit CHSH case are no longer known, the framework naturally allows one to define inequality-dependent guaranteed-violation regions at fixed second-moment budget.

\subsection{Beyond two qubits: Constructing the QC envelopes}
\noindent
We consider an $n$-partite system with ordered dimensions $d_1 \leqslant d_2 \leqslant \dots \leqslant d_n$. We generate the hierarchical QC envelopes by assuming only $m \in [1,\;n-1]$ subsystems are part of a fully quantum state with off-diagonal generators, and the rest are classically simulatable. We obtain the Q$^m$C$^{n-m}$ envelope as follows:
\begin{enumerate}[leftmargin=10pt]
    \item[{\bf\large 1}] \textbf{The peak ($V_{\rm pure}$):} We first determine the point where it touches the pure-state line.
    \begin{enumerate}
        \item If $m= 1$ (i.e., the quantum part cannot be entangled), $V_{\rm pure}$ is the pure product point: 
        \begin{align*}
        B_{\rm L}^{\rm pure} = \sum_{j=1}^{n} (d_j - 1),\  B_{\rm NL}^{\rm pure} = (D - 1) - B_{\rm L}^{\rm pure}.    
        \end{align*}
        \item 
        Otherwise, we assume the smallest $(n-m)$ subsystems to be classical with marginal purities $P_i=1$ with $i=\{1,2,\ldots,n-m\}$, and all $m$ quantum subsystems have the lowest marginal purities $P_j=1/d_j$ with $j\in\{n-m+1,\ldots,n\}$. This gives us
        \begin{align*}
        B_{\rm L}^{\rm pure} = \sum_{i=1}^{n-m} (d_i - 1),\  B_{\rm NL}^{\rm pure} = (D - 1) - B_{\rm L}^{\rm pure}.    
        \end{align*}
    \end{enumerate}

    \item[{\bf\large 2}] 
    \textbf{Maximum correlation} ($V_{\rm cor}$): We then identify the point at which the envelope touches the $B_{\rm L}=0$ axis by assuming that the smallest $m$ subsystems now belong to a fully quantum state. At this point, the smallest classical system of dimension $d_{(m+1)}$ bottlenecks the global purity, and we set $P=1/d_{(m+1)}$ and $P_k=1/d_k$, where $k=\{1,2,\ldots,n\}$. This gives
    \begin{align*}
    B_{\rm L}^{\rm cor} = 0,\  B_{\rm NL}^{\rm cor} = D\frac{1}{d_{(m+1)}} - 1.    
    \end{align*}

    \item[{\bf\large 3}] 
    \textbf{Intermediate hinges} ($V_{\rm hinge}$): These are intermediate bottlenecks that appear in mixed-dimension systems. To find them, we consider all other assignments (permutations) for $m$ subsystems being part of some quantum state and the rest being classical (permutations among the classical subsystems or quantum subsystems do not count). Every new assignment gives a hinge point:
    \begin{align*}
    B_{\rm L}^{\rm hinge} = \sum_{k=1}^{n} \left( \frac{d_k}{d^{\rm C}_{\min}} - 1 \right),
    \  B_{\rm NL}^{\rm hinge} = D\frac{1}{d^{\rm C}_{\min}} - 1-B_{\rm L}^{\rm hinge},    
    \end{align*}    
    where $d^{\rm C}_{\min}$ is the dimension of the smallest classical subsystem in that permutation. We discard all hinges with $B_{\rm L}^{\rm hinge}\leqslant0$ or $B_{\rm L}^{\rm hinge}\geqslant B_{\rm L}^{\rm pure}$.  

    \item[{\bf\large 4}] 
    \textbf{Joining the lines:} Finally we arrange all distinct hinge points in a decreasing $\theta^{\rm hinge} = \tan^{-1}(B_{\rm NL}^{\rm hinge}/B_{\rm L}^{\rm hinge})$ order. We add the $V_{\rm cor}$ (the leftmost point) and $V_{\rm pure}$ (the rightmost point) in the beginning and the end of the set, respectively. We then join the neighbouring points in this ordered set. If the slope of the line segment joining the previous point to a hinge is greater than that of the segment joining the next point, we discard the hinge point to obtain the convex hull.
\end{enumerate}

It is not difficult to see why this algorithm works in the $(B_{\rm L},B_{\rm NL})$ plane. We start from $V_{\rm pure}$ and move inside. To define the maximum budget boundary, we must ensure we get the highest possible $B_{\rm NL}$ return for every unit of $B_{\rm L}$ spent, down to the strict fundamental floor dictated by any chosen classical-quantum partition. In a particular permutation, we attempt to maximise global purity (equivalently $B$) by minimising local purities (we spend $B_{\rm L}$ to increase $B_{\rm NL}$). Because the global state's purity can never exceed the purity of its most mixed classical part, the global purity $P$ hits a hard bottleneck at $1/d^{\rm C}_{\min}$ (where $d^{\rm C}_{\min}$ is the dimension of the smallest classical subsystem in that specific arrangement). At this limit, the subsystems have minimised their local purities as much as the global constraint allows, meaning we cannot extract any more correlation budget. This saturation point is the $V_{\rm hinge}$ for that permutation. Only when all quantum subsystems are smaller than or equal to the smallest classical subsystem can we minimise the quantum marginals completely, reaching the $B_{\rm NL}$ axis. By calculating this capacity limit for all possible structural permutations and plotting them, we map out every local extreme. Finally, by applying the convex hull rule, the algorithm traces the absolute outermost boundary. This guarantees that for any given amount of local polarisation ($B_{\rm L}$), the envelope defines the absolute maximum nonlocal correlation ($B_{\rm NL}$) that the Q$^m$C$^{n-m}$ structure can physically support. From this, it is simple to verify that $B_{\rm NL}=1$ gives the CC and QC envelopes for two-qubit systems.

\begin{figure*}
    \captionsetup[subfigure]{labelformat=empty}
    \centering
    
    \subfloat[{\bf a} Qubit-qutrit phase space]{
    \includegraphics[width=0.32\textwidth]{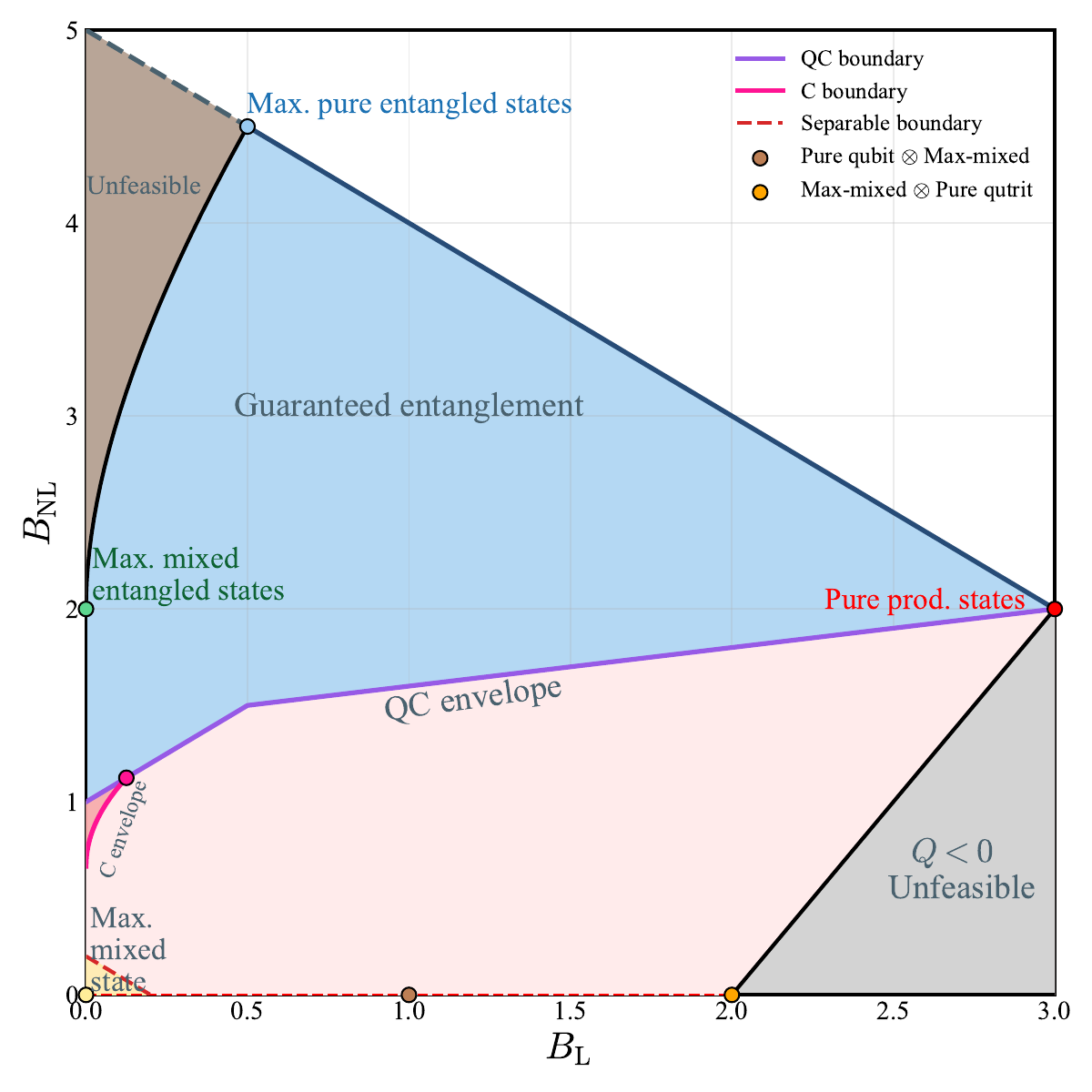}\label{fig:23geometry}}\hfill
    \subfloat[{\bf b} Qutrit-qutrit phase space ]{
    \includegraphics[width=0.32\textwidth]{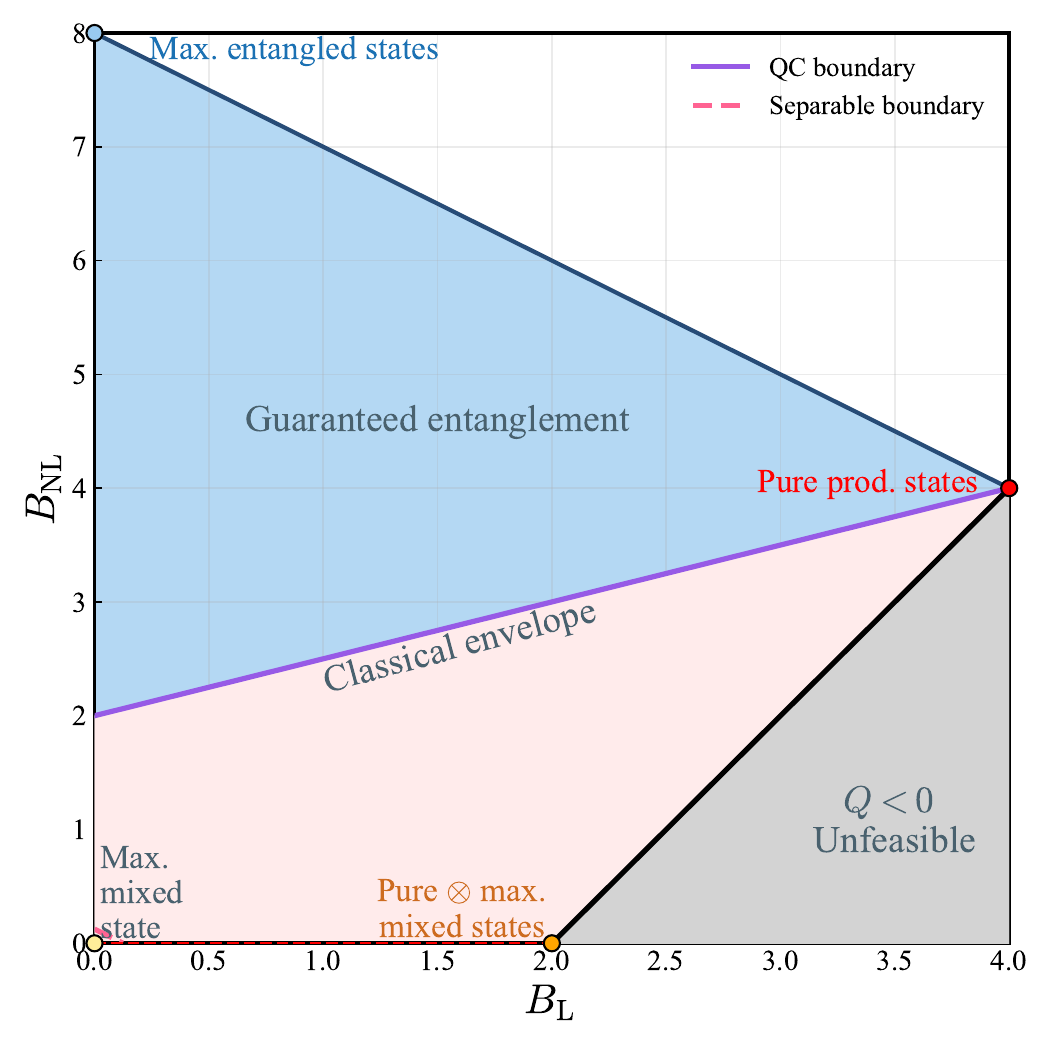}\label{fig:33geometry}}\hfill
    \subfloat[{\bf c} Three-qubit phase space]{
    \includegraphics[width=0.32\textwidth]{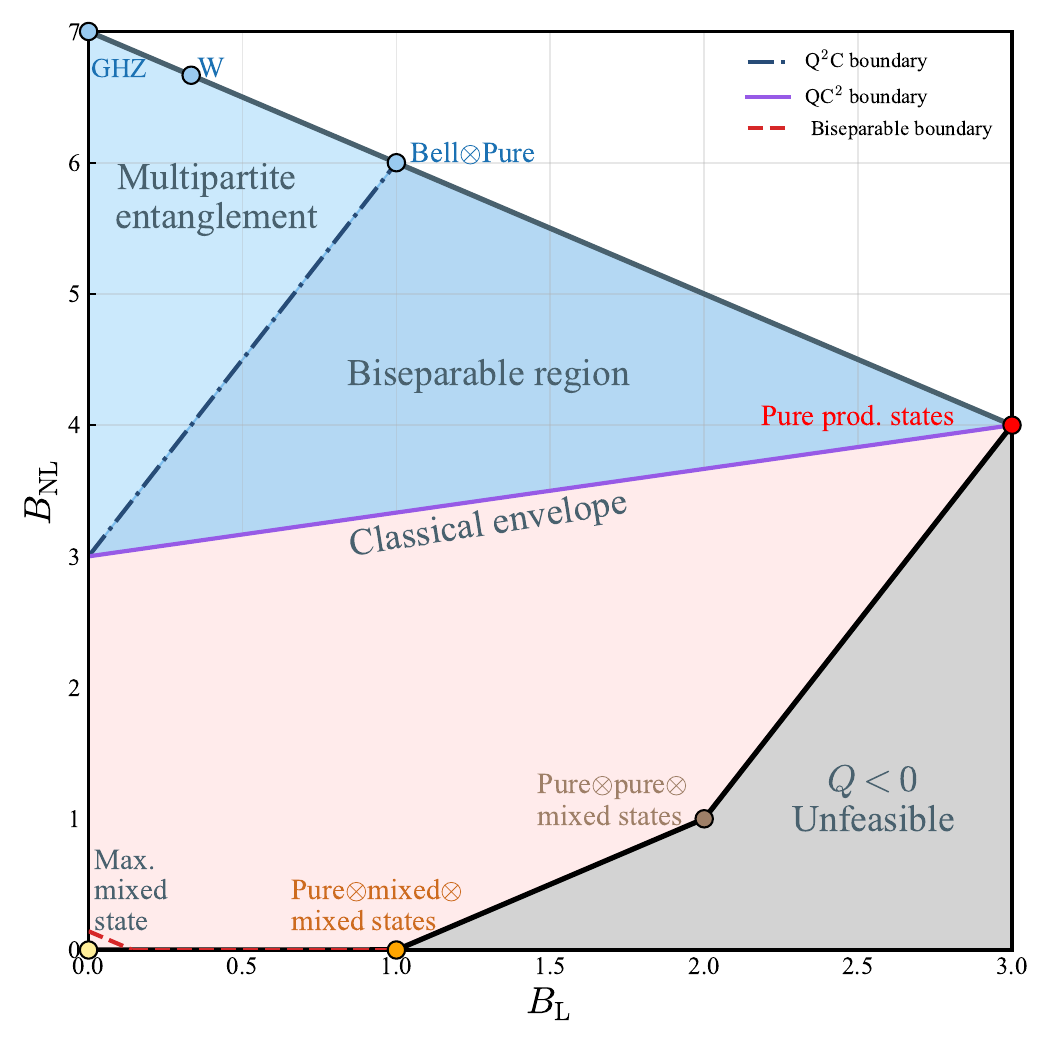}\label{fig:222geometry}}
   \caption{\label{fig:PhaseSpaceBLBNL}\textbf{Dimensional bottlenecks and kinematic envelopes in higher-dimensional purity-budget spaces.} The local ($B_{\rm L}$) and nonlocal ($B_{\rm NL}$) budgets for mixed-dimension and multipartite systems: the grey areas represent unphysical states excluded by time-reversal symmetry constraints ($Q < 0$). \textbf{(a)} Qubit-qutrit ($2 \otimes 3$) phase space. Dimensional asymmetry imposes a strict capacity bottleneck; the smaller qubit lacks the necessary degrees of freedom to entangle with the qutrit maximally, capping the correlation limit below the theoretical peak. It also forces the C and QC envelopes to diverge, producing a guaranteed discordant region (dark pink). \textbf{(b)} Qutrit-qutrit ($3 \otimes 3$) phase space. Symmetric dimensions permit the maximum nonlocal budget to be reached, with the physical space tightly bounded by a continuous elliptical positivity wall corresponding to one-sided depolarisation. \textbf{(c)} Three-qubit ($2 \otimes 2\otimes 2$) phase space. The multipartite structure leads to a multi-segmented geometry where the piecewise positivity wall physically corresponds to a trajectory of sequential, one-by-one local depolarisation. The nested inner boundaries define absolute capacity limits for classical correlations and for entanglement across progressively more parties. Across all systems, traversing above the defined classical-quantum capacity envelopes mathematically forces the activation of asymmetric generators, acting as a universal, basis-independent guarantee of NPT entanglement.} 
\end{figure*}

\begin{figure*}
    \captionsetup[subfigure]{labelformat=empty}
    \centering
    \subfloat[{\bf a} Maximally entangled pure state $(X,Y)$]{
    \includegraphics[width=0.22\textwidth]{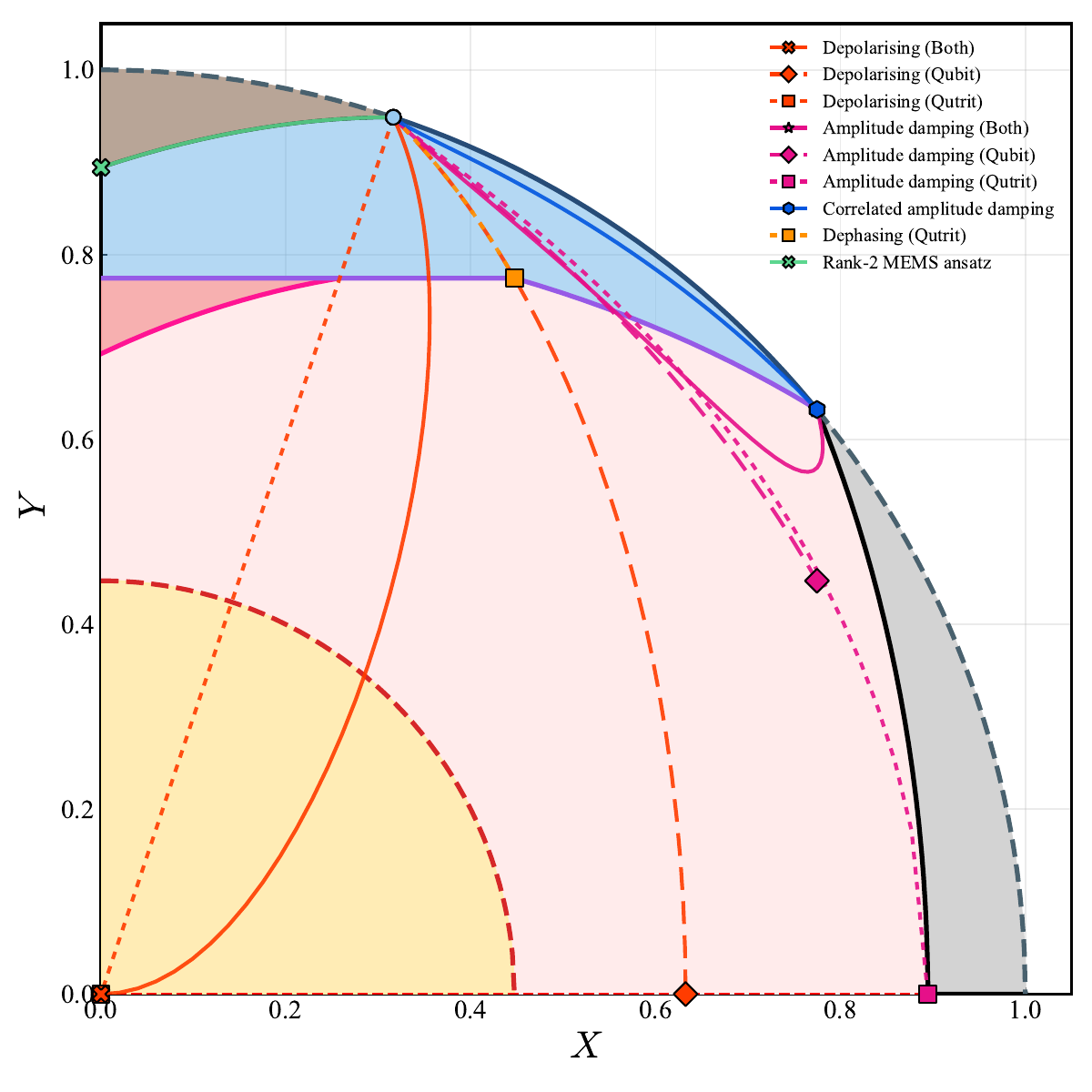}}\hfill
     \subfloat[{\bf b} Maximally entangled pure state $(B_{\rm L}, B_{\rm NL})$]{
    \includegraphics[width=0.22\textwidth]{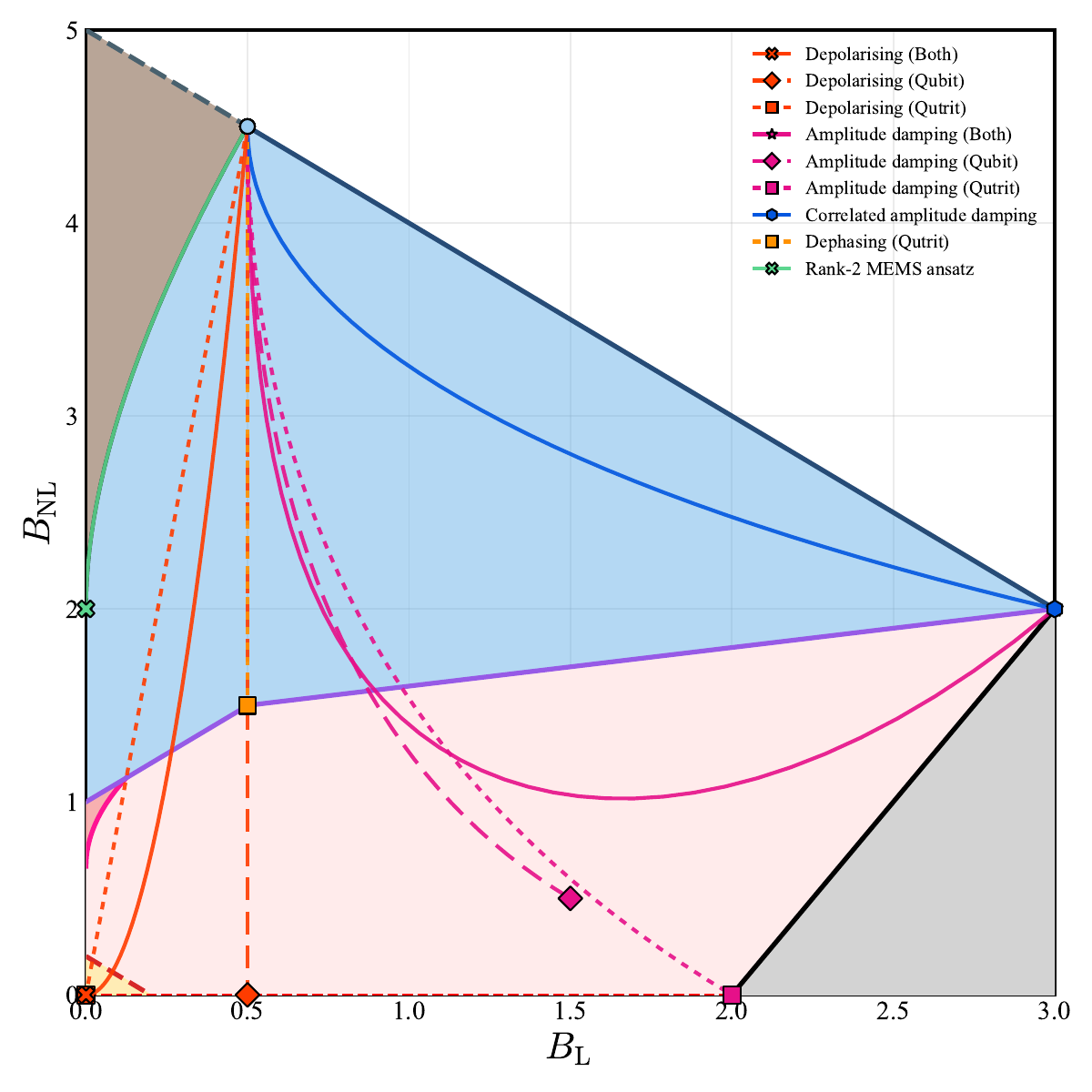}}\hfill
    \subfloat[{\bf c} Maximally mixed entangled state $(X,Y)$]{
    \includegraphics[width=0.22\textwidth]{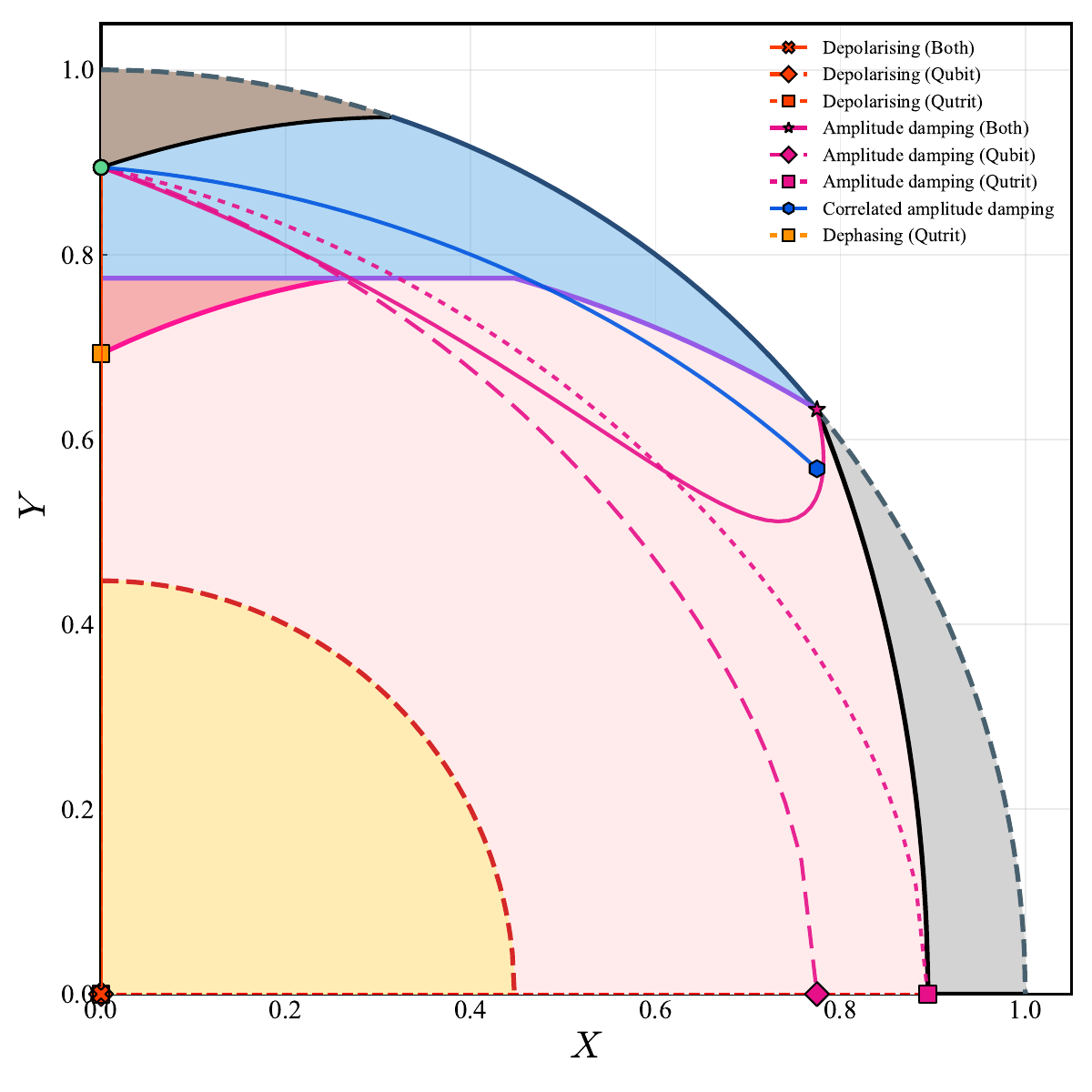}}\hfill
    \subfloat[{\bf d} Maximally mixed entangled state $(B_{\rm L}, B_{\rm NL})$]{
    \includegraphics[width=0.22\textwidth]{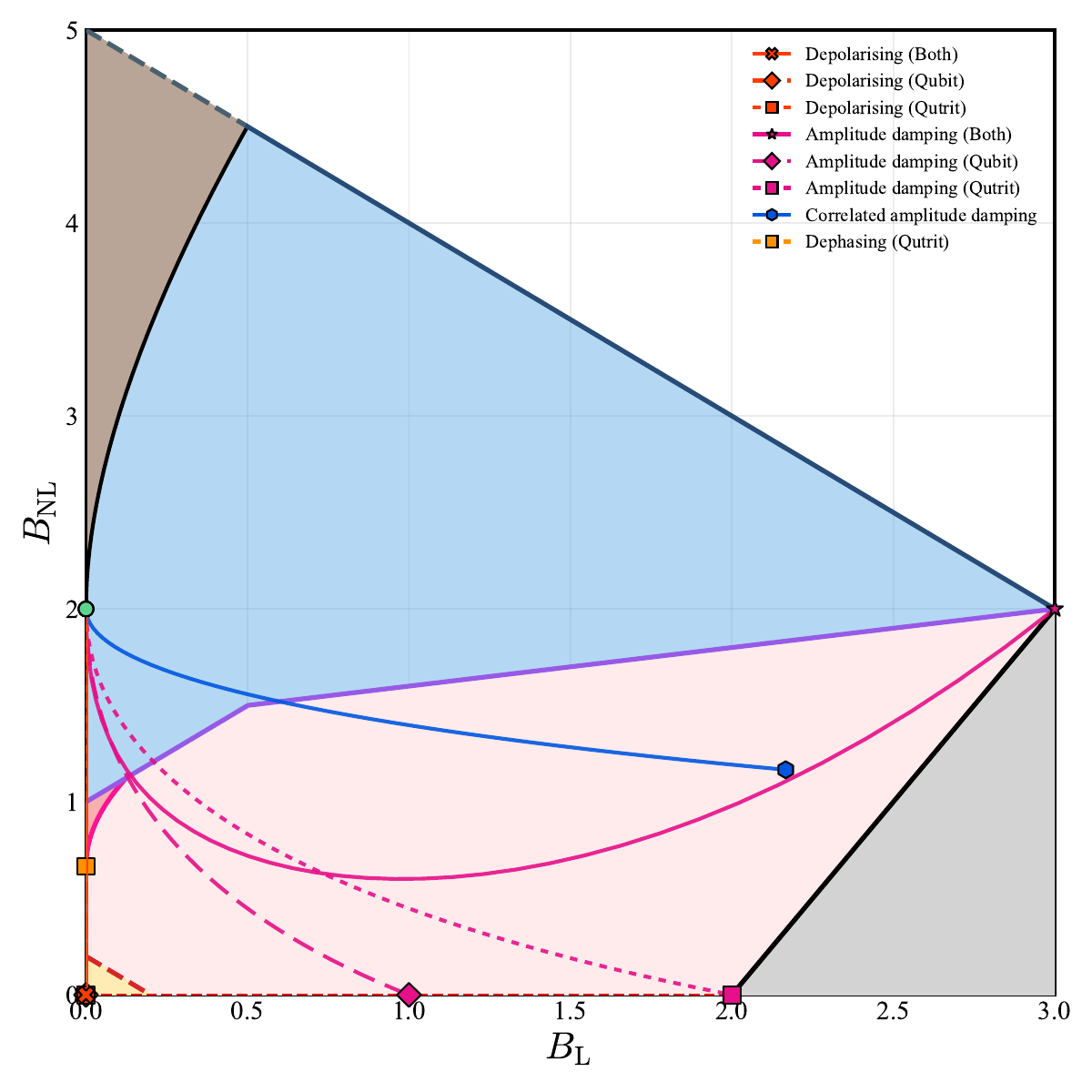}}\\
    \subfloat[{\bf e} Mixed-entangled state $(X,Y)$]{
    \includegraphics[width=0.22\textwidth]
    {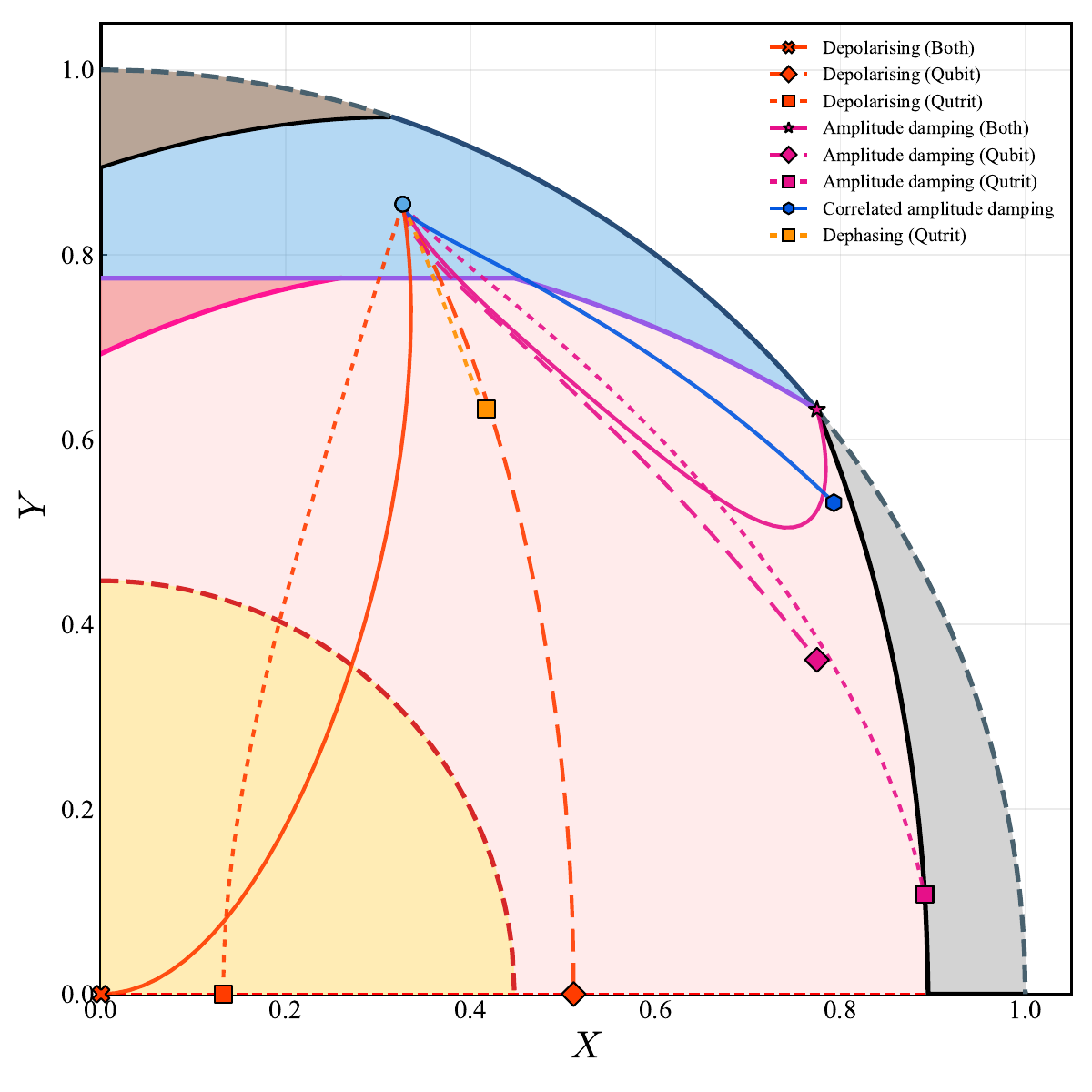}}\hfill
    \subfloat[{\bf f} Mixed-entangled state $(B_{\rm L}, B_{\rm NL})$]{
    \includegraphics[width=0.22\textwidth]{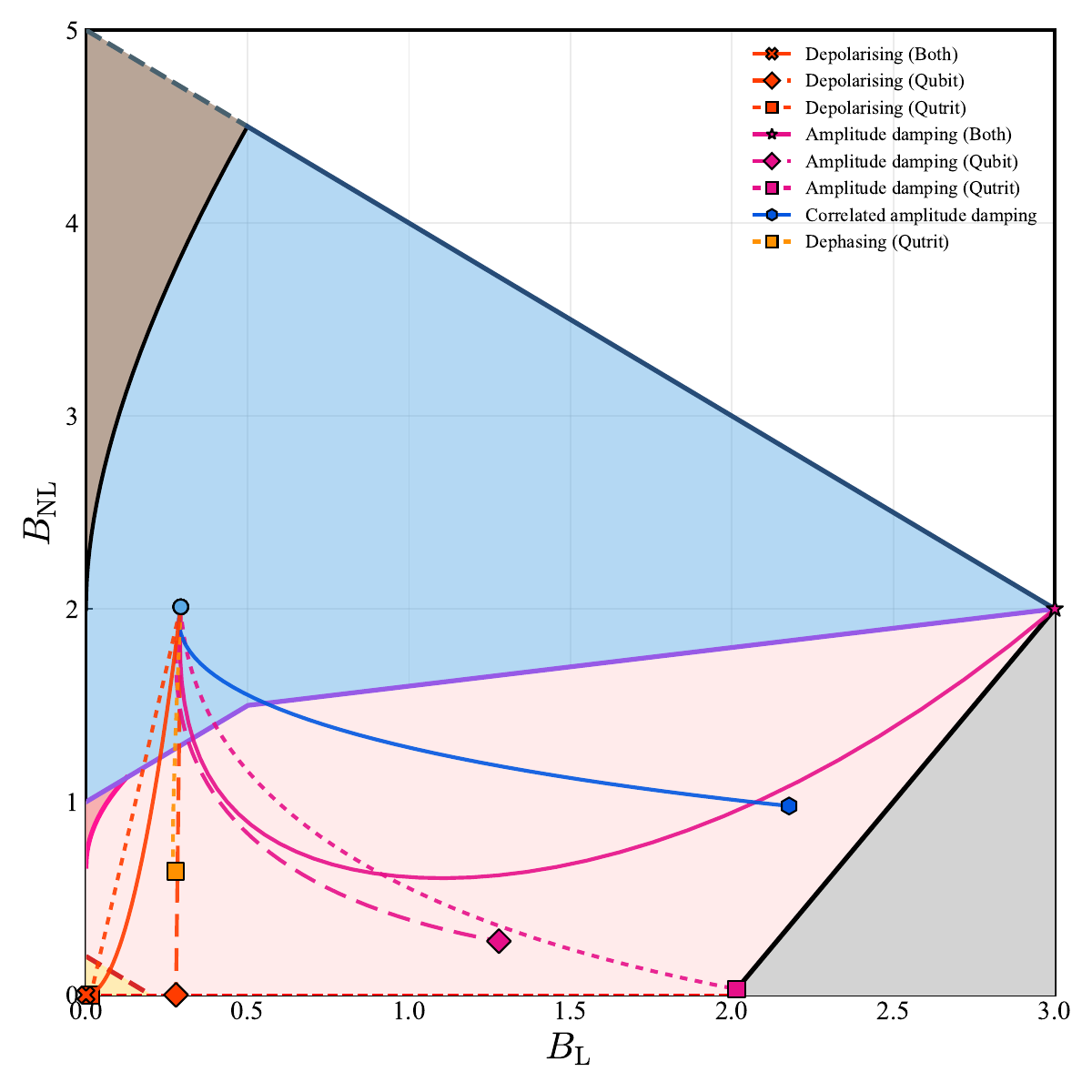}}\hfill
    \subfloat[{\bf g} Mixed-separable state $(X,Y)$ ]{
    \includegraphics[width=0.22\textwidth]{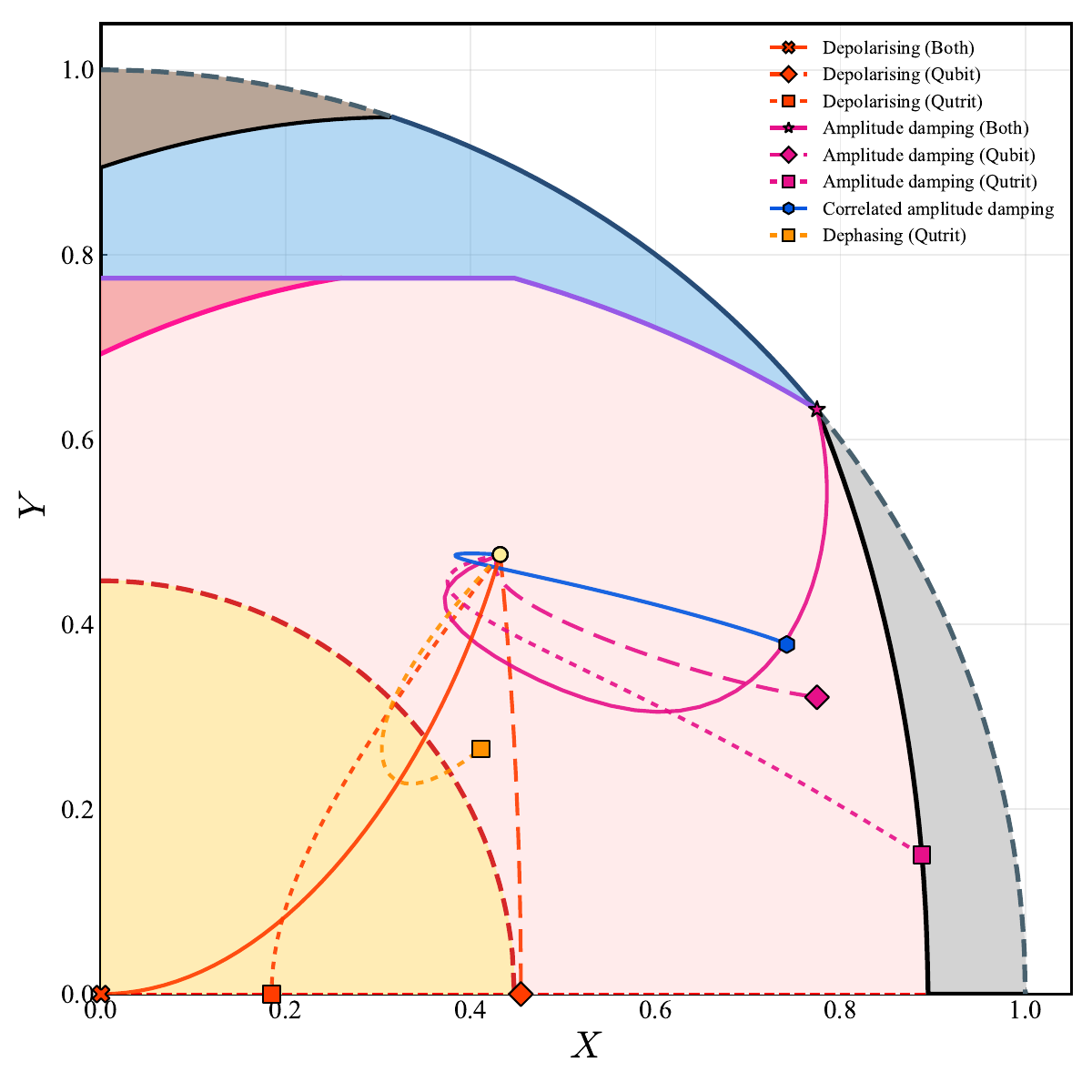}}\hfill
    \subfloat[{\bf h} Mixed-separable state $(B_{\rm L}, B_{\rm NL})$]{
    \includegraphics[width=0.22\textwidth]{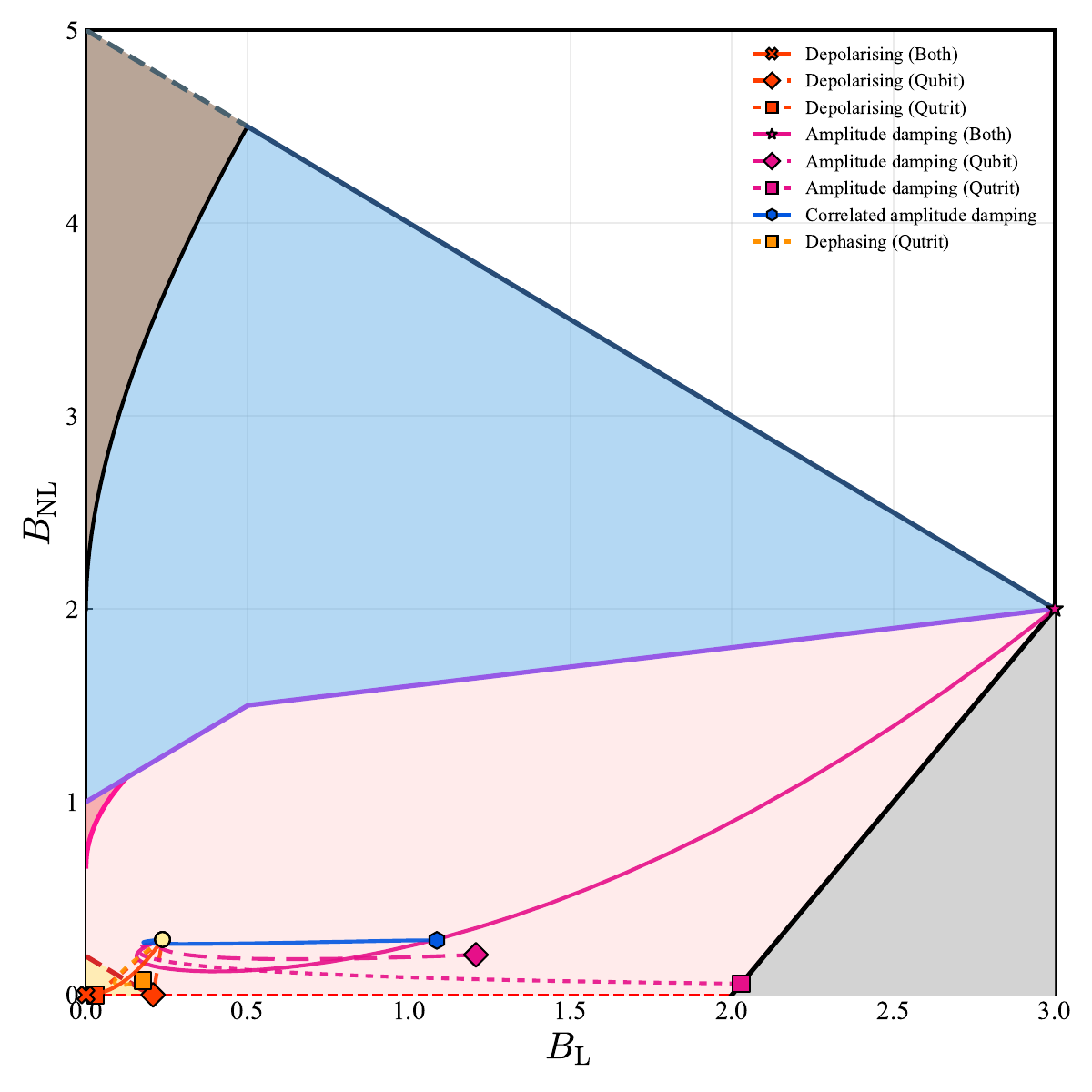}}\\
    \subfloat[{\bf i} Pure-product state $(X,Y)$]{
    \includegraphics[width=0.22\textwidth]{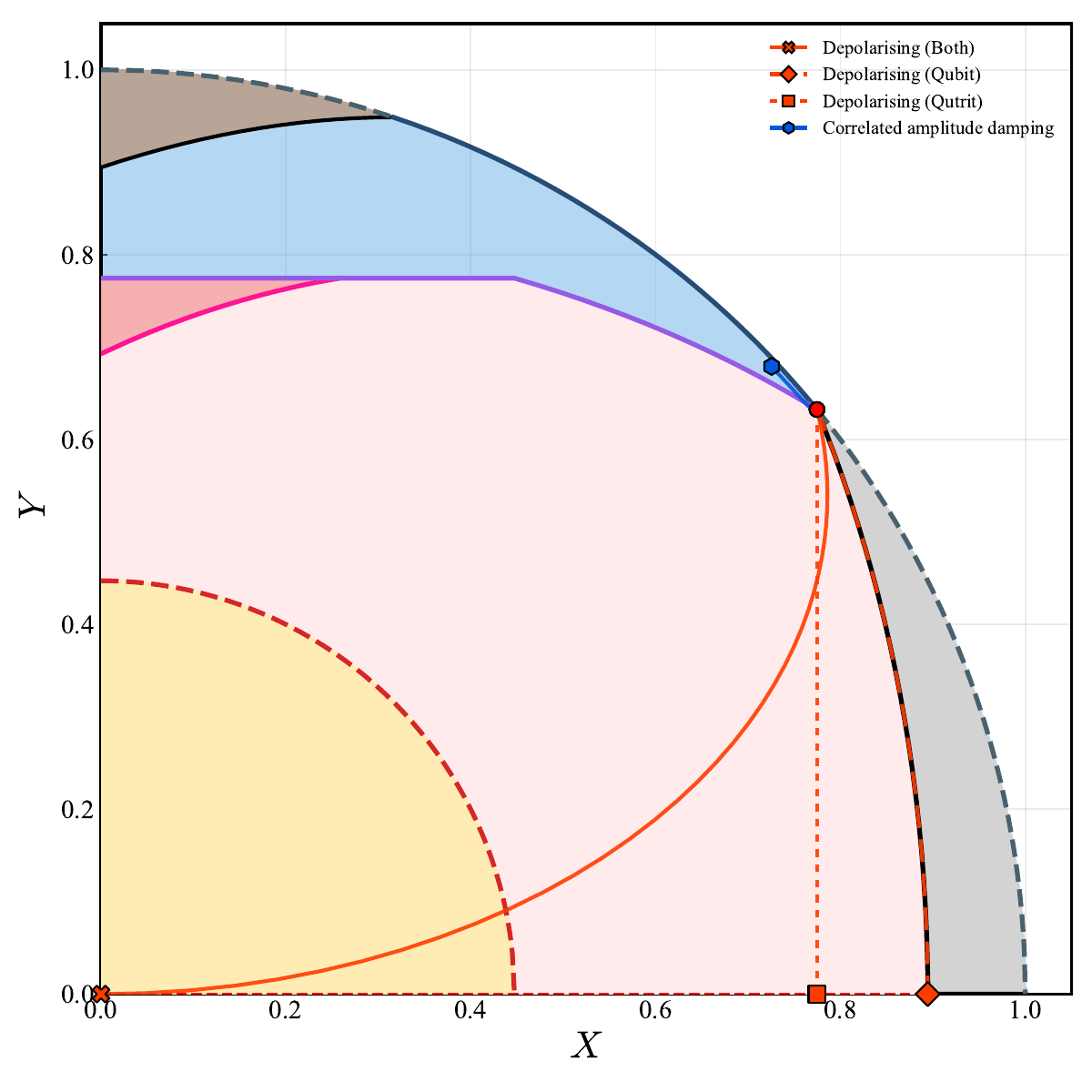}}\hfill
    \subfloat[{\bf j} Pure-product state $(B_{\rm L}, B_{\rm NL})$]{
    \includegraphics[width=0.22\textwidth]{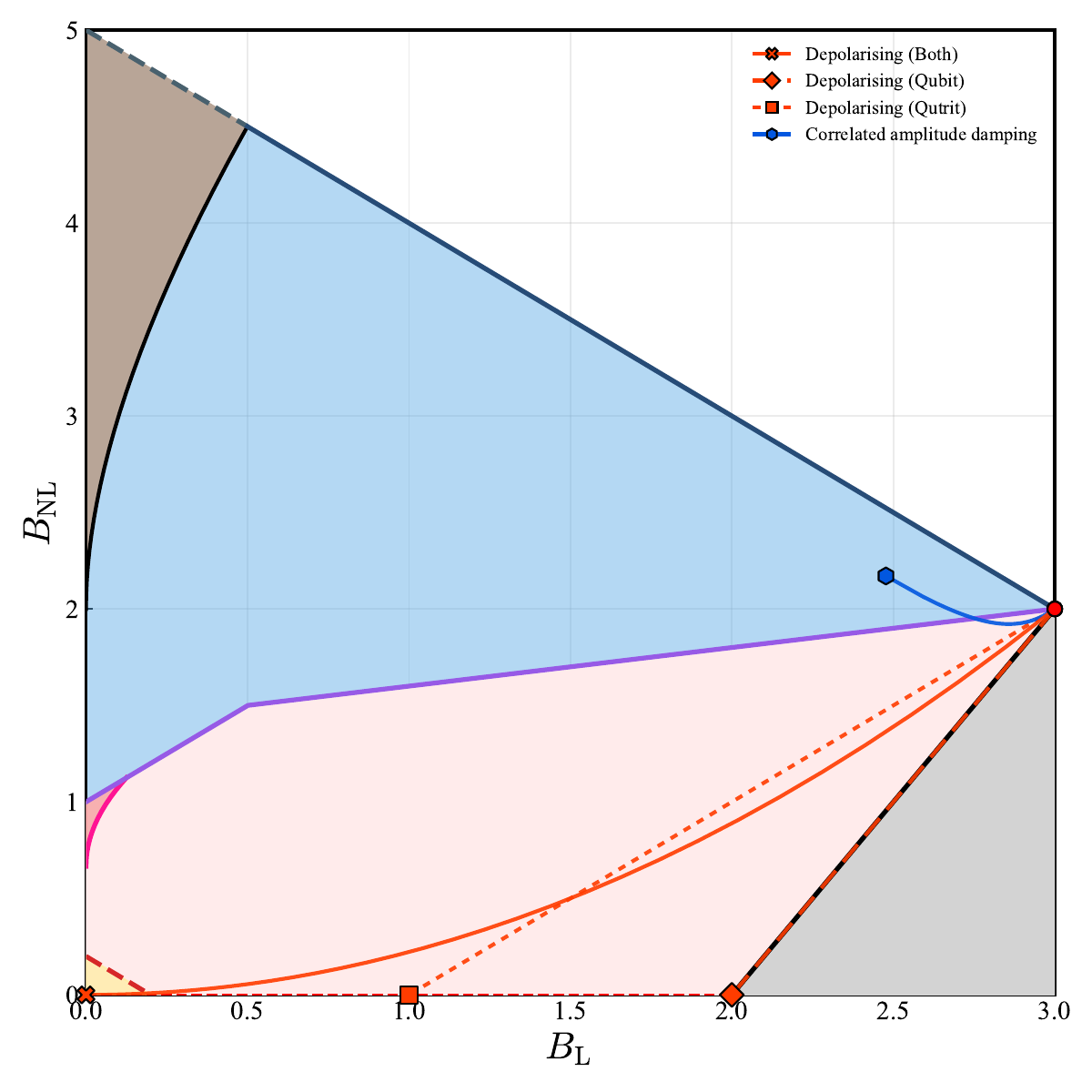}}\hfill
    \subfloat[{\bf k} Classical state $(X,Y)$]{
    \includegraphics[width=0.22\textwidth]{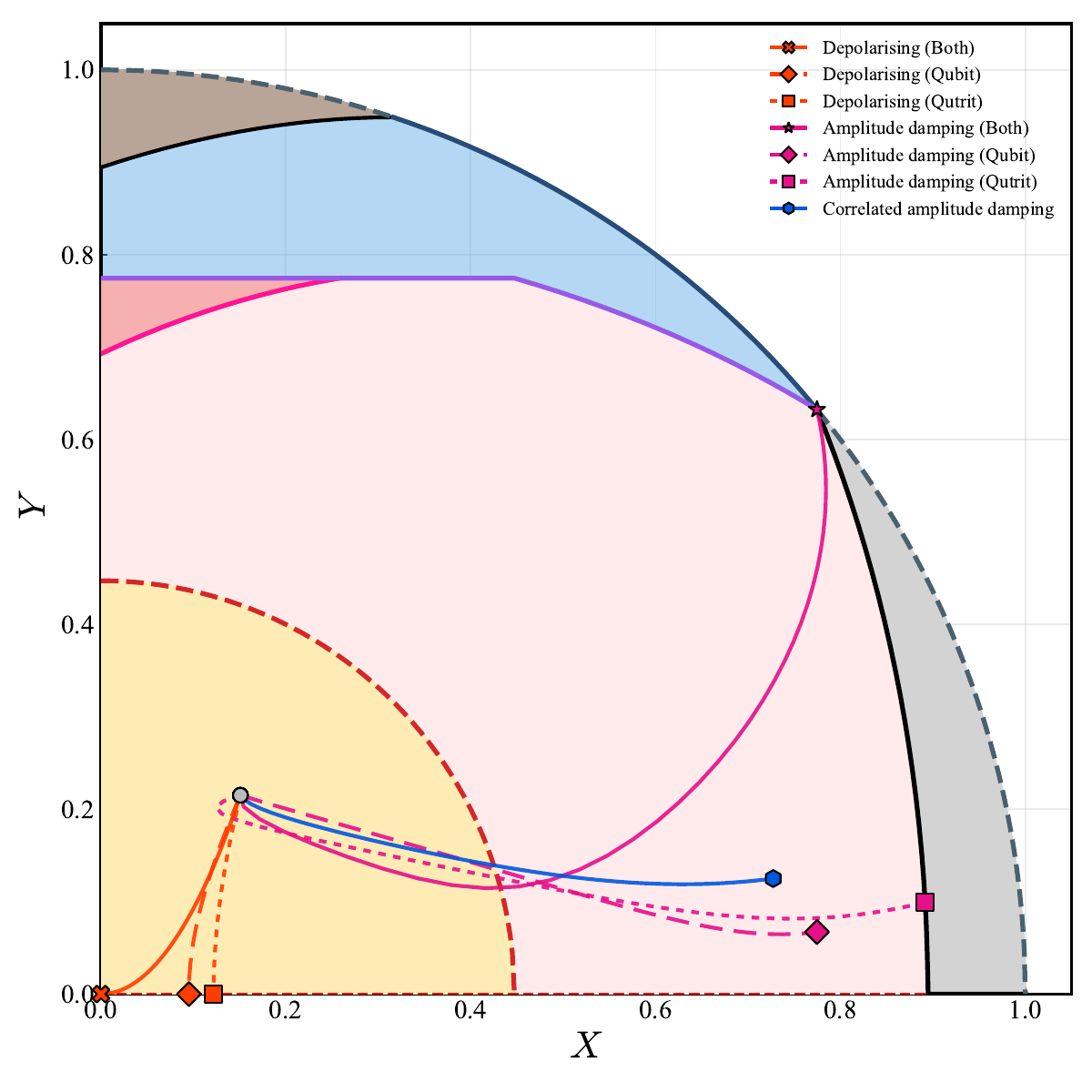}}\hfill
    \subfloat[{\bf l} Classical state $(B_{\rm L}, B_{\rm NL})$]{
    \includegraphics[width=0.22\textwidth]{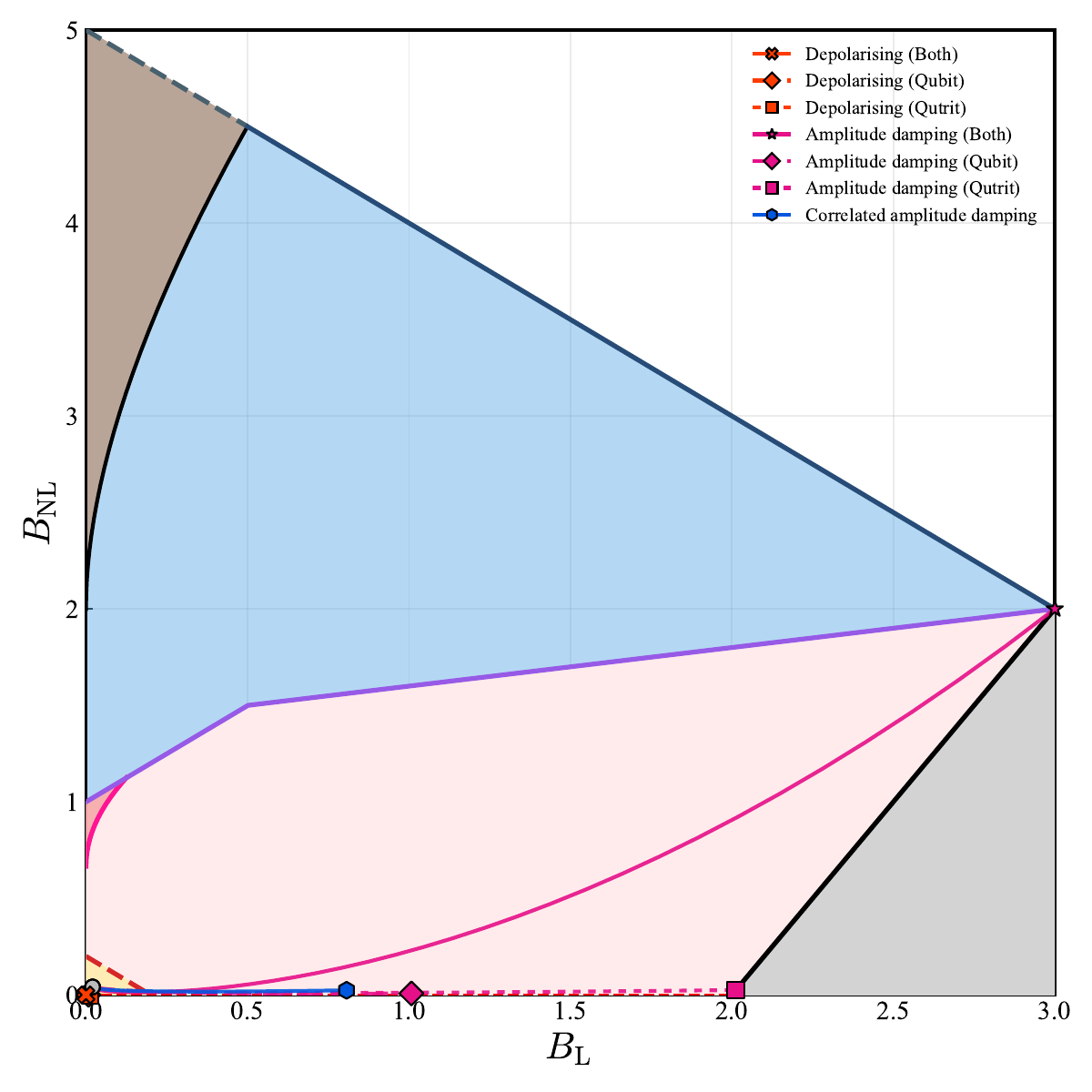}}
    \caption{\label{fig:QubitQutritNoise}\textbf{Asymmetric kinematic flows in qubit-qutrit open-system dynamics.} \textbf{a--l}, Parametric decoherence trajectories for bipartite qubit-qutrit ($2 \otimes 3$) systems, projected onto the $(X,Y)$ plane (odd panels) and the $(B_{\rm L},B_{\rm NL})$ plane (even panels). We show evolution for six distinct initial states: (\textbf{a, b}) a maximally entangled pure state, (\textbf{c, d}) a maximally mixed entangled state, (\textbf{e, f}) an entangled state, (\textbf{g, h}) a separable state, (\textbf{i, j}) a pure-product state, and (\textbf{k, l}) a classically correlated state. Dimensional asymmetry dictates that noise trajectories depend on which subsystem is targeted. For one-sided local noises, anchoring the pure qubit while depolarising the qutrit drives the state down to a rigid vertical boundary ($X^2=0.6$). Conversely, depolarising the qubit while anchoring the massive local budget of the qutrit pushes the trajectory outward along a stretched ellipse to the floor ($X^2=0.8$). This structural skew alters the endpoints and geometric signatures of one-sided amplitude damping and one-sided depolarising channels based on the targeted subsystem. The trajectories for local noises now behave differently, e.g., local amplitude damping on the qubit no longer takes the system to the $Y=0$ line, but the same on the qutrit does. The effects of both local and correlated two-sided noise are still similar to those of the two-qubit case.}
\end{figure*}

\begin{figure*}
    \captionsetup[subfigure]{labelformat=empty}
    \subfloat[{\bf a} Maximally entangled state $(X,Y)$]{
    \includegraphics[width=0.22\textwidth]{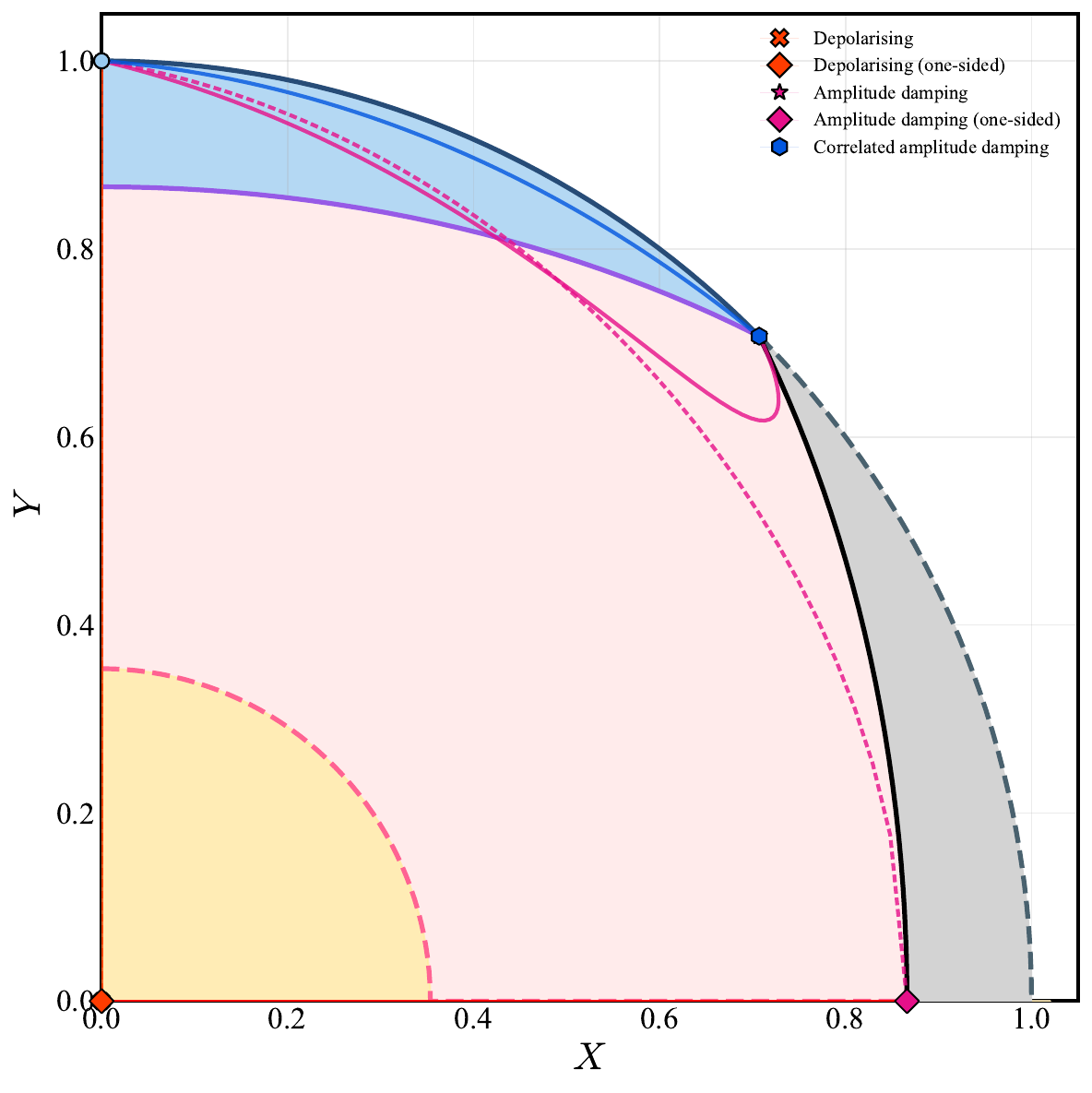}}\hfill
    \subfloat[{\bf b} Maximally entangled state $(B_{\rm L}, B_{\rm NL})$]{
    \includegraphics[width=0.22\textwidth]{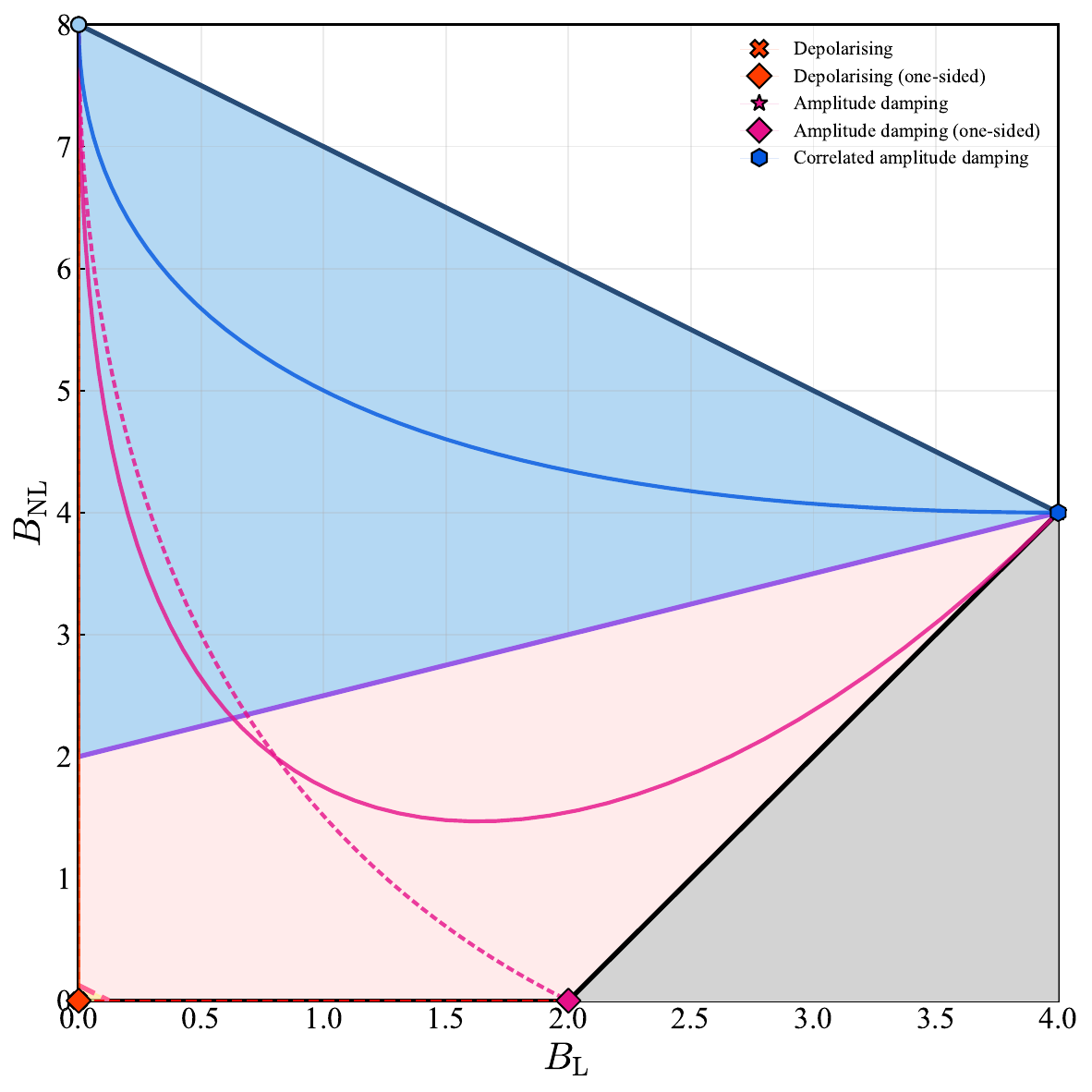}}\hfill
    \subfloat[{\bf c} Pure-product state $(X,Y)$]{
    \includegraphics[width=0.22\textwidth]{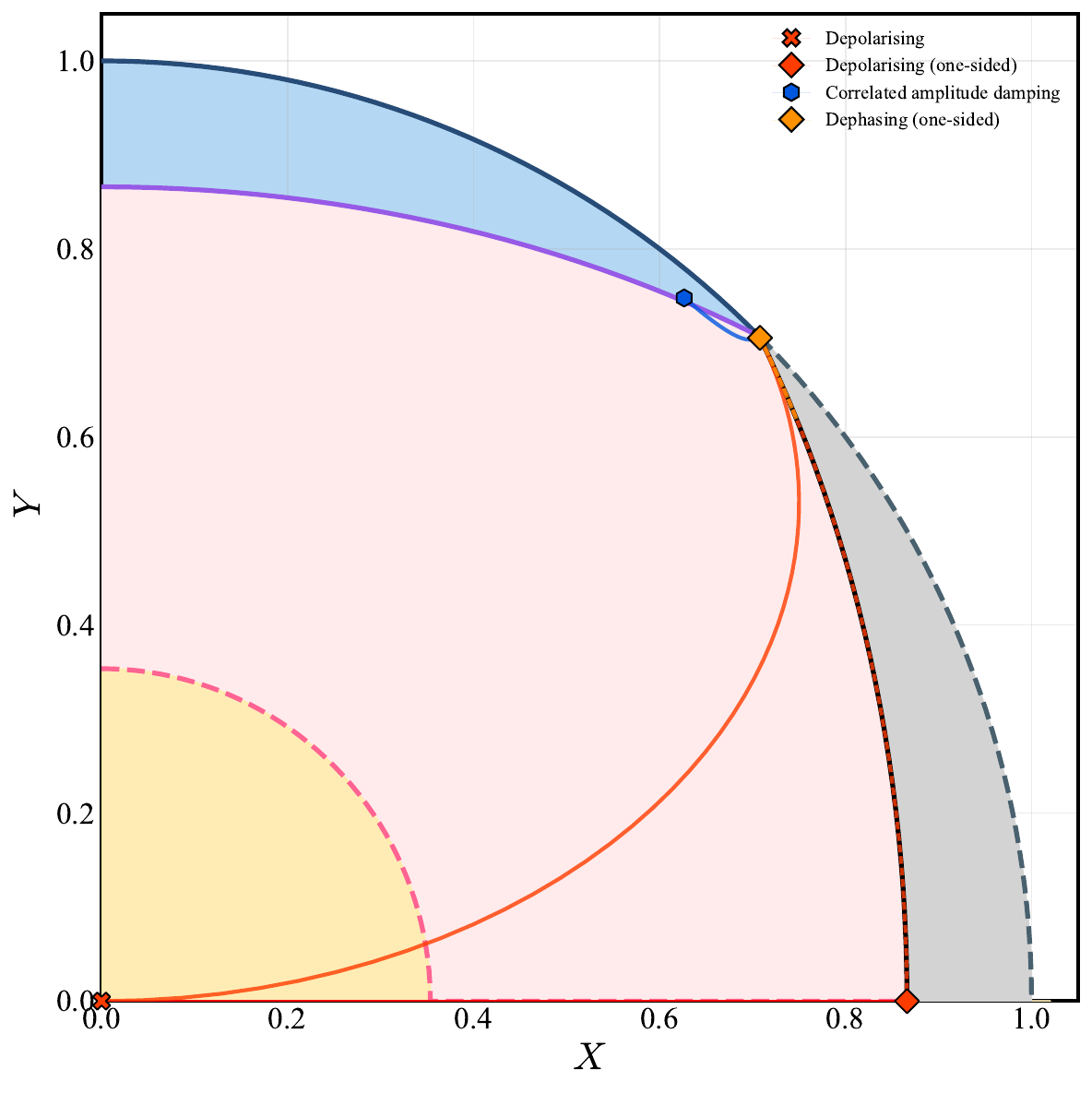}}\hfill
    \subfloat[{\bf d} Pure-product state $(B_{\rm L}, B_{\rm NL})$]{
    \includegraphics[width=0.22\textwidth]{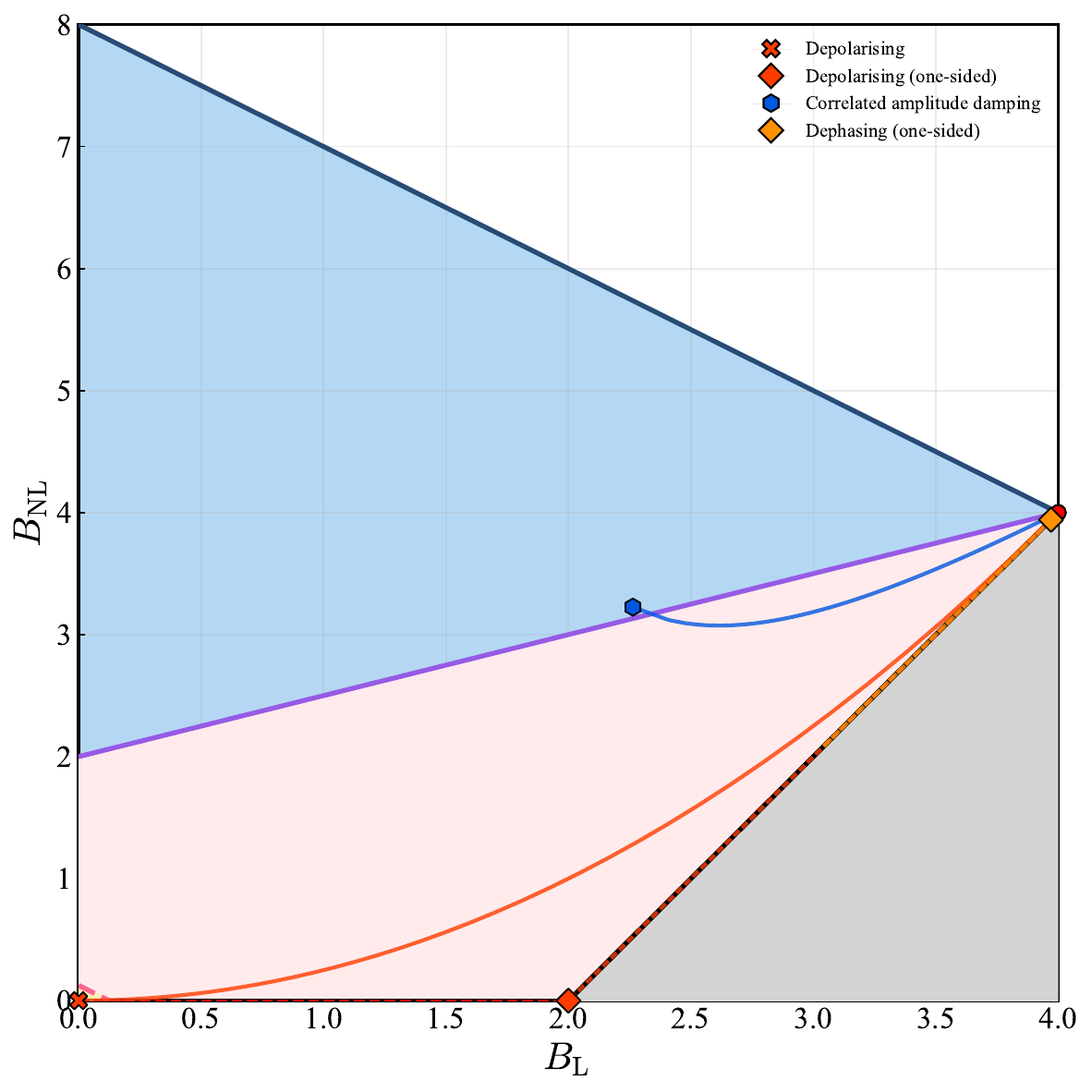}}\\
    \subfloat[{\bf e} PPT-separable state $(X,Y)$]{
    \includegraphics[width=0.22\textwidth]{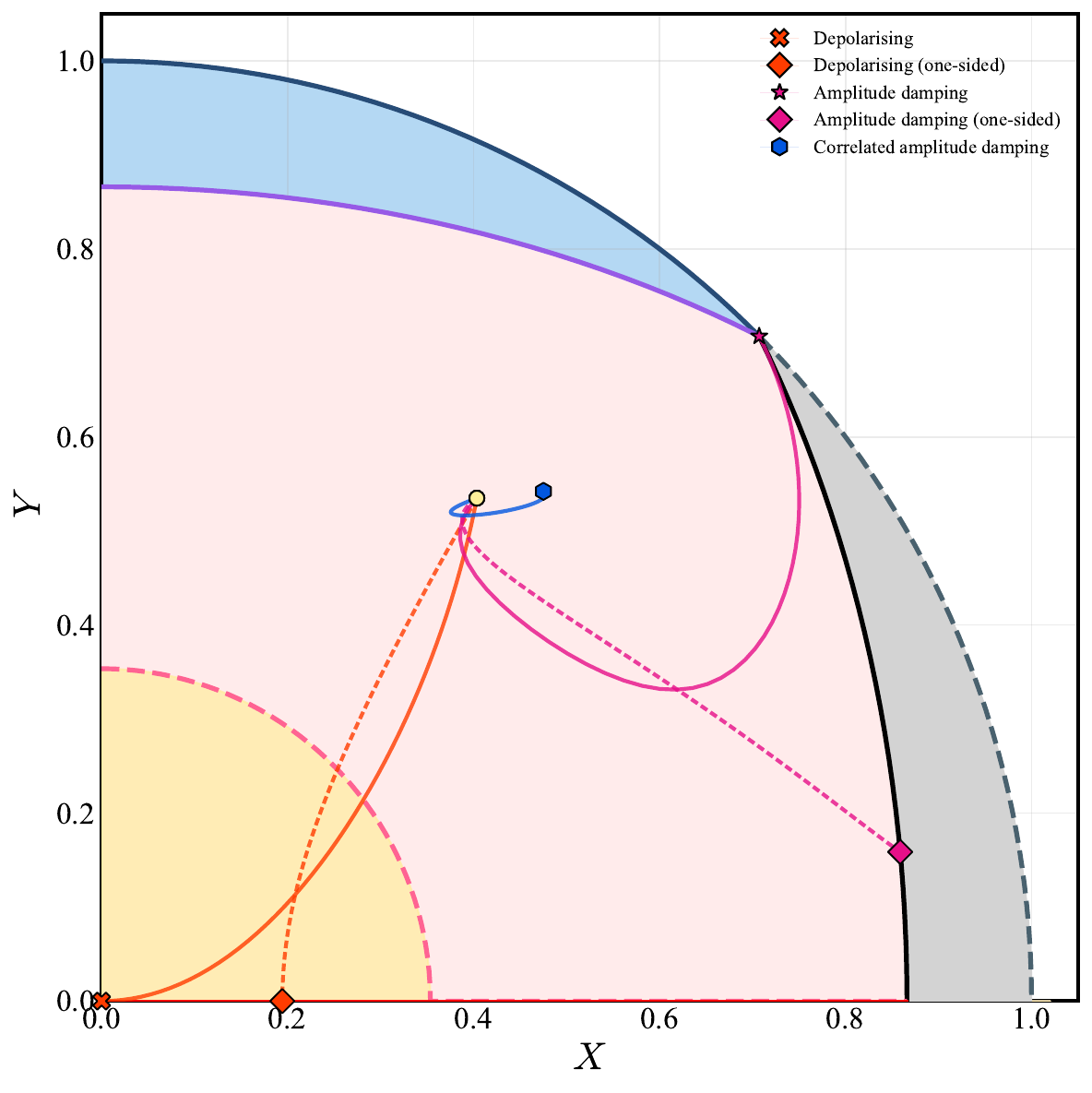}}\hfill
    \subfloat[{\bf f} PPT-separable state $(B_{\rm L}, B_{\rm NL})$]{
    \includegraphics[width=0.22\textwidth]{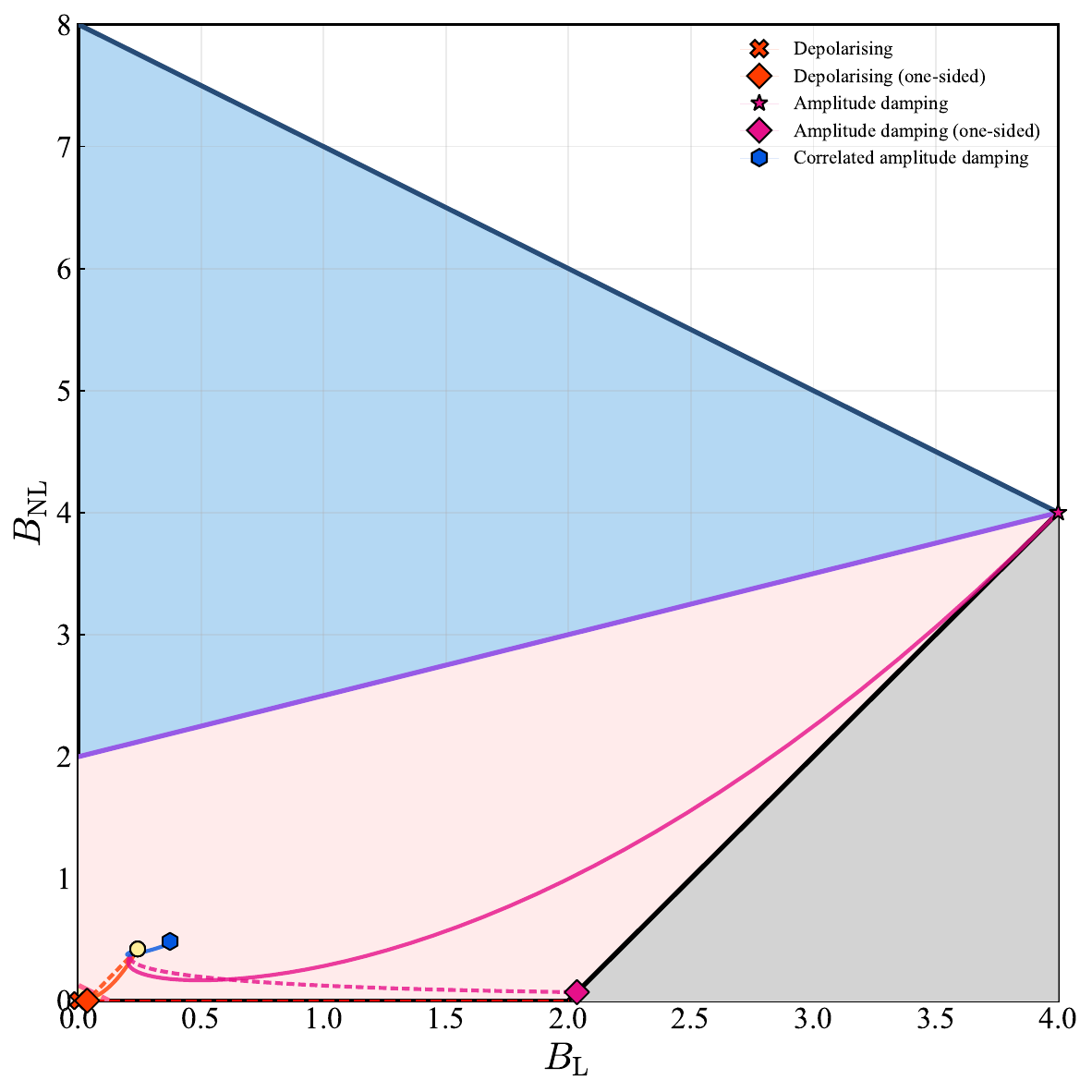}}\hfill
    \subfloat[{\bf g} NPT-entangled state $(X,Y)$]{
    \includegraphics[width=0.22\textwidth]{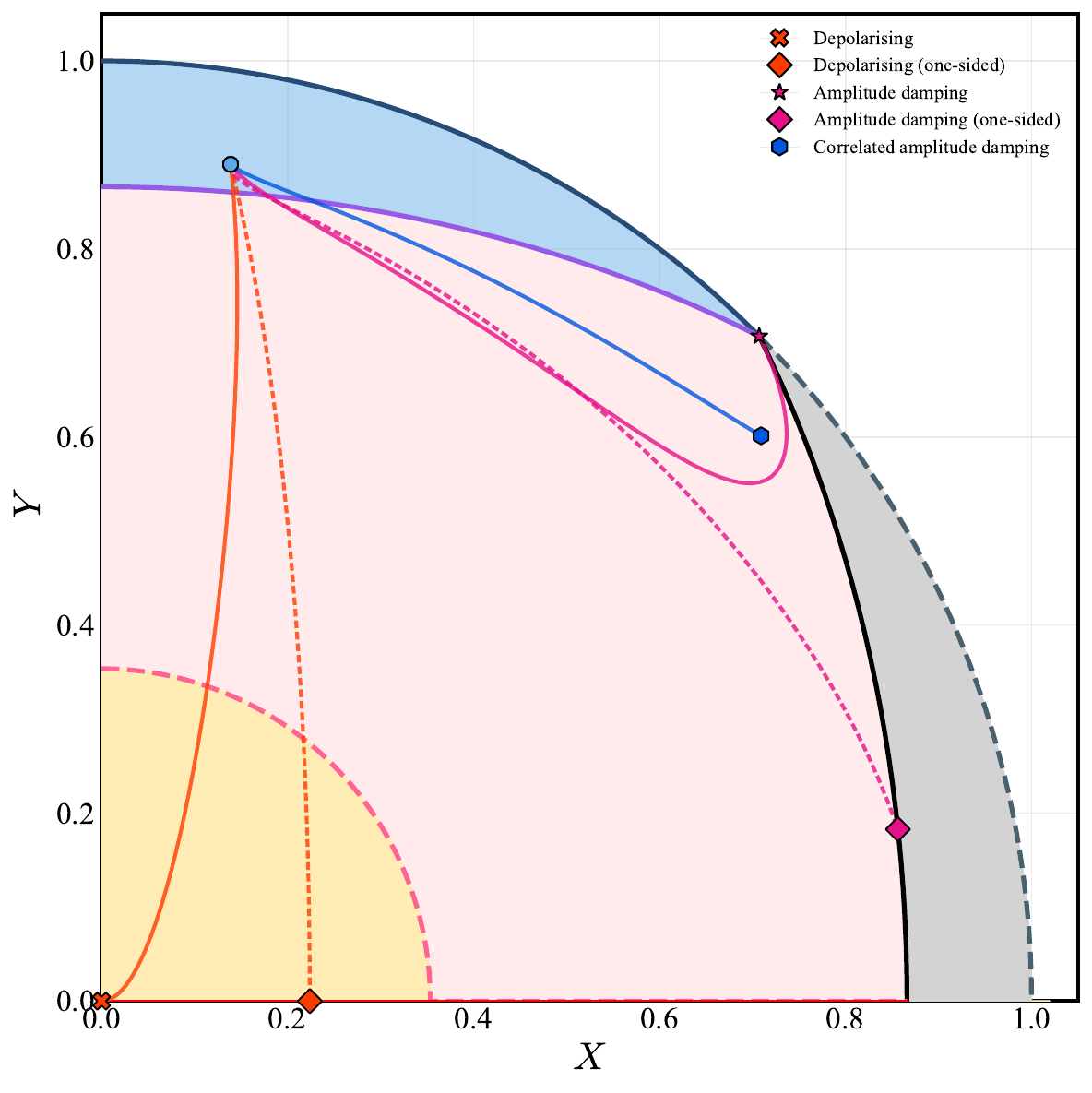}}\hfill
    \subfloat[{\bf h} NPT-entangled state $(B_{\rm L}, B_{\rm NL})$]{
    \includegraphics[width=0.22\textwidth]{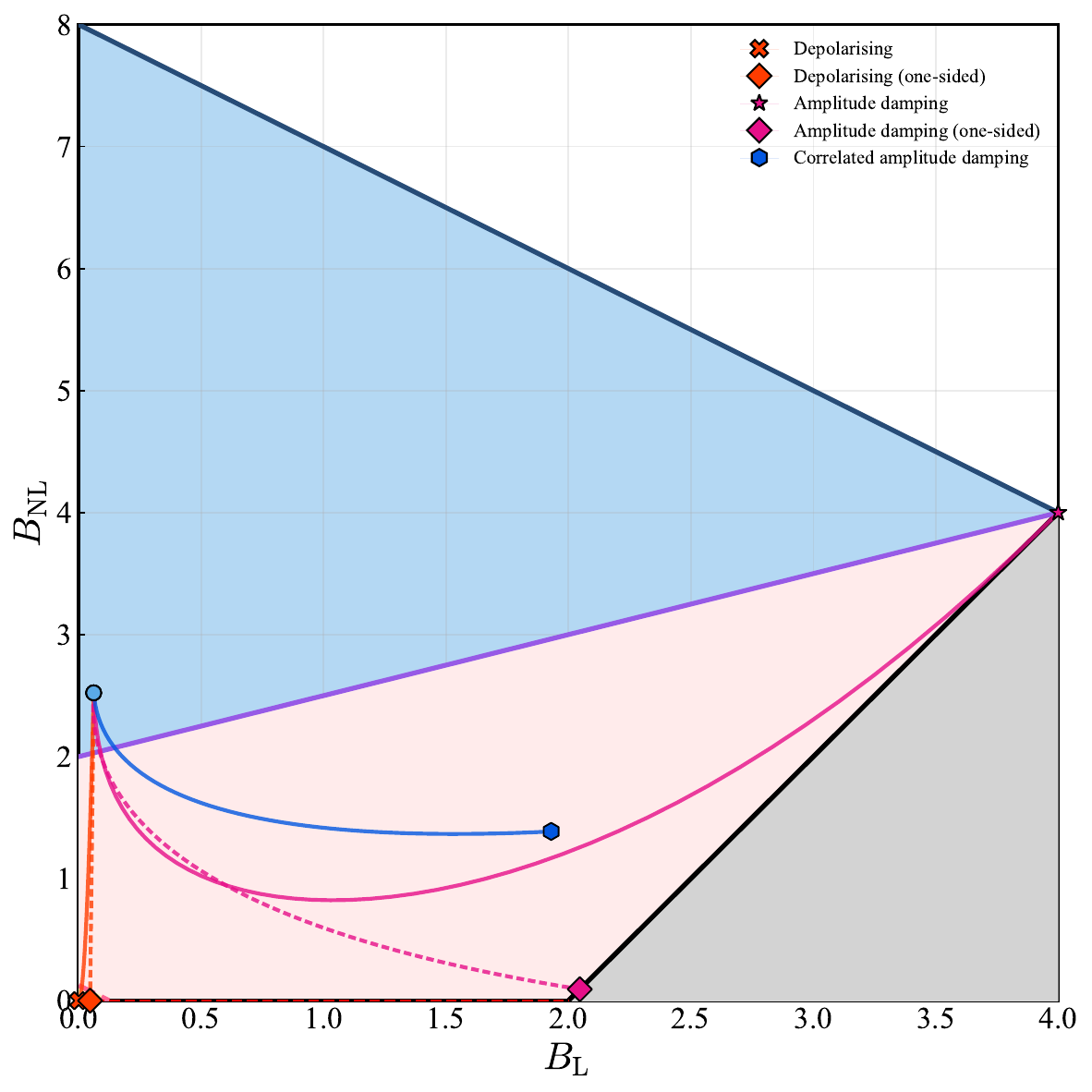}}\\
    \subfloat[{\bf i} Classical state $(X,Y)$]{
    \includegraphics[width=0.22\textwidth]{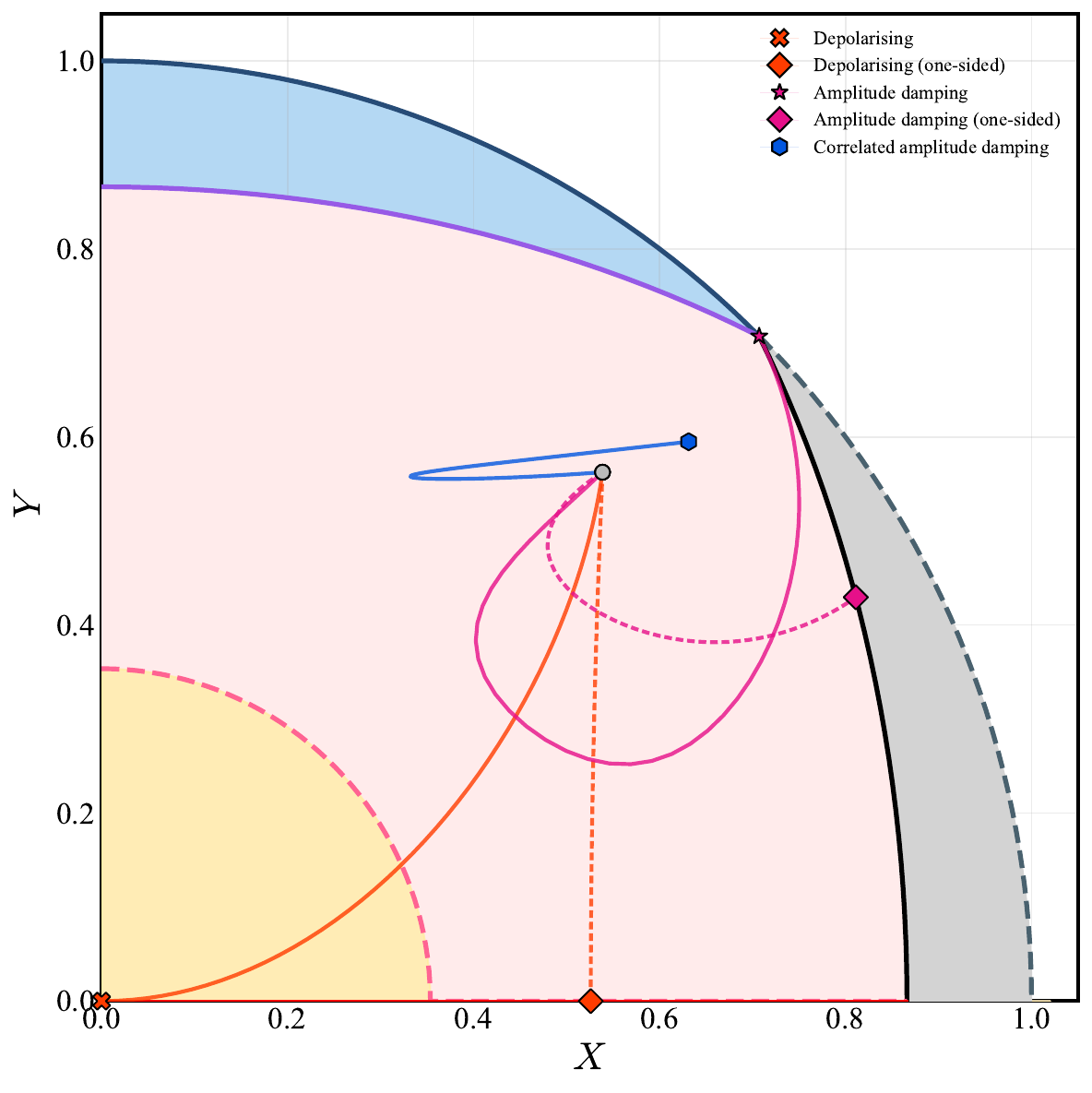}}\hspace{15pt}
    \subfloat[{\bf j} Classical state $(B_{\rm L}, B_{\rm NL})$]{
    \includegraphics[width=0.22\textwidth]{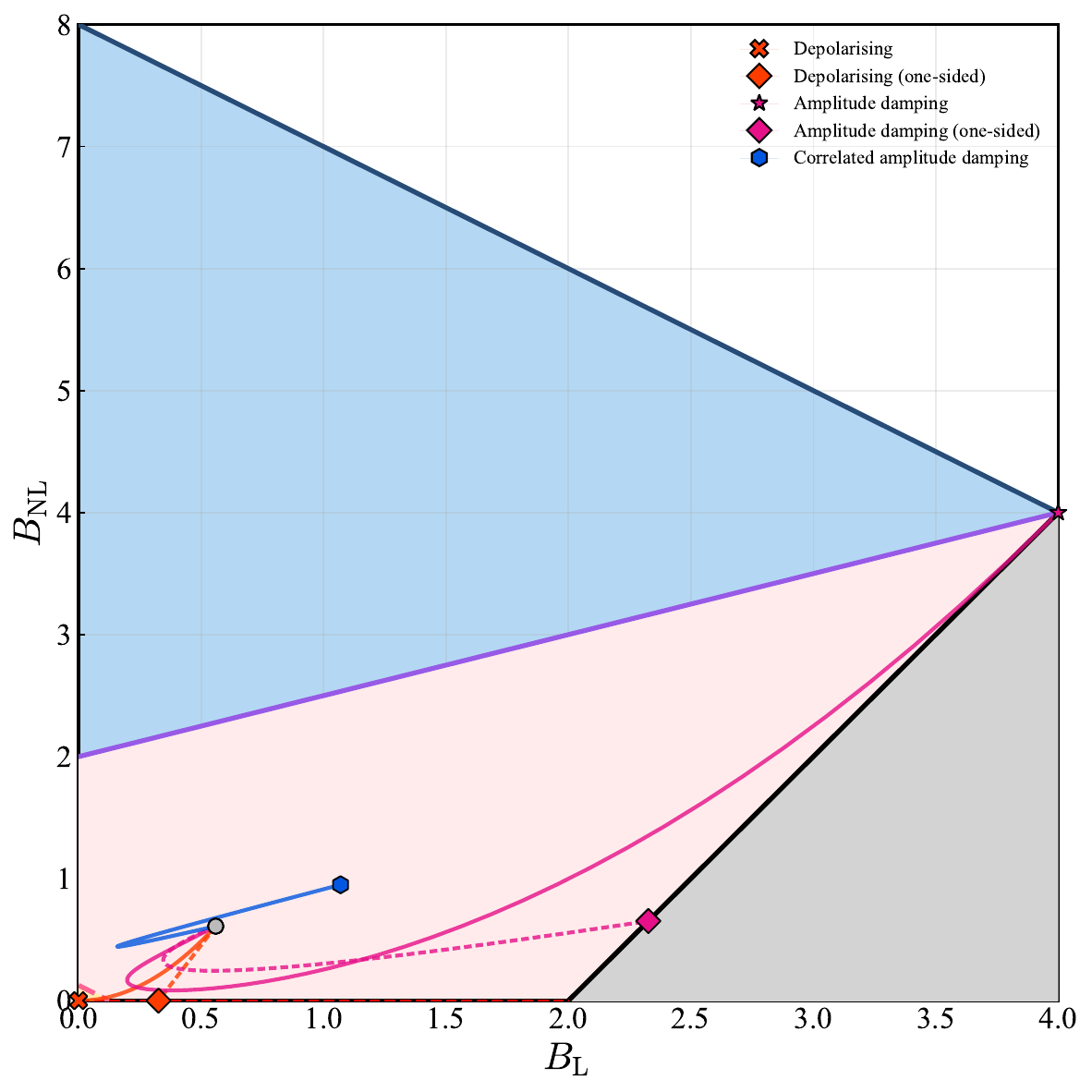}}\hfill\null\\
    \caption{\label{fig:QutritQutritNoise}\textbf{Kinematic flows of qutrit open-system dynamics.} \textbf{a--j}, Parametric decoherence trajectories for bipartite qutrit systems projected onto the $(X,Y)$ plane (odd panels) and the $(B_{\rm L}, B_{\rm NL})$ plane (even panels). The evolution is captured for five distinct initial states: (\textbf{a, b}) a pure maximally entangled state, (\textbf{c, d}) a pure-product state, (\textbf{e, f}) a PPT-separable state, (\textbf{g, h}) a NPT-entangled state and (\textbf{i, j}) a classically correlated state. Open-system dynamics generated by local and correlated qutrit noise channels~\cite{Xiao:2024tuh,Sebastian:2023iak} manifest as dimensionally invariant flows constrained by the overarching purity-budget geometry. Notably, these noise trajectories exhibit the same fundamental flow patterns observed in two-qubit systems, confirming that these geometric constraints are dimensionally invariant.}
\end{figure*}

\begin{figure*}
    \captionsetup[subfigure]{labelformat=empty}
    \centering
    \subfloat[{\bf a} GHZ state $(X,Y)$]{
    \includegraphics[width=0.22\textwidth]{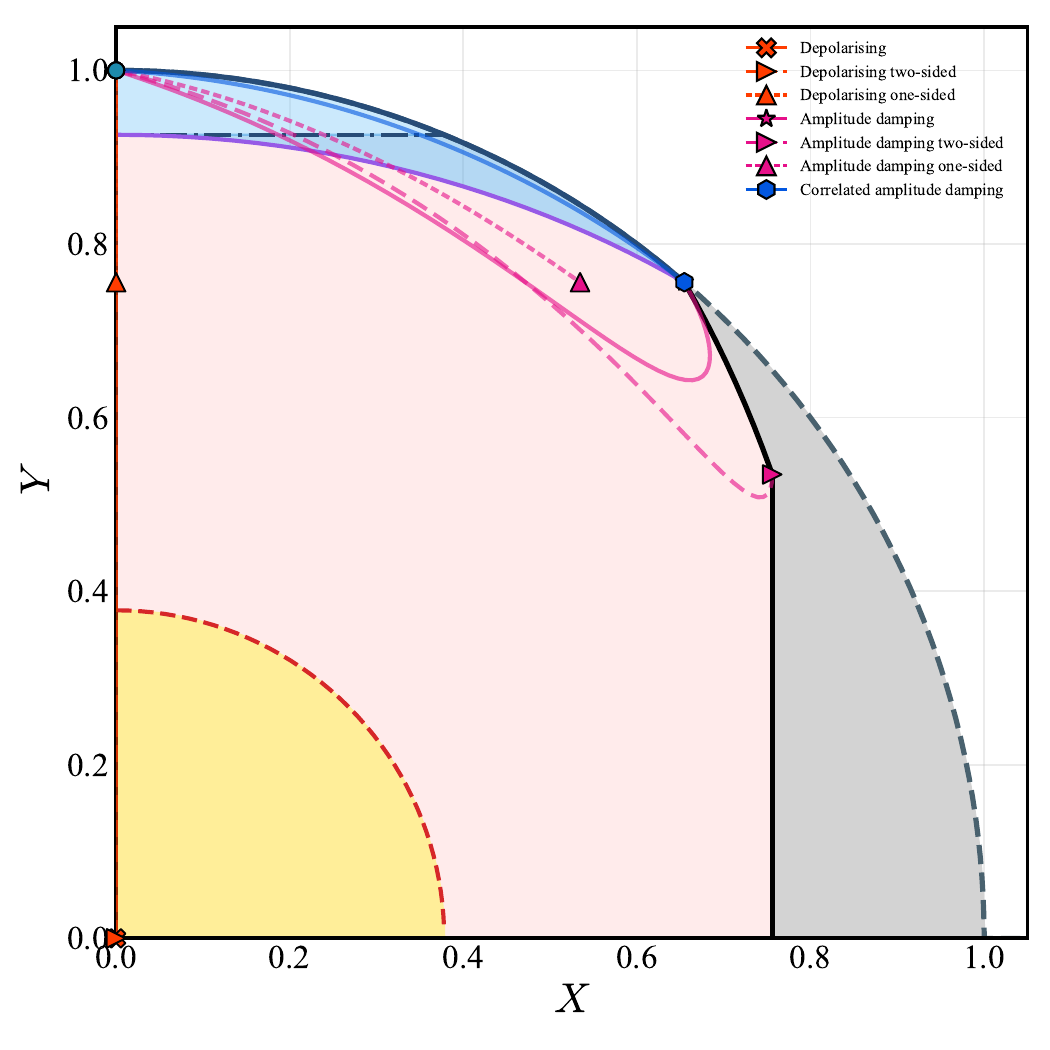}}\hfill
     \subfloat[{\bf b} GHZ state $(B_{\rm L}, B_{\rm NL})$]{
    \includegraphics[width=0.22\textwidth]{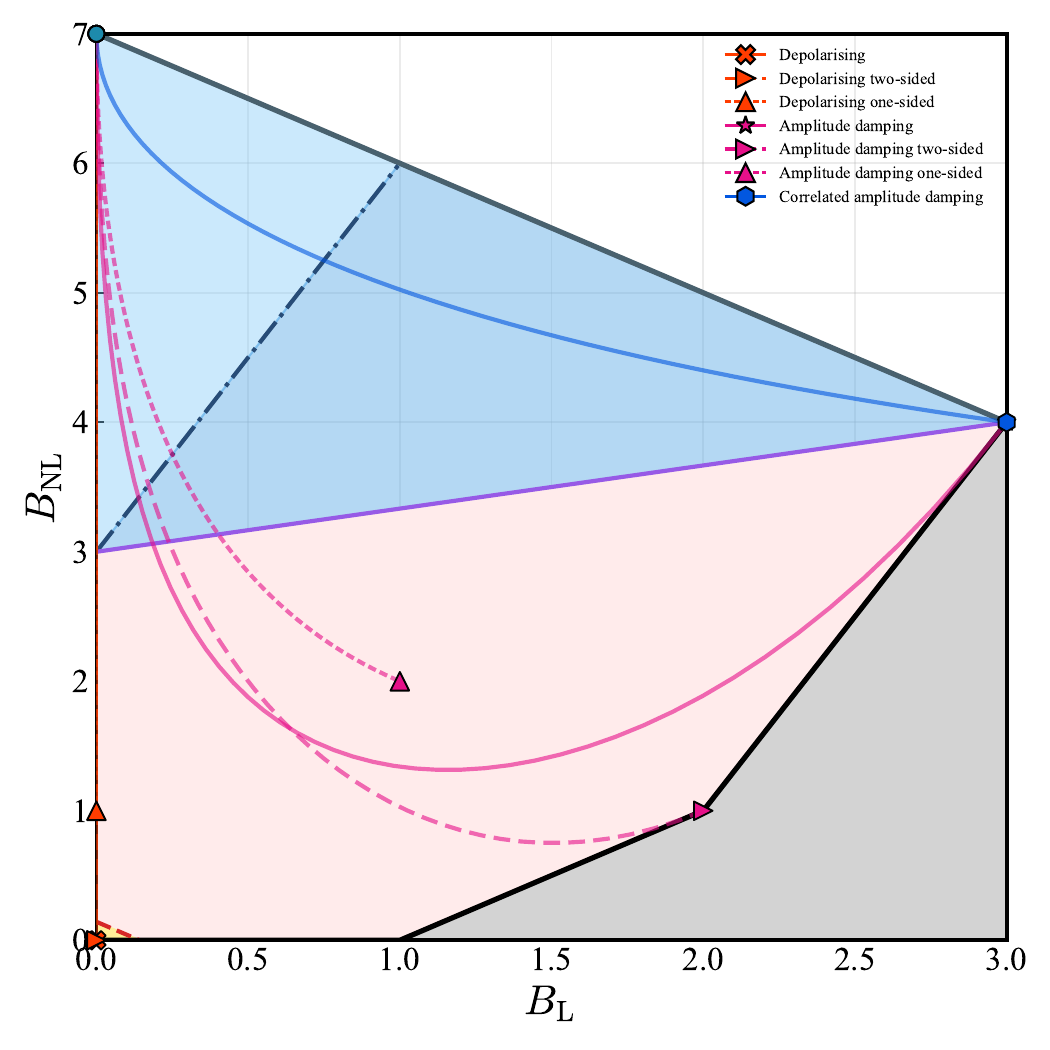}}\hfill
    \subfloat[{\bf c} W state $(X,Y)$]{
    \includegraphics[width=0.22\textwidth]{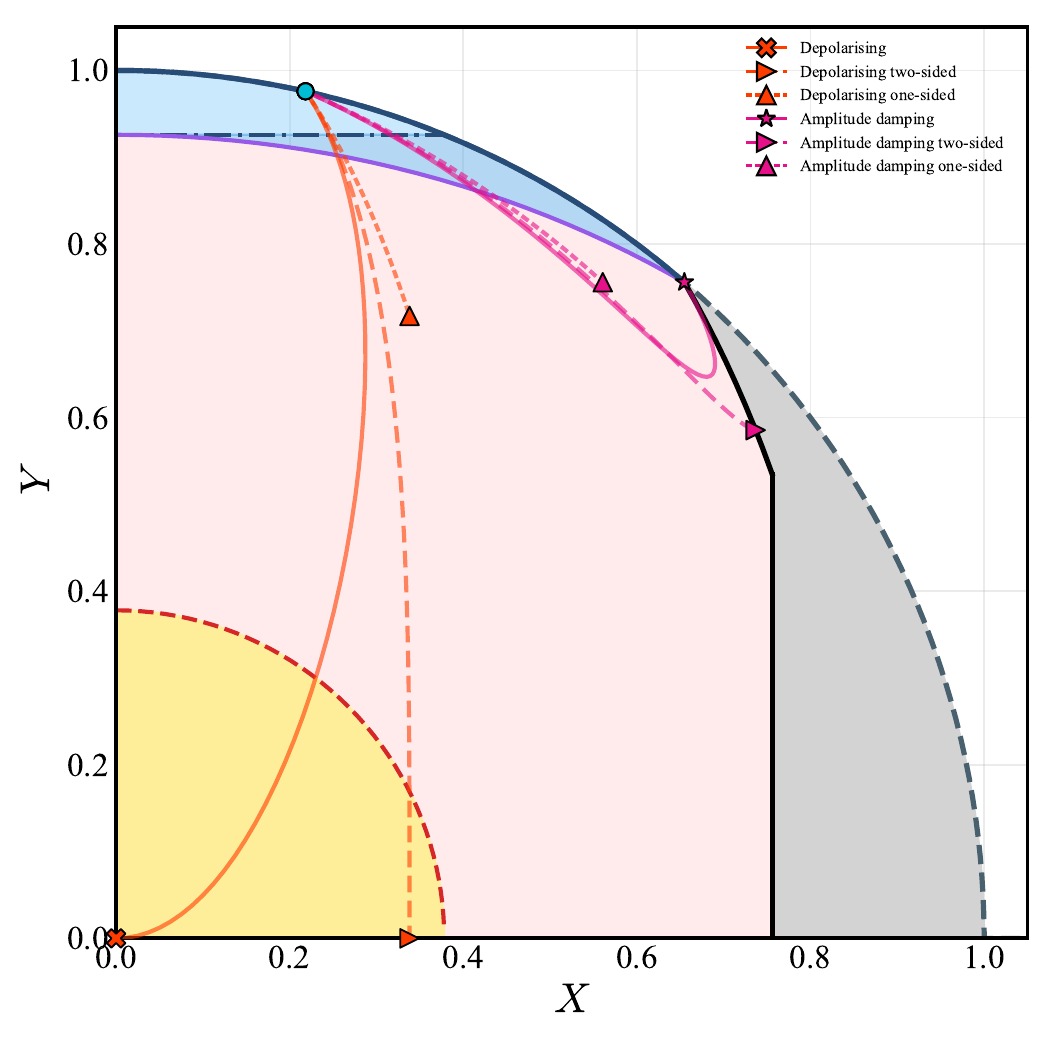}}\hfill
    \subfloat[{\bf d} W state $(B_{\rm L}, B_{\rm NL})$]{
    \includegraphics[width=0.22\textwidth]{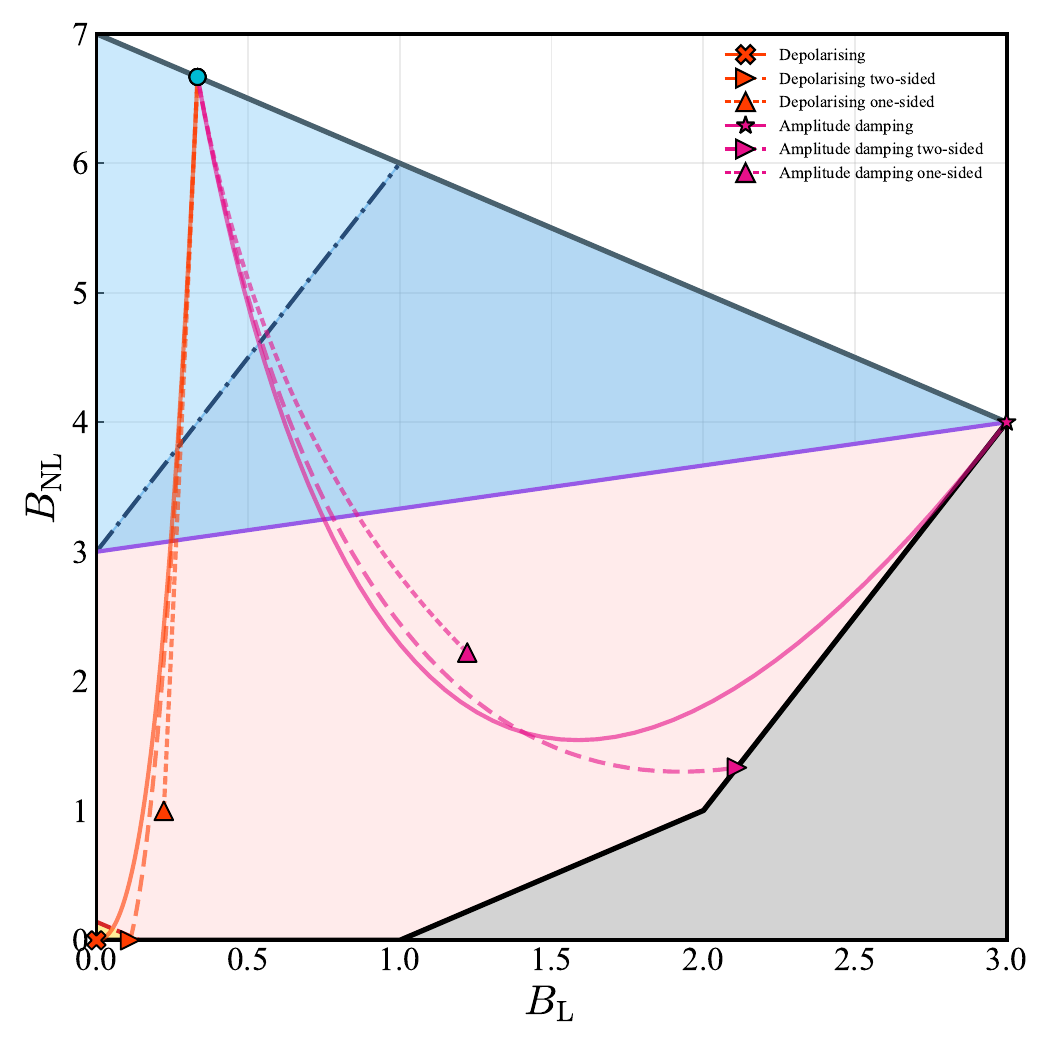}}\\
    \subfloat[{\bf e} Mixed-GME state $(X,Y)$]{
    \includegraphics[width=0.22\textwidth]
    {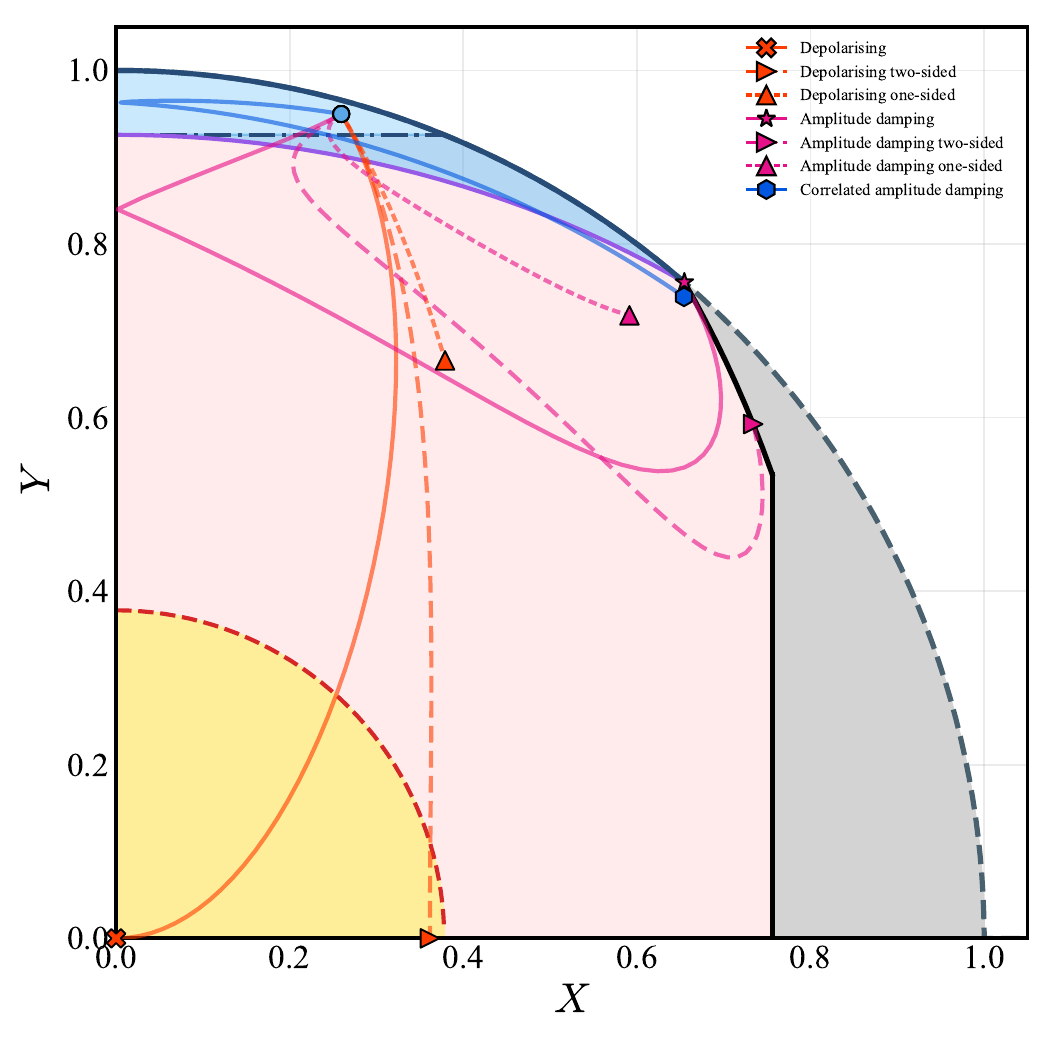}}\hfill
    \subfloat[{\bf f} Mixed-GME state $(B_{\rm L}, B_{\rm NL})$]{
    \includegraphics[width=0.22\textwidth]{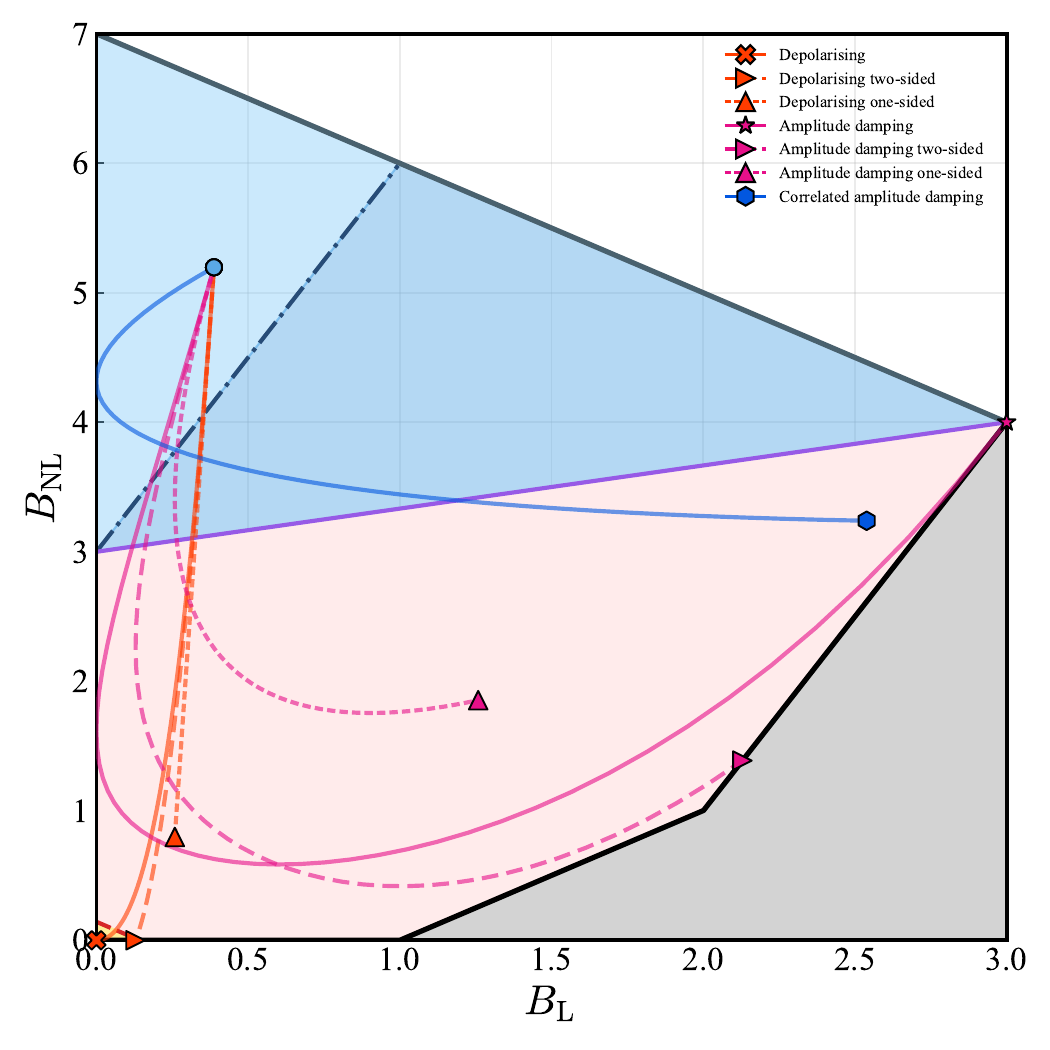}}\hfill
    \subfloat[{\bf g} Bell $\otimes$ Pure state $(X,Y)$ ]{
    \includegraphics[width=0.22\textwidth]{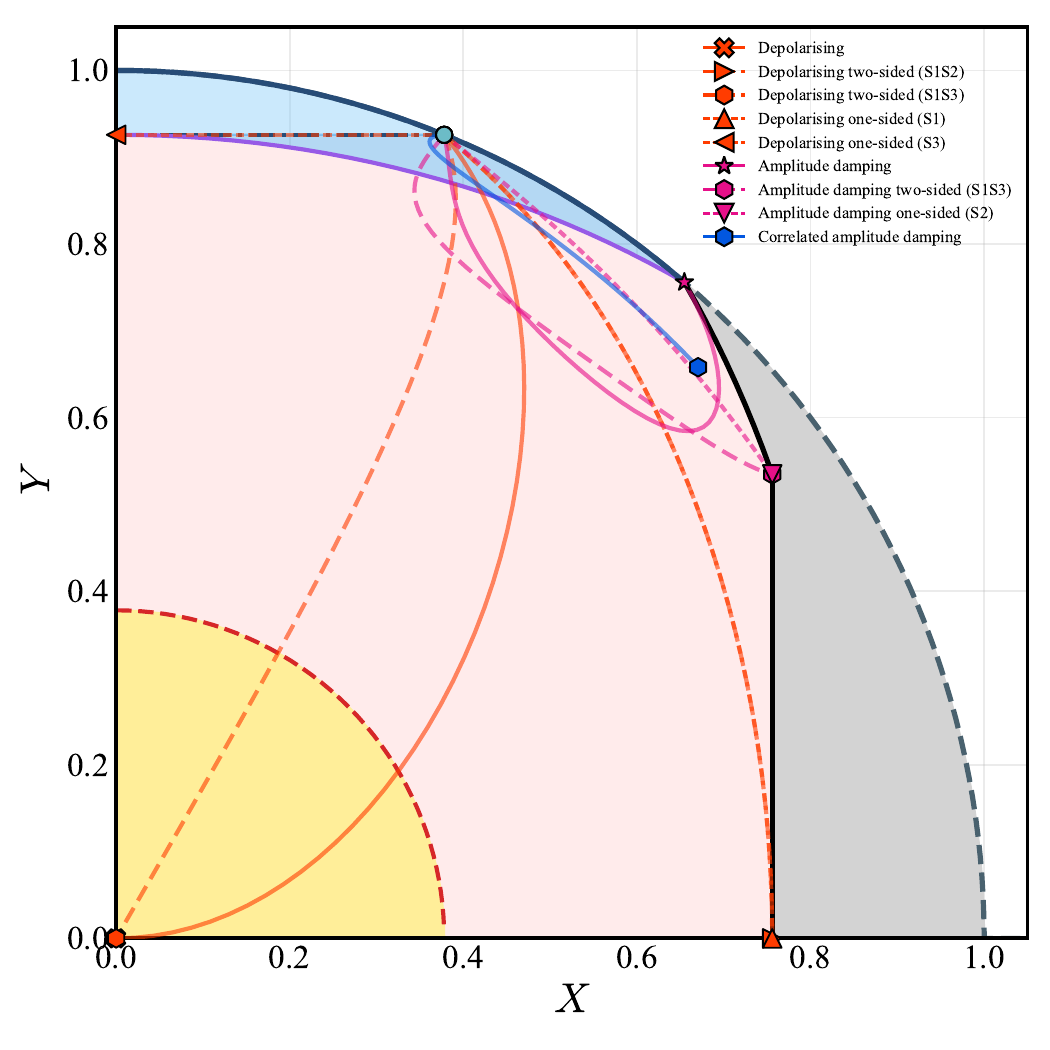}}\hfill
    \subfloat[{\bf h} Bell $\otimes$ Pure state $(B_{\rm L}, B_{\rm NL})$]{
    \includegraphics[width=0.22\textwidth]{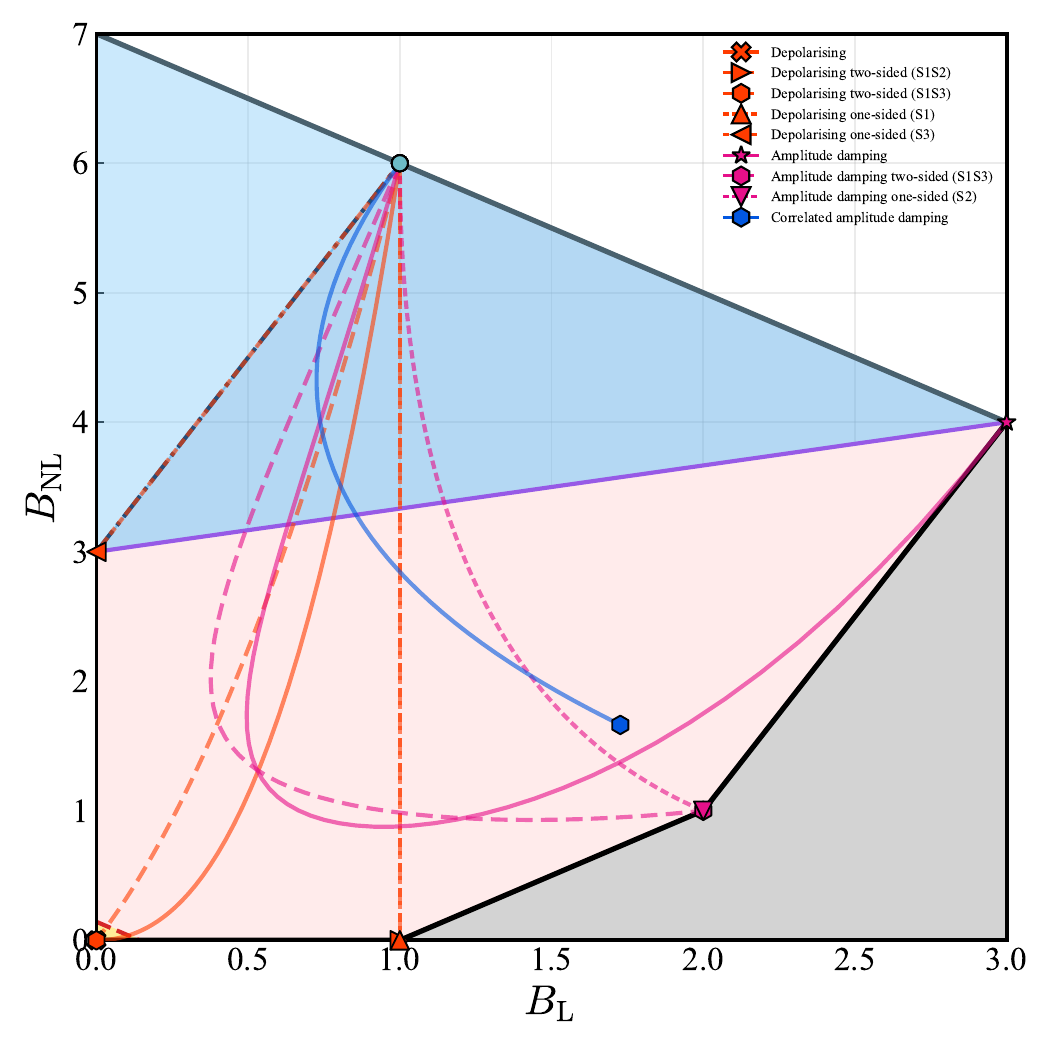}}\\
    \subfloat[{\bf i} Pure-biseparable state $(X,Y)$]{
    \includegraphics[width=0.22\textwidth]{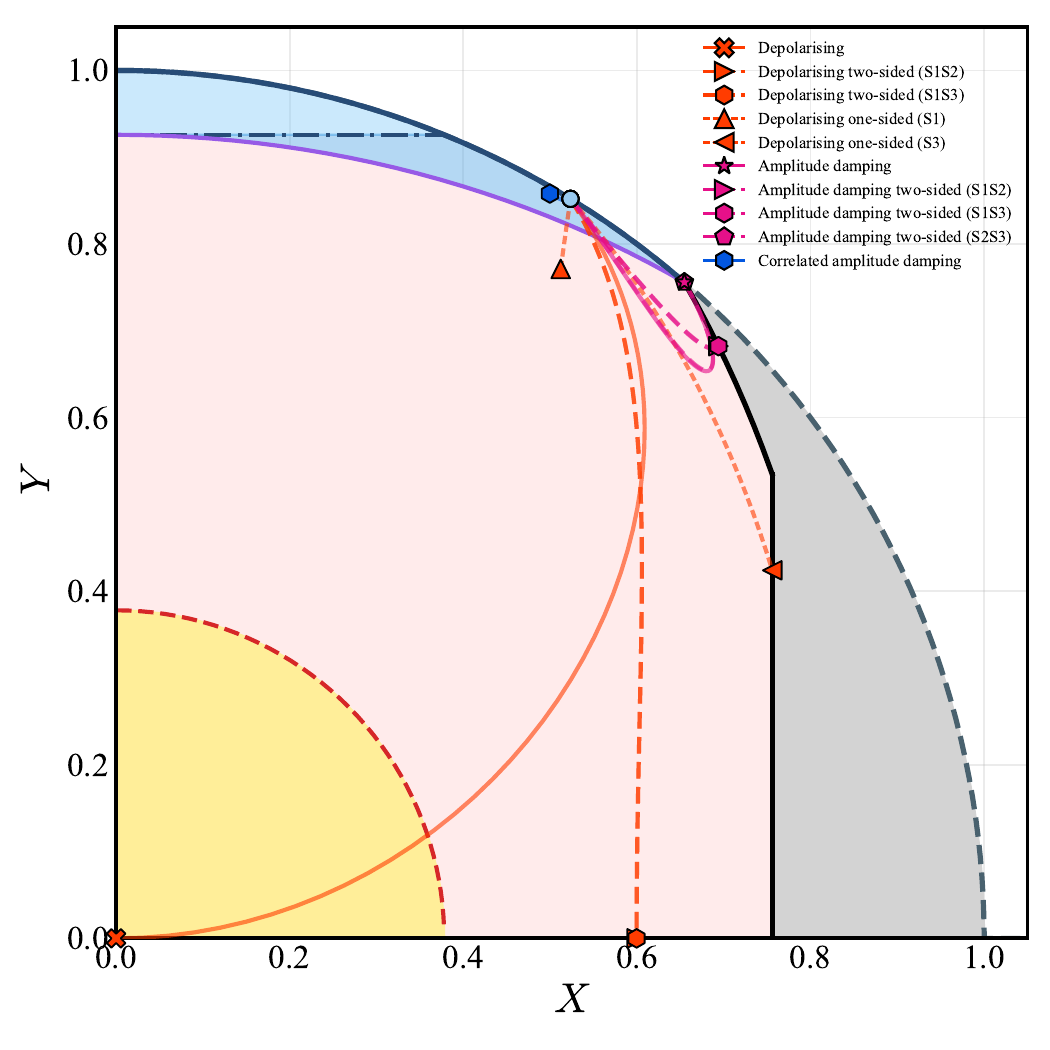}}\hfill
    \subfloat[{\bf j} Pure-biseparable state $(B_{\rm L}, B_{\rm NL})$]{
    \includegraphics[width=0.22\textwidth]{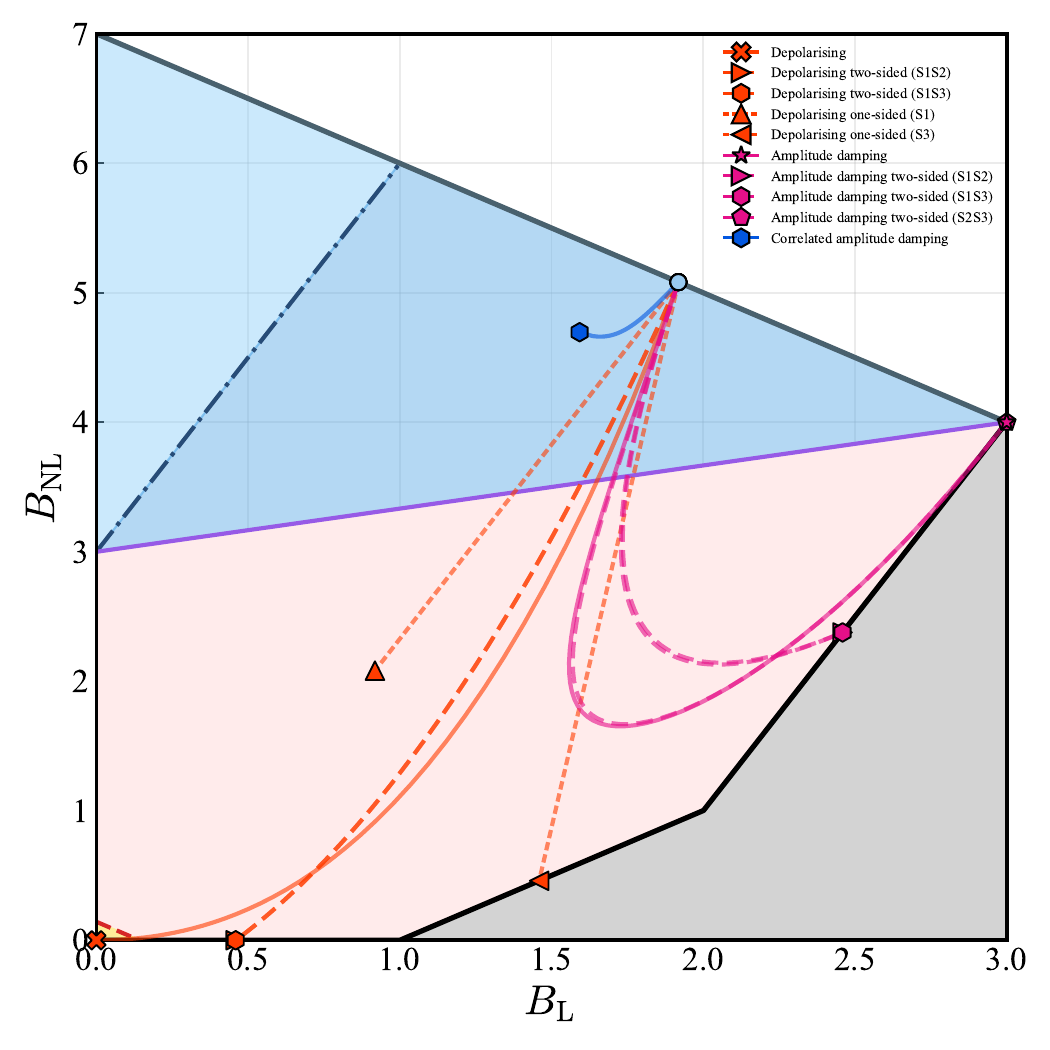}}\hfill
    \subfloat[{\bf k} Mixed-biseparable state $(X,Y)$]{
    \includegraphics[width=0.22\textwidth]{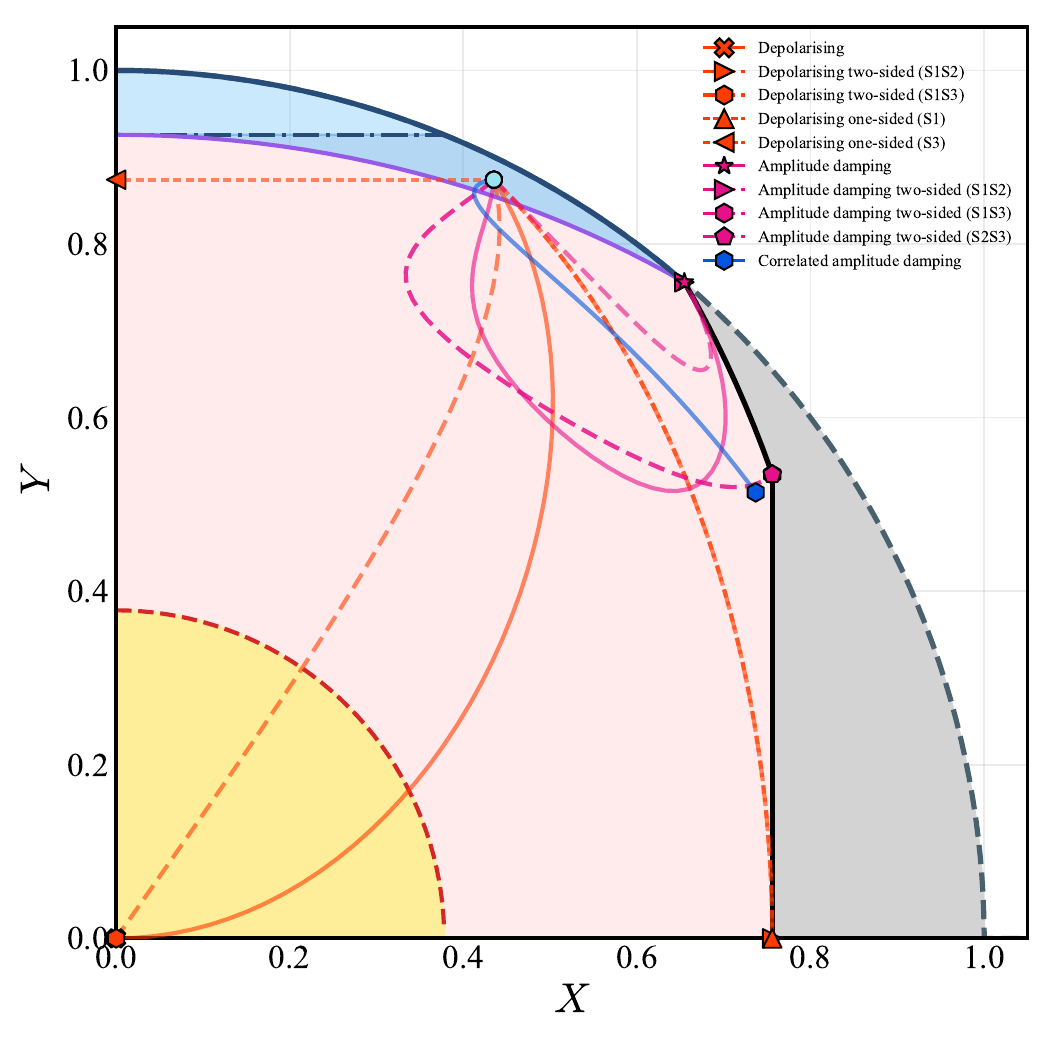}}\hfill
    \subfloat[{\bf l} Mixed-biseparable state $(B_{\rm L}, B_{\rm NL})$]{
    \includegraphics[width=0.22\textwidth]{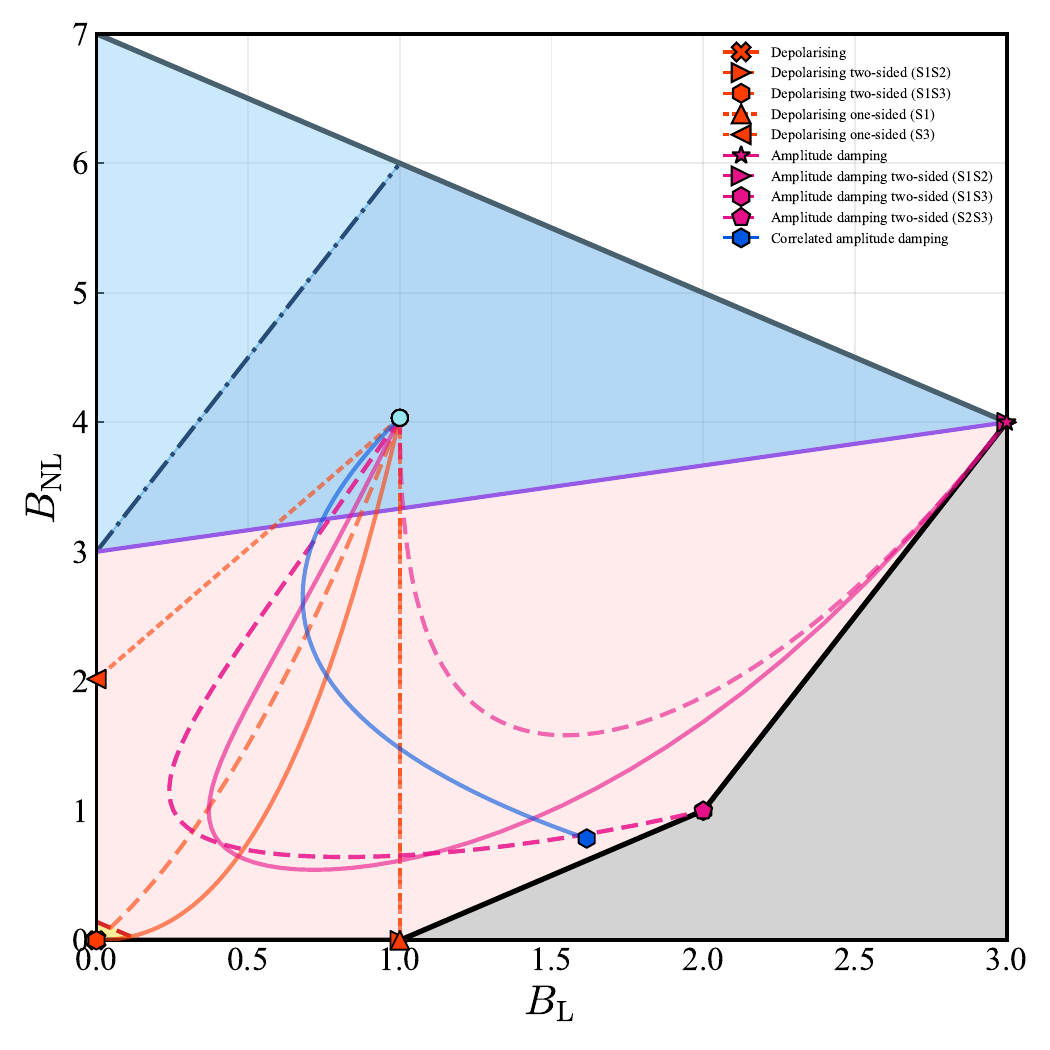}}\hfill
     \subfloat[{\bf m} Fully factorisable-pure state $(X,Y)$]{
    \includegraphics[width=0.22\textwidth]{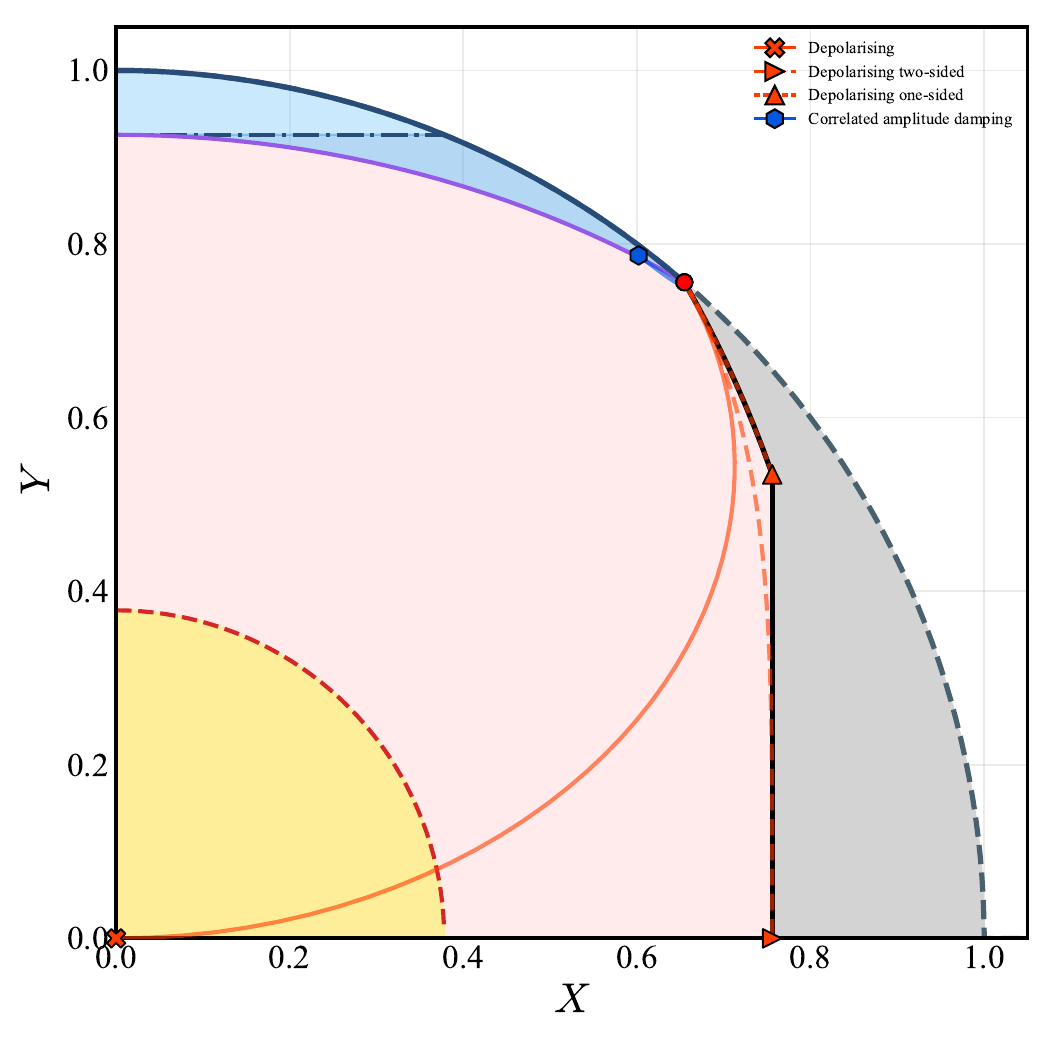}}\hfill
     \subfloat[{\bf n} Fully factorisable-pure state $(B_{\rm L}, B_{\rm NL})$]{
    \includegraphics[width=0.22\textwidth]{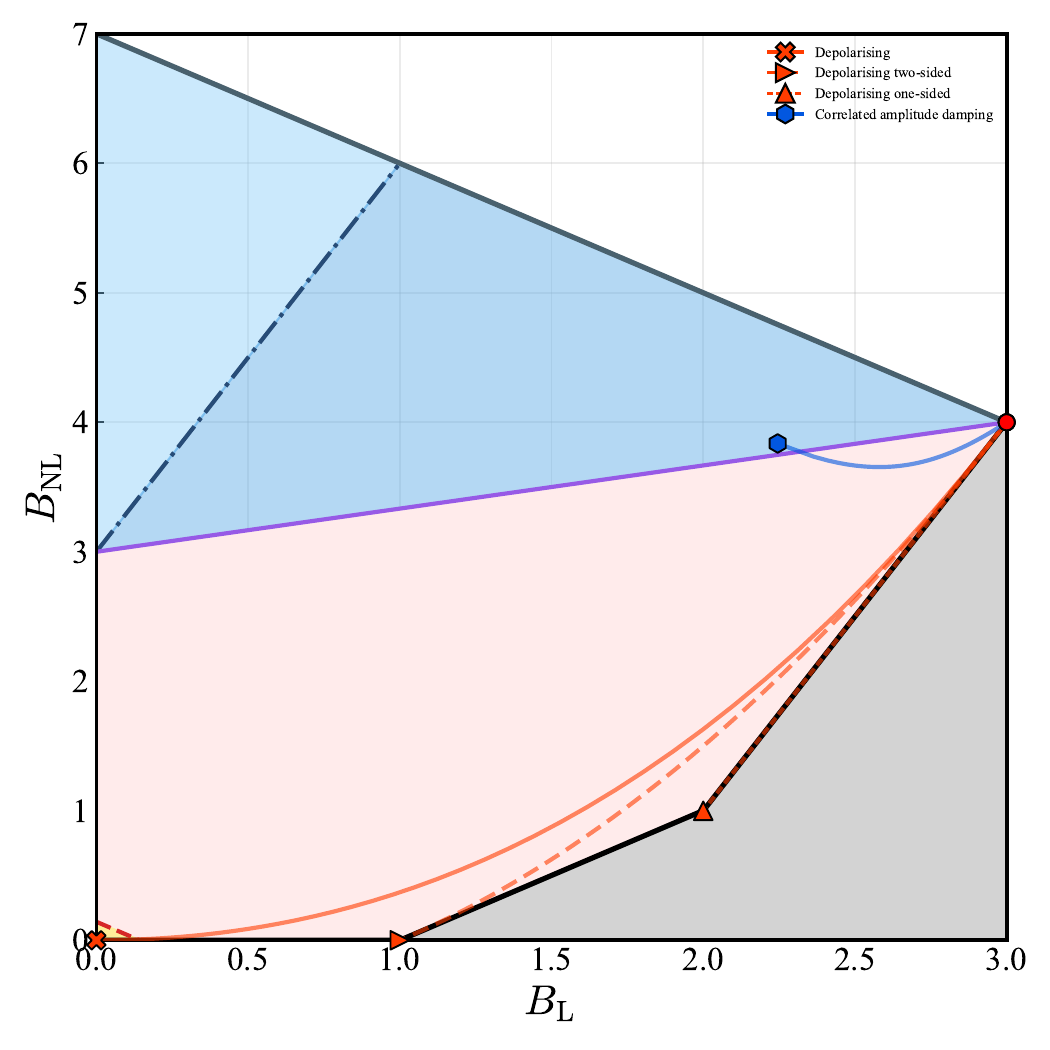}}\hfill
     \subfloat[{\bf o} Pure $\otimes$ pure $\otimes$ mixed state $(X,Y)$ ]{
    \includegraphics[width=0.22\textwidth]{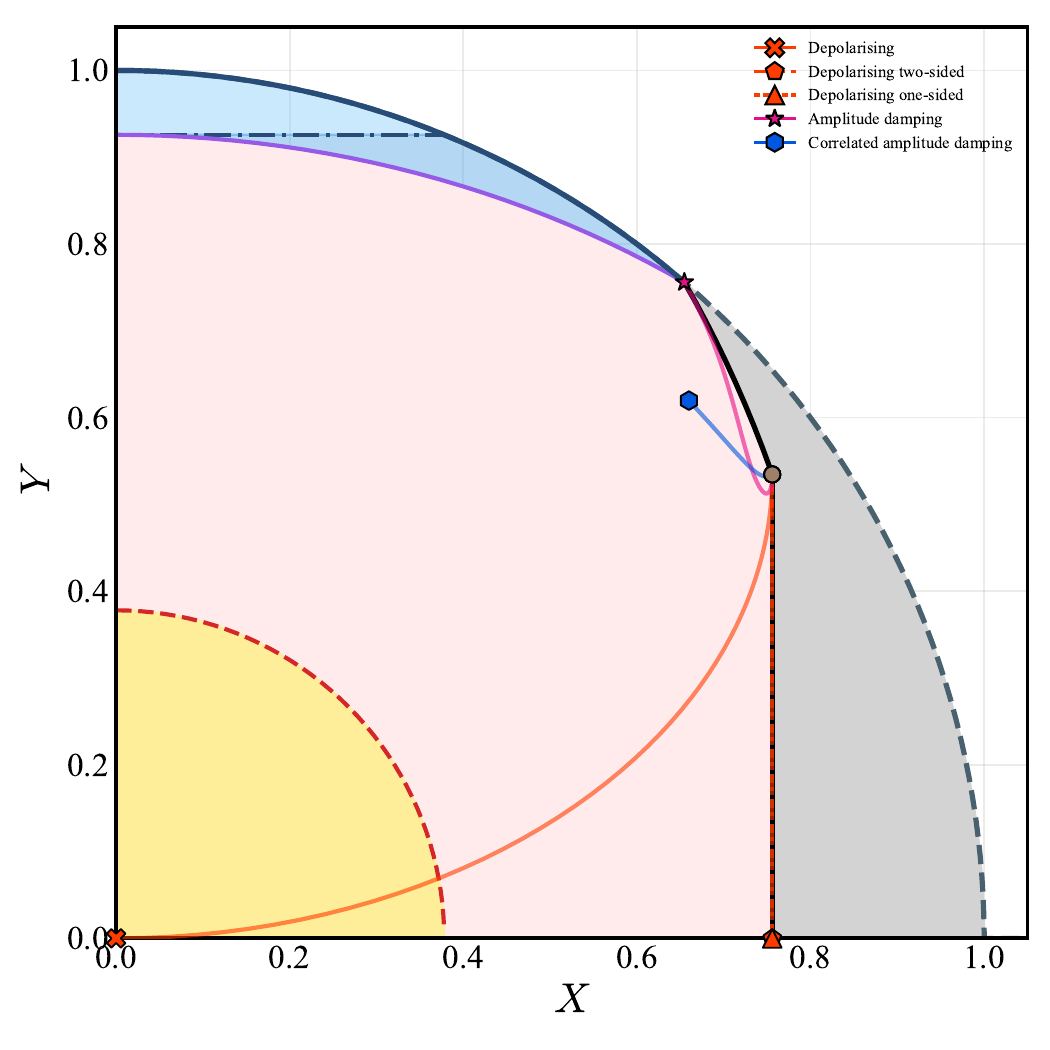}}\hfill
     \subfloat[{\bf p} Pure $\otimes$ pure $\otimes$ mixed state $(B_{\rm L}, B_{\rm NL})$]{
    \includegraphics[width=0.22\textwidth]{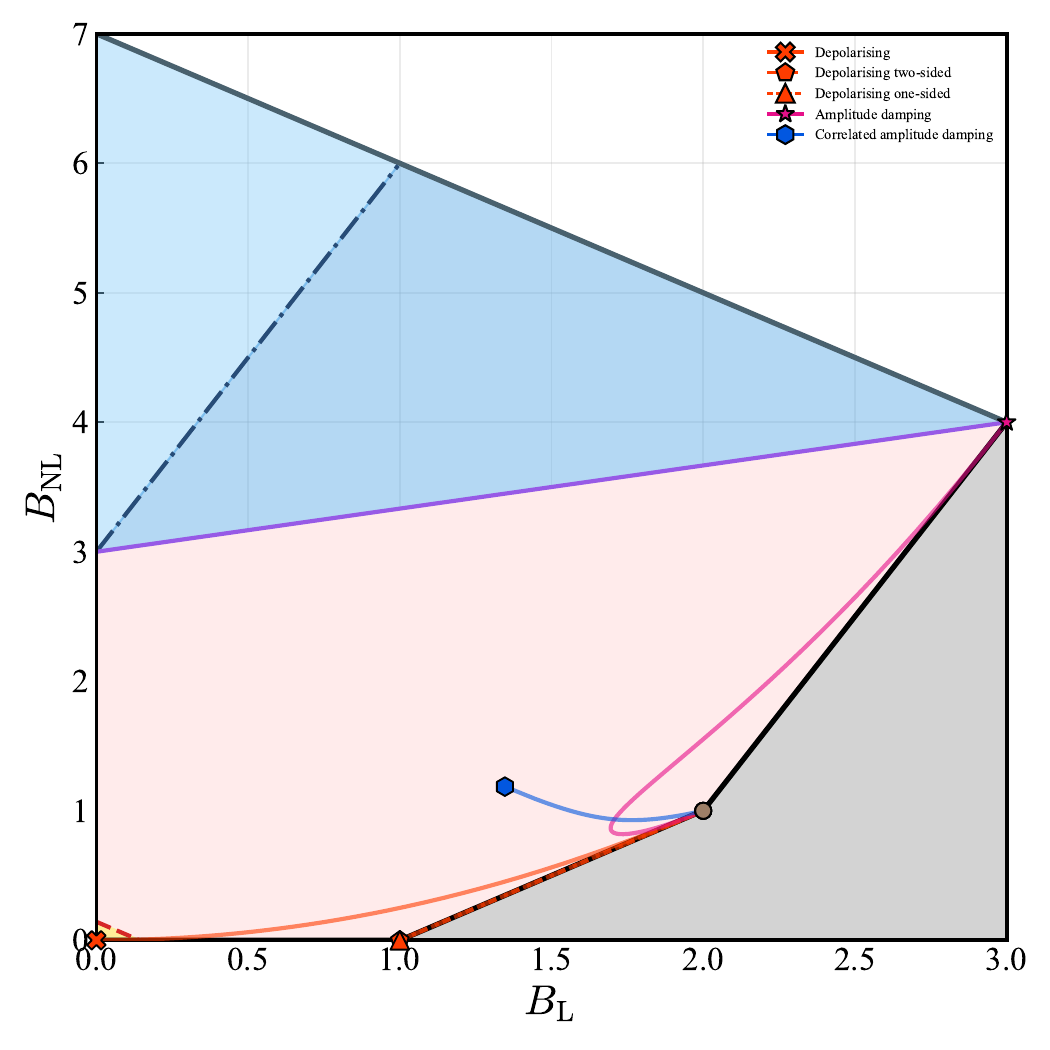}}\hfill
    \subfloat[{\bf q} Fully factorisable-mixed state $(X,Y)$]{
    \includegraphics[width=0.22\textwidth]{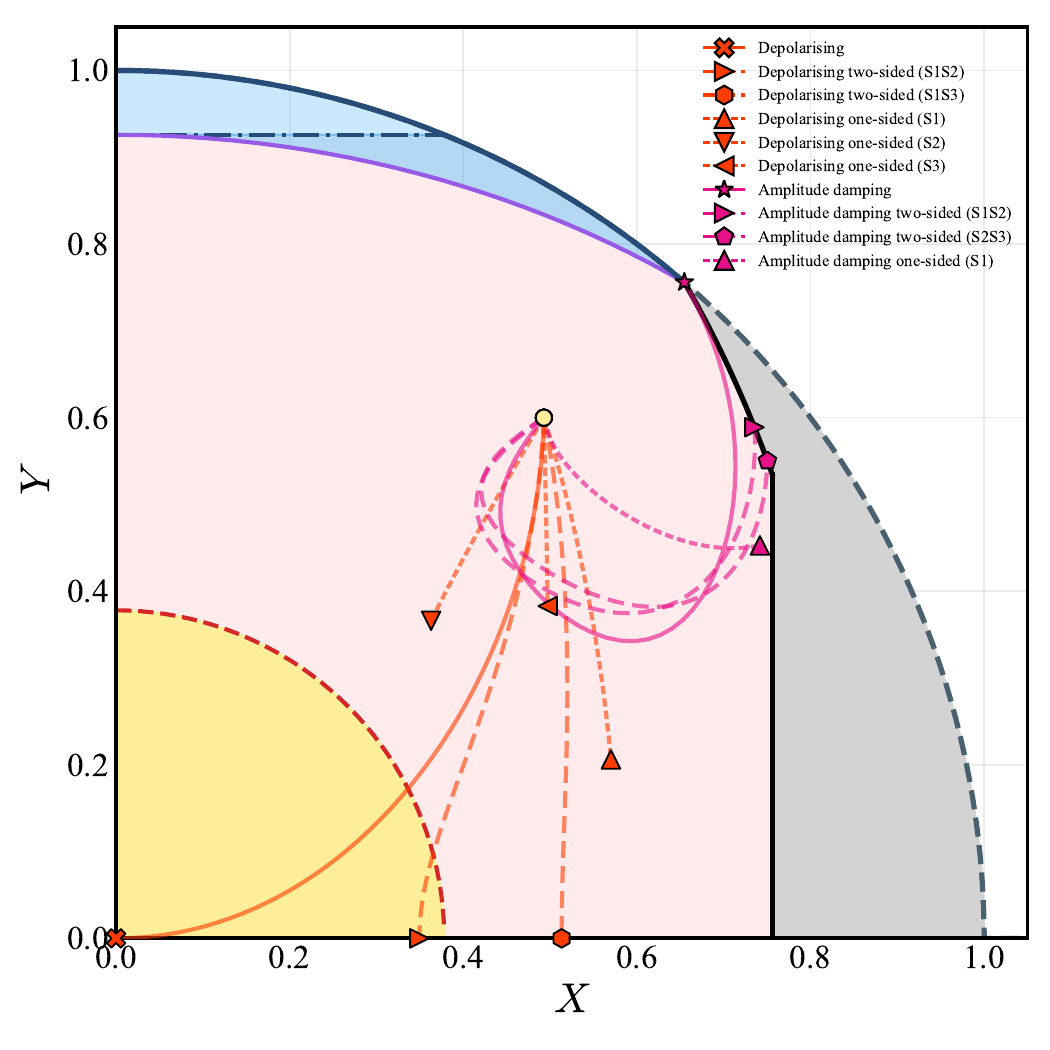}}\hfill
     \subfloat[{\bf r} Fully factorisable-mixed state $(B_{\rm L}, B_{\rm NL})$]{
    \includegraphics[width=0.22\textwidth]{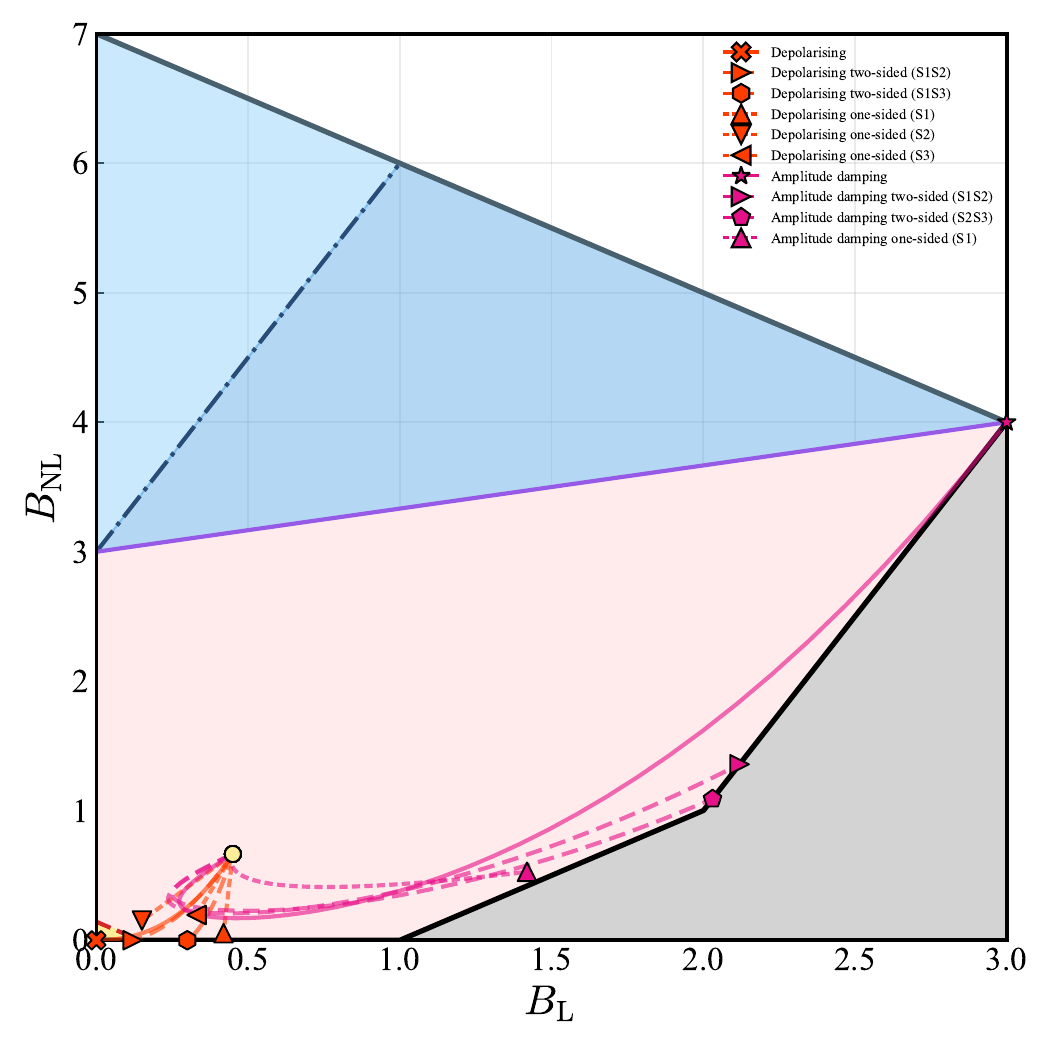}}\hfill
     \subfloat[{\bf s} Classical (C$^3$) state $(X,Y)$ ]{
    \includegraphics[width=0.22\textwidth]{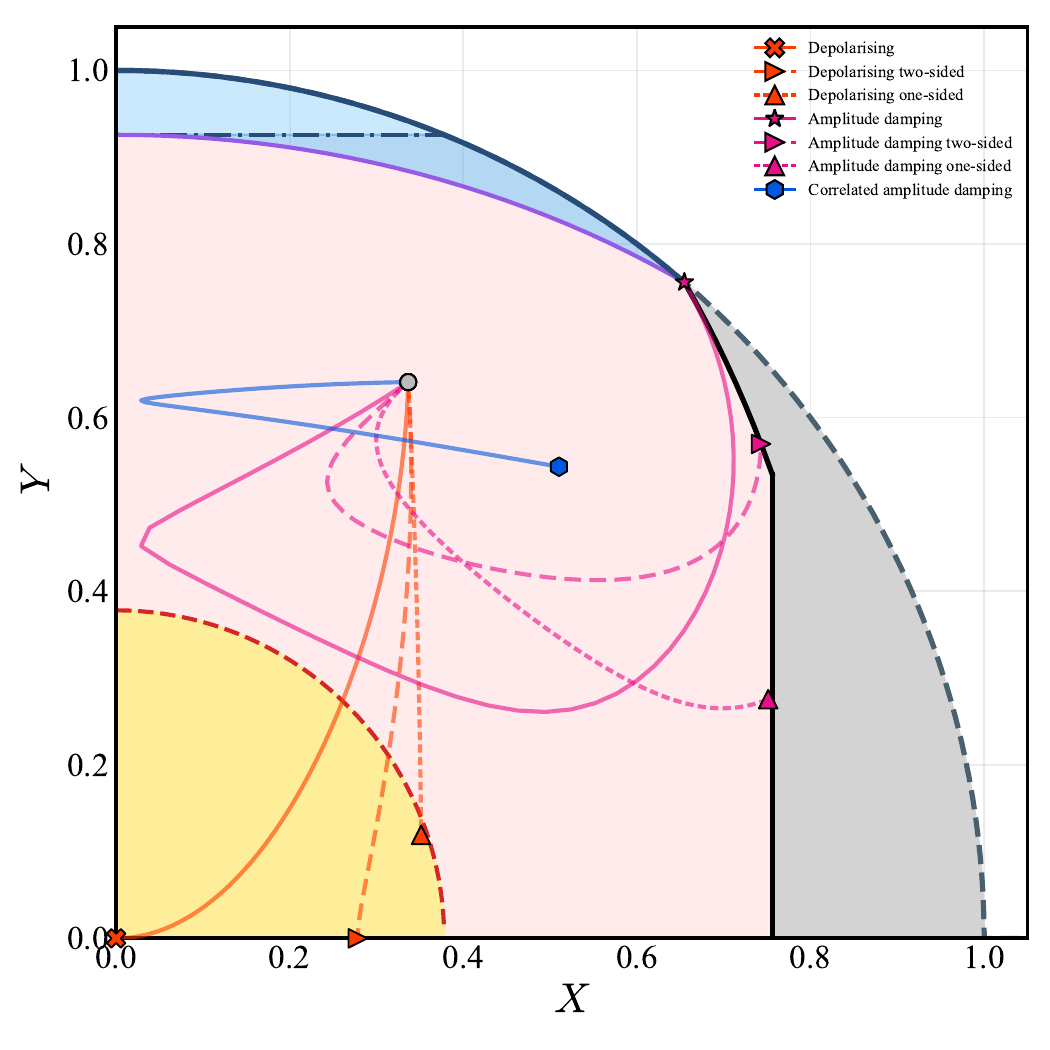}}\hfill
     \subfloat[{\bf t} Classical (C$^3$) state $(B_{\rm L}, B_{\rm NL})$]{
    \includegraphics[width=0.22\textwidth]{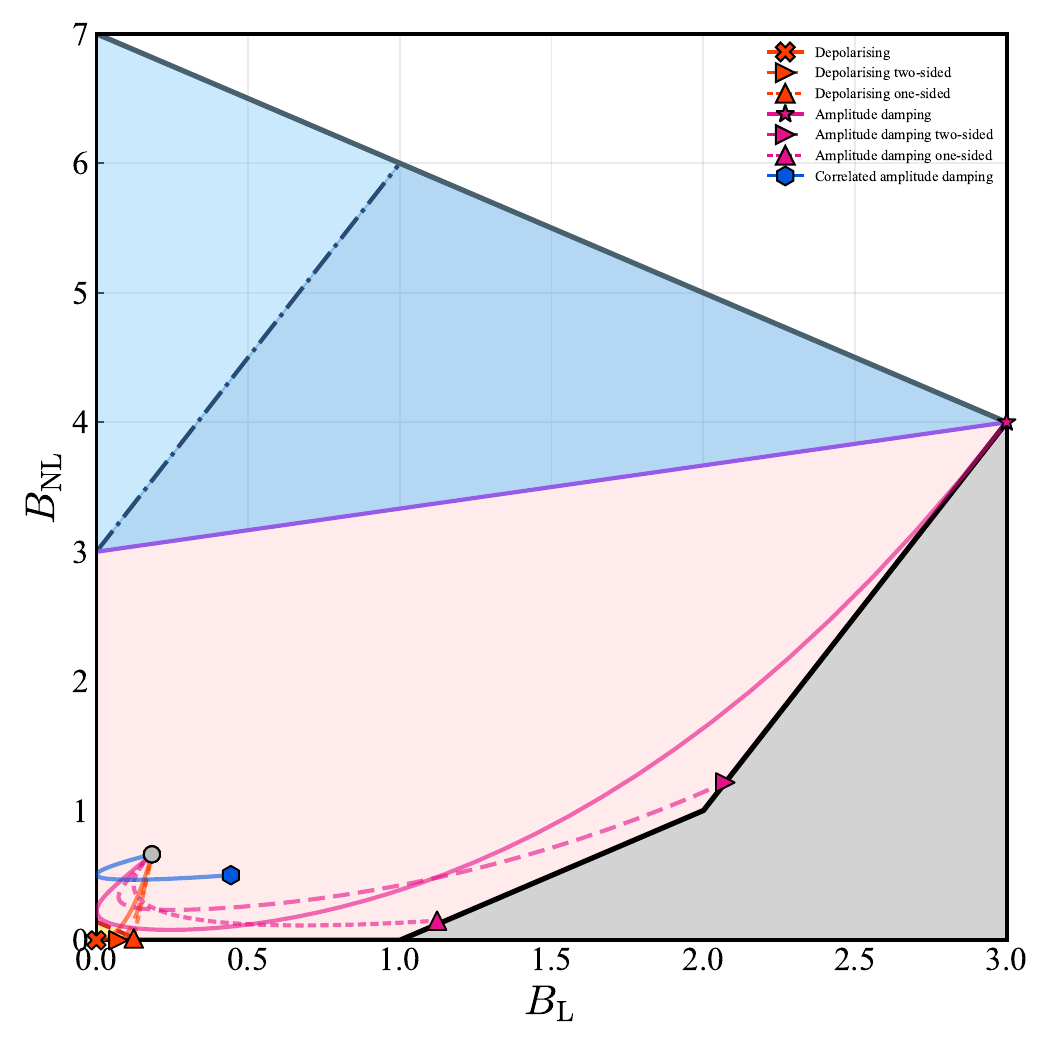}}\hfill
    \caption{\label{fig:ThreeQutritNoise}\textbf{Kinematic flows in three-qubit open-system dynamics.} \textbf{a--t}, Parametric decoherence trajectories for three-qubit ($2 \otimes 2\otimes2$) systems, projected onto the $(X,Y)$ plane (odd panels) and the $(B_{\rm L}, B_{\rm NL})$ plane (even panels). Evolution is shown for $10$ distinct initial states. For biseparable entanglement, we label the subsystems as $S_1$, $S_2$, $S_3$ and take $S_3$ to be the separable qubit. The trajectories show similar behaviours as in the two-qubit case.}
\end{figure*}

\begin{figure*}
    \captionsetup[subfigure]{labelformat=empty}
    \centering
    \subfloat[{\bf a} $(X,Y)$]{
    \includegraphics[width=0.32\textwidth]{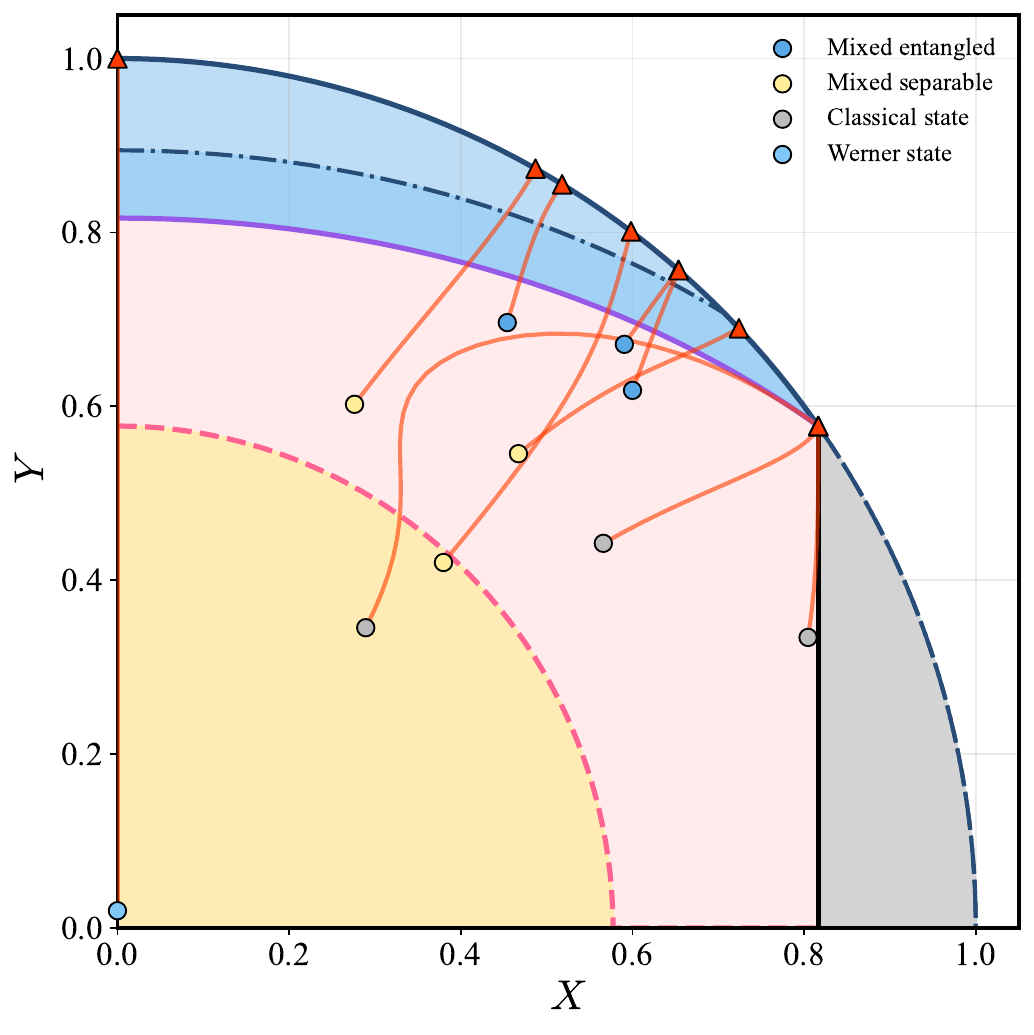}}\hfill
    \subfloat[{\bf b} $(B_{\rm L}, B_{\rm NL})$]{
    \includegraphics[width=0.32\textwidth]{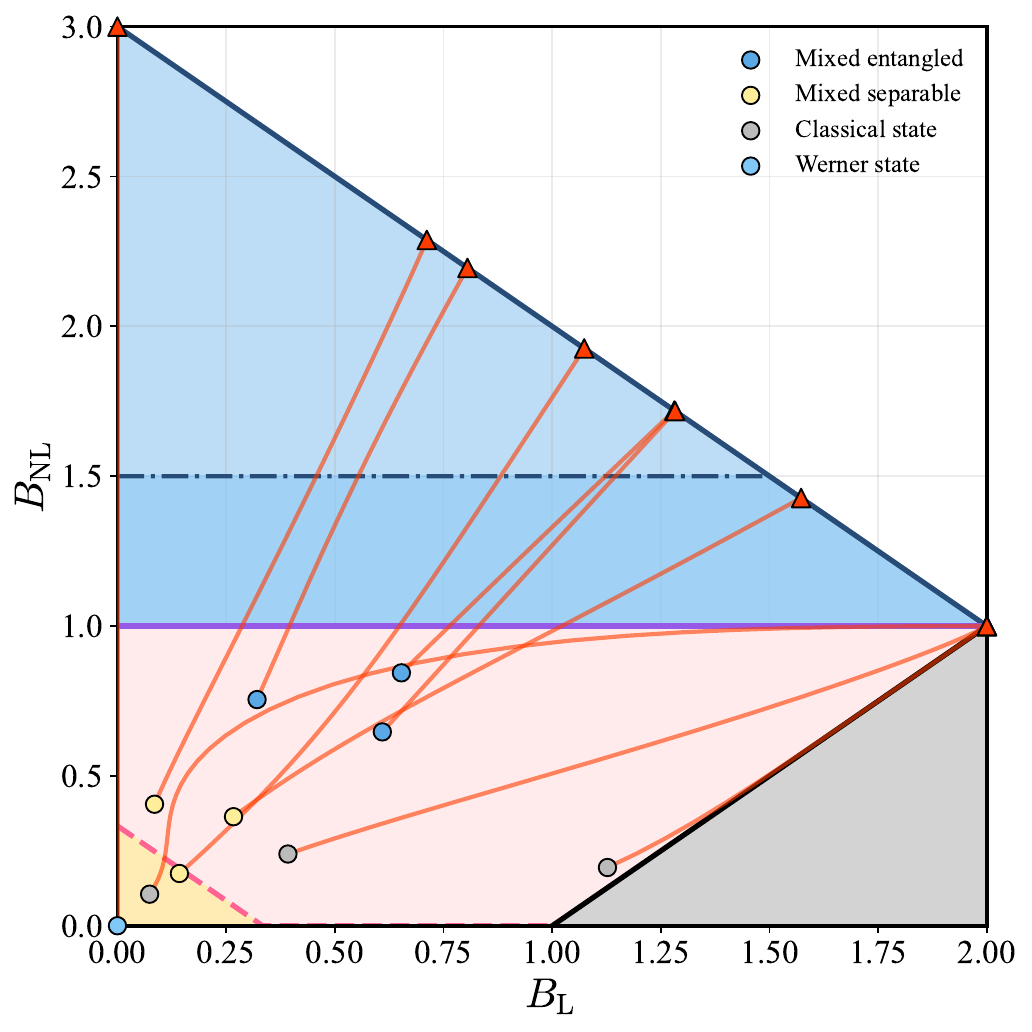}}\hfill
    \subfloat[{\bf c} $(P,Q)$]{
    \includegraphics[width=0.32\textwidth]{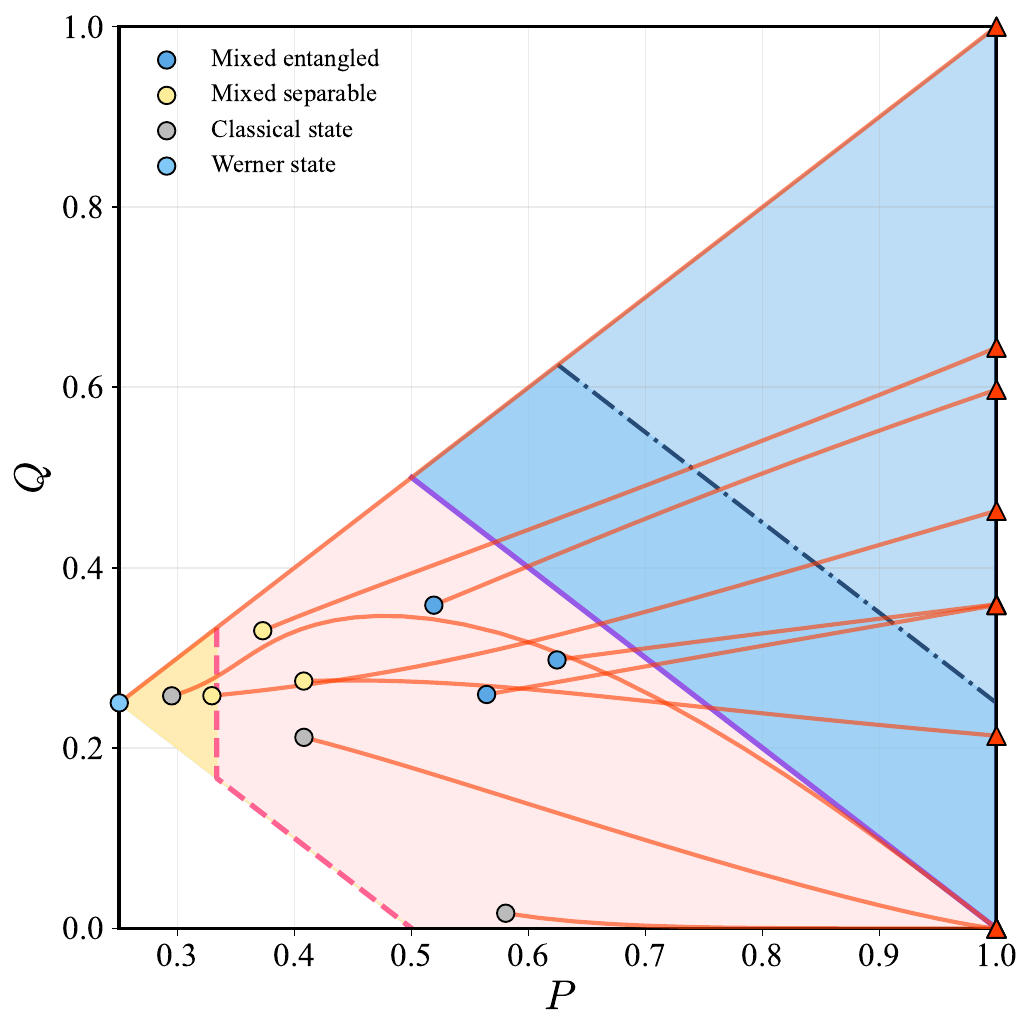}}
    \caption{\textbf{Active kinematic flow of quantum purification.} \textbf{a--c}, Active error mitigation and purification protocols can reverse the natural arrow of decoherence (the empirical \emph{no-go}). Here, geometric trajectories of algorithmic purification are projected across the macroscopic (\textbf{a}) $(X,Y)$ plane, (\textbf{b}) the $(B_{\rm L}, B_{\rm NL})$ plane, and (\textbf{c}) the $(P,Q)$ plane for two-qubit systems. The simulated flows originate from representative mixed-entangled, mixed-separable, and classical states. The nonlinear map $\rho \mapsto \rho^n / \mathrm{tr}(\rho^n)$, which effectively couples the system to a fictitious zero-temperature bath, increases global purity by amplifying the statistical weight of the largest eigenvalue (i.e., a purity pump). This process drives the state outward toward the $R=1$ pure-state boundary, restoring second-moment capacity until it terminates at the coordinates of the dominant pure eigenvector. The $(P,Q)$ projection (\textbf{c}) confirms that such active interventions (executable via virtual distillation) violate the empirical no-go rule by simultaneously increasing both global purity ($P$) and time-reflection overlap ($Q$).}
    \label{fig:Purification}
\end{figure*}

\subsection{The C envelope}
\noindent
To isolate the fully classical C ($\equiv$ C$^n$) envelope for the same system, the budget maximisation must be restricted entirely to the discrete classical marginal problem, thereby explicitly forbidding the real-symmetric $S$ generators. For that, we set the global purity as: $ P = \sum_{x=1}^D p_x^2$ (with $ p_x \ge 0, \quad \sum_{x=1}^D p_x = 1$)
and the local purity $P_k$ for the $k$th subsystem as the sum of its squared marginal probabilities: $P_k = \sum_{y=1}^{d_k} \big(p^{(k)}_y \big)^2$.
Here, each marginal probability $p^{(k)}_y$ is obtained via the standard classical summation of the joint distribution $p_x$ over all other subsystems. Tracing the maximum $B_{\rm NL}$ for $0\leqslant B_{\rm L}\leqslant \sum_{k=1}^n (d_k - 1)$ yields the $C^n$ envelope. Because this classical probability space is strictly less permissive than the continuous $H \oplus S$ Lie algebra utilised by the main algorithm, the resulting C$^n$ boundary mathematically must lie at or below the QC$^{n-1}$ envelope. Only when $d_1=d_2=\cdots=d_n$ (fully symmetric system), does the C envelope completely overlap with the QC$^{n-1}$ envelope. In asymmetric systems, the C$^n$ envelope separates from the QC$^{n-1}$ envelope near the $B_{\rm NL}$ axis, as seen in Fig.~\ref{fig:23system} for $2\otimes 3$ systems. Near the axis, the curve follows a square-root law ($B_{\rm NL} \sim \sqrt{B_{\rm L}}$) because of a basic cost-benefit mismatch: creating a small probability peak generates correlation gain ($B_{\rm NL}$) linearly while consuming local budget ($B_{\rm L}$) only quadratically.

\subsection{Stabiliser R\'enyi-$\mathbf{2}$ magic}
\noindent
The stabiliser R\'enyi-$2$ magic $\widetilde M_2$ is defined as~\cite{Leone:2021rzd} 
\begin{align*}
\widetilde{M}_2= -\log_2 \left(\frac{1+S_4}{1+B}\right),
\end{align*}
where $S_4$ is the fourth moment of the state, $S_4=\sum_{k=1}^{n_{\rm P}}x_k^4$ with $\{x_k\}$ denoting the set of all Pauli coefficients. Since the fourth moment is bounded by the second moment as $B^2/n_{\rm P} \leqslant S_4 \leqslant x_{\max}^2 B$, we get
\begin{align*}
M_B-\log_2 \left(
1 + x_{\max}^2\, B
\right) \leqslant \widetilde M_2 \leqslant 
M_B-\log_2\left(
1 + B^2/n_{\rm P}\right),
\end{align*}
where $M_B=\log_2(1+B)$ and $x_{\max}$ is the Pauli coefficient with the largest magnitude. If the correlation matrix has rank one, in the canonical 
basis, a single Pauli axis appears through which all 
classical correlations propagate. The local Bloch vectors 
are forced to align with it. Consequently, the nonlocal 
Pauli sector collapses onto a single stabiliser direction, 
suppressing the nonlocal contribution to magic (though 
the local sector can still sustain nonzero stabiliser 
R\'enyi entropy).

\subsection{Random-state generation}
\noindent 
For Fig.~\ref{fig:budgetXY_right}, we generate large ensembles of two-qubit density matrices from physically relevant families of quantum states.
\begin{enumerate}[leftmargin=10pt]
    \item[--] {\bf Pure-product states:} We sample two random single-qubit pure states $\rho_A$ and $\rho_B$ uniformly with respect to the Haar measure on the Bloch sphere. Their tensor product, $\rho = \rho_A \otimes \rho_B$, is always separable.

    \item[--] {\bf CC states:} We draw a random probability vector $p = (p_{00}$, $p_{01}$, $p_{10}$, $p_{11})$ from a Dirichlet distribution, $\mathrm{Dir}(1,\,1,\,1,\,1)$, defining a diagonal state in the computational basis: $\rho = \mathrm{diag}(p_{00},\,p_{01},\,p_{10},\,p_{11})$.
    
    \item[--] {\bf Pure-entangled states:}  We obtain a Haar-random state vector $|\psi\rangle \in \mathbb{C}^4$ by normalising a complex Gaussian vector. The associated density matrix $\rho = |\psi\rangle\langle\psi|$ is almost surely entangled.
    
    \item[--] {\bf  Werner-like states:} We use mixtures of a Bell state, $|\Phi^+\rangle = (|00\rangle + |11\rangle)/\sqrt{2}$, with white noise, 
    \begin{align*}
    \rho = p\,|\Phi^+\rangle\langle\Phi^+| + (1-p)\,\mathbb{I}_4/4,    
    \end{align*}
    where $p \sim U[0,1]$. This family interpolates between a maximally entangled state and the completely mixed state.
    
    \item[--] {\bf Mixed-entangled states:} A random complex Ginibre matrix $G$ yields a Wishart state, $\rho = GG^\dagger/\mathrm{tr}(GG^\dagger)$, sampled according to the Hilbert-Schmidt measure. We retain only states with strictly positive negativity $\mathcal{N}(\rho) > 10^{-8}$, indicating violation of the positive PT (PPT) criterion.
    
    \item[--] {\bf Mixed-separable states:} Using the same Wishart ensemble, we instead select states satisfying $\mathcal{N} < 10^{-8}$ and $P<0.999$, ensuring PPT separability and non-purity.
\end{enumerate}
Though we generate large samples per family ($\sim 10^5$), we only show a small representative sample set in the figure for clarity.

\subsection{Noise models}
\noindent
We model open-system dynamics using Kraus operators. We consider three standard local quantum channels, applied either to a single qubit or independently to both subsystems $(\Lambda \otimes \Lambda)$: (1) \textit{dephasing} (phase flip) with strength $p$, which suppresses off-diagonal elements by a factor $1-p$; (2) \textit{depolarising noise}, $\rho \mapsto (1-p)\rho + p\,\mathbb{I}/2$; and (3) \textit{amplitude damping} with excitation-loss probability $p$. We also consider two correlated noise models: (4) \textit{correlated phase flip}, $\rho \mapsto (1-p)\rho + p(\sigma_3 \otimes \sigma_3)\rho(\sigma_3 \otimes \sigma_3)$; and (5) \textit{correlated amplitude damping}, in which only the joint excitation decays, $|11\rangle \xrightarrow{p} |00\rangle$, implemented via Kraus operators $K_0 = \mathrm{diag}(1,1,1,\sqrt{1-p})$ and $K_1 = \sqrt{p}\,|00\rangle\langle11|$. For each channel and initial state, we vary the noise strength on a uniform grid $p \in [0,1]$ and compute the evolved state $\rho(p) = \sum_k K_k(p)\,\rho(0)\,K_k^\dagger(p)$. We then extract the relevant quantities from $\rho(p)$ and visualise their evolution as parametrised decoherence trajectories in the purity-budget plane in Fig.~\ref{fig:noise}.\bigskip

\subsection{Code availability}
\noindent
The codes to generate the figures are available at \href{https://github.com/mitra-subhadip/Kinematic-Budget-of-Quantum-Correlations}{https://github.com/mitra-subhadip/Kinematic-Budget-of-Quantum-Correlations}.

\subsection{Acknowledgements}
\noindent
We thank Samyadeb Bhattacharya, Shantanav Chakraborty, Mayank Goel, and Harjinder Singh for their comments on the manuscript.

\section{Appendix}
\setcounter{section}{0}
\renewcommand{\theequation}{A\arabic{equation}}
\renewcommand{\thesection}{\Alph{section}}

\noindent
Below, we illustrate some beyond-two-qubit budget geometries and discuss an active process that can overcome the natural arrow of decoherence.

\subsection{Appendix A: Budget geometry beyond two qubits}\label{app:a}
\subsection{Qubit-qutrit budget geometry}
\noindent 
For a bipartite qubit-qutrit ($S_1$-$S_2$) system ($D=6$), the total budget $B=6P-1$ splits into local ($B_{\rm L}=2P_{S_1}+3P_{S_2}-2$) and nonlocal ($B_{\rm NL}$) components (see Fig.~\ref{fig:23geometry}). So the rationalised coordinates are $X=\sqrt{B_{\rm L}/5P}$ and $Y=\sqrt{B_{\rm NL}/5P}$, and $R=X^2+Y^2=(6P-1)/5P$. Setting $B_{\rm L}=0$ strictly necessitates maximally mixed marginals: $\rho_{S_1}=\mathbb{I}_2/2$, $\rho_{S_2}=\mathbb{I}_3/3$. 
\begin{itemize}[leftmargin=10pt]
    \item[--] {\bf The QC envelope:} When the qutrit acts as a classical control, maximising global purity forces the conditional qubit states into a symmetric \emph{trine} configuration on the Bloch sphere, setting a strict nonlocal capacity limit of $B_{\rm NL}^{\rm cor}=1$. The absolute classical-quantum boundary forms a piecewise linear convex hull spanning the trine state ($B_{\rm L}=0, B_{\rm NL}=1$), a rank-two dimer mixture of orthogonal pure product states $(1/2, 3/2)$, and the pure product peak $(3,2)$.

    \item[--] {\bf The C$^2$ envelope:} It starts at $B_{\rm NL}=2/3$ on the $B_{\rm NL}$ axis and follows a curve that increases as $\sqrt{B_{\rm L}}$ until $B_{\rm L}=1/8$, where it merges with the QC envelope.

    \item[--] {\bf Frustrated-entanglement envelope:} Dimensional asymmetry ($d_{S_1}\neq d_{S_2}$) bottlenecks this system from reaching the correlation apex. The qubit lacks the necessary degrees of freedom required to maximally entangle with the larger space of the qutrit. This continuous maximisation generates an analytic boundary in the budget plane, tracing $B_{\rm L}=B_{\rm NL}+2-\sqrt{8B_{\rm NL}}$ for $B_{\rm NL}\in[2,9/2]$. This is achieved by the rank-two mixed-state ansatz $\rho_{ent}=w|\Phi^+\rangle\langle\Phi^+|+(1-w)\sigma$, where $|\Phi^+\rangle\langle\Phi^+|$ is a Bell state in the $2\otimes2$ subspace of the qutrit and $\sigma=I_2/2\otimes|2\rangle\langle2|$ rigidly weighted at $w=2/3$ to satisfy the marginals and to trace the absolute correlation capacity. This yields a maximum nonlocal budget of $B_{\rm NL}^{\max}=2$ at $B_{\rm L}=0$. 

    \item[--]{\bf The feasibility wall:} For obtaining the outermost $Q=0$ boundary, trajectories traced by one-sided local depolarisation depend on which subsystem acts as the pure anchor satisfying $Q=0$. Fixing a pure qubit anchor while depolarising the qutrit generates a rigid vertical boundary at $X^2=3/5$. Conversely, anchoring the qutrit in a time-odd pure state and depolarising the qubit forces the state along a stretched ellipse: $2X^2+Y^2=8/5$. Because this odd-qutrit-anchored ellipse extends to $X^2=4/5$ on the $B_{\rm L}$ axis, it defines the outer positivity boundary. The massive local budget of the pure qutrit physically shields the mixed qubit, pushing the feasible geometry further outward. This illustrates how one can access the budget geometry using physical processes.
\end{itemize}

\subsection{Two-qutrit budget geometry}
\noindent
For a symmetric two-qutrit ($S_1$-$S_2$) system ($D=9$), we have $B=9P-1$ and $B_{\rm L}=3P_{S_1}+3P_{S_2}-2$ (see Fig.~\ref{fig:33geometry}). The rationalised coordinates are $X=\sqrt{B_{\rm L}/8P}$ and $Y=\sqrt{B_{\rm NL}/8P}$, with $R=X^2+Y^2=(9P-1)/8P$. The geometry spans structurally from the pure product anchor at $B_{\rm L}=B_{\rm NL}=4$ ($X=Y=\sqrt{1/2}$) to the maximally entangled state at the theoretical apex ($X=0, Y=1$).
\begin{itemize}[leftmargin=10pt]
    \item[--] {\bf The C$^2$/QC envelopes:} Because the system is dimensionally symmetric, the all-classical (C$^2$) and the quantum-classical (QC) capacity limits fully coincide. This boundary is a straight line connecting the pure product state to the absolute correlation floor. At the floor, maximal classical correlation with zero local information forces $P=1/3$, yielding a total budget strictly confined to the nonlocal sector ($0,2$). Analytically, the envelope can be expressed as $B_{\rm NL}=B_{\rm L}/2+2$, which translates to the ellipse $2X^2+4Y^2=3$. 

    \item[--] {\bf The feasibility wall :} Unlike the rigid vertical wall in two-qubit systems, the expanded $SU(3)$ algebra forces the qutrit positivity boundary to curve outward, defining an absolute feasibility limit governed by the ellipse $2X^2+Y^2=3/2$. This elliptical boundary physically manifests as the dynamic trajectory of one-sided depolarisation. Because the reflection overlap factorises for product states, anchoring one qutrit in a perfectly odd ($Q_{S_1}=0$) pure state globally locks the state to the positivity threshold. Depolarising the unanchored subsystem forces the bipartite state to slide exactly along this maximally protruding wall, pushing the feasible geometry to the widest point ($X=\sqrt{3/4}, Y=0$) or ($B_{\rm L}=2, B_{\rm NL}=0$), where mixed product states lie.
\end{itemize}

\subsection{Three-qubit budget geometry}
\noindent
For a tripartite three-qubit ($S_1$-$S_2$-$S_3$) system, we have $D=8$. Hence, $B=8P-1$, $B_{\rm L}=2(P_{S_1}+P_{S_2}+P_{S_3})-3$ (see Fig.~\ref{fig:222geometry}), $X=\sqrt{B_{\rm L}/7P}$, and $Y=\sqrt{B_{\rm NL}/7P}$ with $R=X^2+Y^2=(8P-1)/7P$. The geometric manifold is anchored structurally by pure-product states at $(B_{\rm L}=3, B_{\rm NL}=4)$ and spans outward to the maximally entangled Greenberger-Horne-Zeilinger (GHZ) state~\cite{Greenberger:1989tfe} at $(B_{\rm L}=0, B_{\rm NL}=7)$ or $(X=0, Y=1)$.
\begin{itemize}[leftmargin=10pt]
    \item[--] {\bf The C$^3$/QC$^2$ envelopes:} In this case, the all-classical and the QC$^2$ envelopes are the same. The boundary connects the pure product anchor to a maximally correlated classical mixture ($0,3$), tracing the line $B_{\rm NL}=B_{\rm L}/3+3$ in the $(B_{\rm L},B_{\rm NL})$  plane. It translates to the ellipse $3Y^2+2X^2=18/7$. Operationally, this boundary is traced by a global synchronisation channel -- a convex mixing strategy defined as $\Lambda_p(\rho)=(1-p)\rho_{\rm pure}+p\sigma_{\rm sync}$ for $p\in[0,1]$ -- which physically interpolates between the fully factorised pure state and the absolute classical correlation floor. States above this line are entangled between one pair of qubits.

    \item[--] {\bf The Q$^2$C envelope:} Restricting one qubit to act as a classical control while permitting bipartite entanglement between the remaining two dictates the absolute Q$^2$C capacity limit. This boundary is a straight line connecting the pure product anchor ($3, 4$) to the biseparable peak ($1, 6$) in the $(B_{\rm L},B_{\rm NL})$ plane. Physically, this boundary can be traced by subjecting the unentangled qubit of a pure biseparable state (e.g., $|0\rangle_{S_1}\otimes|\Phi^+\rangle_{S_2S_3}$) to local depolarising noise, driving the state to the mixed entangled floor. Above the line, all three qubits are GME.

    \item[--] {\bf The feasibility wall :} This outermost positivity wall mathematically manifests as a piecewise boundary that corresponds to the physical trajectory of sequential, one-by-one local depolarisation of a pure, fully factorised state where every qubit satisfies $Q_{S_i}=0$. Depolarising any qubit (say, $S_1$) while locking $S_2$ and $S_3$ traces an elliptical roof: $2X^2+Y^2=10/7$, mapping the segment $B_{\rm NL}=3B_{\rm L}-5$. Subsequently, mixing $S_2$ takes the slope to exactly unity, $B_{\rm NL}=B_{\rm L}-1$. This causes the $Y$-dependence to cancel out perfectly, dropping a rigid vertical cliff straight down to the $X$-axis at $X^2=4/7$. Any geometric excursion beyond this multi-segmented wall is unphysical.
\end{itemize}

In $2 \otimes 2 \otimes 2$ systems, crossing the QC$^2$ envelope prevents the excess nonlocal correlation from hiding entirely within the three-body tensor. Instead, it manifests in the bipartite marginals -- and the two-qubit guarantees apply directly to these marginals, enforcing both steering and violation of the CHSH inequality. This capacity spillover presents a compelling quantum marginal problem for future analytical study.

\subsection{Open-system dynamics and decoherence flows}
\noindent
Open-system dynamics become kinematic flows within the purity-budget projection. Extending our trajectory analysis to asymmetric ($2\otimes3$) and symmetric ($3\otimes3$ and $2\otimes2\otimes2$)  higher-dimensional systems confirms that the kinematic constraints are not dimension-specific (see Figs.~\ref{fig:QubitQutritNoise}, \ref{fig:QutritQutritNoise}, \ref{fig:ThreeQutritNoise}, respectively). Under local noise (e.g., depolarising or pure dephasing), trajectories are degradative: pure states contract radially inward toward the maximally mixed origin, or collapse downward by suppressing nonlocal coherences. Conversely, correlated amplitude damping redistributes resources. By dissipating local weight faster than nonlocal weight, this channel induces pronounced trajectory curvature, shielding nonclassical correlations and allowing states to temporarily survive or, in some cases, pass into the guaranteed entanglement regions despite an overall loss of global purity.

\setcounter{subsection}{1}

\subsection{Appendix B: Purification as active-kinematic flow}\label{app:b}
\noindent
While environmental coupling typically induces decoherence by contracting the purity budget radially inward, active purification and error-mitigation protocols forcefully reverse this empirical arrow. Algorithmic interventions, such as thermal purification, dynamically reclaim second-moment capacity, driving states outward toward absolute kinematic boundaries. Simulating coupling to a fictitious zero-temperature bath via the non-linear map $\rho\mapsto\rho^n/\mathrm{tr}(\rho^n)$ strictly inflates the statistical weight of the largest eigenvalues, functioning as a global purity pump. This exponentiation drives the state nonisotropically outward to the $R=1$ pure-state perimeter, curving through the projection to actively redistribute the budget until terminating exactly at the coordinates of the dominant pure eigenvector. While physically unrealisable as a deterministic single-copy operation, this exact mathematical flow is practically executed via virtual distillation~\cite{Huggins:2020pgg}. The engineered process shows violations of the empirical no-go rule, see Fig.~\ref{fig:Purification}.

\bibliography{budget}
\end{document}